\documentclass[aps,prd,reprint,preprintnumbers,superscriptaddress,showpacs,nofootinbib]{revtex4-2}
\usepackage[utf8]{inputenc}
\usepackage{parskip}
\usepackage{amssymb}
\usepackage{hhline}
\usepackage{amsmath}
\usepackage{bm}
\usepackage{xcolor}
\usepackage{xspace}
\usepackage{multirow,tabularx}
\usepackage{siunitx}
\usepackage{textgreek}
\usepackage{upgreek}
\usepackage{multirow}
\usepackage{graphicx}
\usepackage{pifont}
\usepackage{xstring}
\usepackage{etoolbox}
\usepackage{stmaryrd}

\DeclareMathAlphabet{\mathscr}{OT1}{pzc}{m}{it}

\newcolumntype{V}{>{\centering\arraybackslash} m{0.5cm} }
\DeclareMathOperator{\sgn}{sgn}
\DeclareSIUnit\parsec{pc}

\newrobustcmd{\numbercosmicclass}{14}%
\newrobustcmd{\cosmicclass}[1]{%
\IfEqCase{#1}{%
{1}{Class \textsuperscript{3}E\xspace}%
{2}{Class \textsuperscript{2}A\xspace}%
{8}{Class \textsuperscript{4}H\xspace}%
{9}{Class \textsuperscript{5}M\xspace}%
{10}{Class \textsuperscript{5}M\xspace}%
{11}{Class \textsuperscript{4}I\xspace}%
{12}{Class \textsuperscript{4}J\xspace}%
{14}{Class \textsuperscript{4}H\xspace}%
{13}{Class \textsuperscript{4}K\xspace}%
{16}{Class \textsuperscript{3}C\xspace}%
{15}{Class \textsuperscript{3}D\xspace}%
{21}{Class \textsuperscript{3}F\xspace}%
{20}{Class \textsuperscript{3}F\xspace}%
{23}{Class \textsuperscript{4}L\xspace}%
{22}{Class \textsuperscript{3}F\xspace}%
{26}{Class \textsuperscript{4}N\xspace}%
{27}{Class \textsuperscript{3}E\xspace}%
{24}{Class \textsuperscript{3}F\xspace}%
{25}{Class \textsuperscript{3}F\xspace}%
{30}{Class \textsuperscript{3}E\xspace}%
{31}{Class \textsuperscript{3}G\xspace}%
{28}{Class \textsuperscript{4}N\xspace}%
{29}{Class \textsuperscript{2}A\xspace}%
{35}{Class \textsuperscript{3}E\xspace}%
{34}{Class \textsuperscript{3}G\xspace}%
{33}{Class \textsuperscript{4}N\xspace}%
{32}{Class \textsuperscript{4}N\xspace}%
{39}{Class \textsuperscript{3}G\xspace}%
{38}{Class \textsuperscript{3}G\xspace}%
{37}{Class \textsuperscript{2}A\xspace}%
{36}{Class \textsuperscript{2}B\xspace}%
{40}{Class \textsuperscript{2}B\xspace}%
{41}{Class \textsuperscript{2}A\xspace}%
{null}{Class \textsuperscript{3}C*\xspace}%
}[\packageError{cosmicclass}{Unidentified Critical Case: #1}{}]%
}
\newrobustcmd{\criticalcase}[1]{%
\IfEqCase{#1}{%
{1}{Case 1\xspace}%
{2}{Case 2\xspace}%
{8}{Case 8\xspace}%
{9}{Case \textsuperscript{*1}9\xspace}%
{10}{Case \textsuperscript{*3}10\xspace}%
{11}{Case \textsuperscript{*4}11\xspace}%
{12}{Case 12\xspace}%
{14}{Case 14\xspace}%
{13}{Case \textsuperscript{*2}13\xspace}%
{16}{Case 16\xspace}%
{15}{Case 15\xspace}%
{21}{Case 21\xspace}%
{20}{Case 20\xspace}%
{23}{Case 23\xspace}%
{22}{Case 22\xspace}%
{26}{Case \textsuperscript{*6}26\xspace}%
{27}{Case 27\xspace}%
{24}{Case 24\xspace}%
{25}{Case \textsuperscript{*5}25\xspace}%
{30}{Case \textsuperscript{*7}30\xspace}%
{31}{Case \textsuperscript{*8}31\xspace}%
{28}{Case 28\xspace}%
{29}{Case 29\xspace}%
{35}{Case \textsuperscript{*9}35\xspace}%
{34}{Case 34\xspace}%
{33}{Case 33\xspace}%
{32}{Case 32\xspace}%
{39}{Case 39\xspace}%
{38}{Case 38\xspace}%
{37}{Case 37\xspace}%
{36}{Case \textsuperscript{*10}36\xspace}%
{40}{Case 40\xspace}%
{41}{Case 41\xspace}%
}[\packageError{criticalcase}{Unidentified Critical Case: #1}{}]%
}

\usepackage{hyperref}
\hypersetup{
     colorlinks = true,
     linkcolor = black,
     citecolor = black,
     filecolor = blue,
     urlcolor = blue
     }
\usepackage[capitalize]{cleveref}

\begin{document}

\title{Systematic study of background cosmology in unitary Poincar\'e gauge theories with application to emergent dark radiation and $H_0$ tension}

\author{W.E.V. Barker}
\email{wb263@cam.ac.uk}
\affiliation{Astrophysics Group, Cavendish Laboratory, JJ Thomson Avenue, Cambridge CB3 0HE, UK}
\affiliation{Kavli Institute for Cosmology, Madingley Road, Cambridge CB3 0HA, UK}
\author{A.N. Lasenby}
\email{a.n.lasenby@mrao.cam.ac.uk}
\affiliation{Astrophysics Group, Cavendish Laboratory, JJ Thomson Avenue, Cambridge CB3 0HE, UK}
\affiliation{Kavli Institute for Cosmology, Madingley Road, Cambridge CB3 0HA, UK}
\author{M.P. Hobson}
\email{mph@mrao.cam.ac.uk}
\affiliation{Astrophysics Group, Cavendish Laboratory, JJ Thomson Avenue, Cambridge CB3 0HE, UK}
\author{W.J. Handley}
\email{wh260@cam.ac.uk}
\affiliation{Astrophysics Group, Cavendish Laboratory, JJ Thomson Avenue, Cambridge CB3 0HE, UK}
\affiliation{Kavli Institute for Cosmology, Madingley Road, Cambridge CB3 0HA, UK}

\date{\today}

\begin{abstract}
  We propose a one-parameter extension to \textLambda CDM, expected to strongly affect cosmological tensions. An effective dark radiation component in the early universe redshifts away as hot dark matter, then quintessence, tracking the dominant equation-of-state parameter and leaving a falsifiable torsion field in the current epoch. 
  This picture results from a new Poincar\'e gauge theory (PGT), one of the most promising among the latest batch of 58 PGTs found to be both power-counting renormalisable and free from ghosts and tachyons. 
  We systematically categorise the cosmologies of 33 of these PGTs, as special cases of the most general parity-preserving, Ostrogradsky-stable PGT with a purely Yang-Mills action. 
  The theory we consider contains two propagating massless gravitons, which may be $J^P=2^+$ (long-range gravitation and gravitational waves). 
  A conspiracy among the coupling constants eliminates the spatial curvature $k\in\{\pm 1,0\}$ from the field equations. 
  We show that this `$k$-screening' is not restricted to conformal gravity theories. 
  The flat Friedmann equations are then emergent, with potentially tension-resolving freedom at the early scale-invariant epoch that reliably gives way to an attractor-like state of modern \textLambda CDM evolution.
  We compare with related theories and promising special cases, such as $k$-screened theories with negative-definite effective $k$, and more traditional theories with effective $\Lambda$ and a $J^P=0^-$ massive graviton (dark matter candidate).
  As a bonus, we analyse similarly constrained actions in the new extended Weyl gauge theory (eWGT). 
  We show that in cosmology, PGT and eWGT span exactly the same classical phenomenology up to a linear map between their coupling constants, hinting at a deeper relationship between the two.
\end{abstract}


\maketitle
\section{Introduction}\label{intro}
Once constrained by the strong cosmological principle, the geometry of the universe is free to vary in two ways according to the FRW metric
\begin{equation}\label{frw}
  \mathrm{d}s^2=\mathrm{d}t^2-\frac{R^2\mathrm{d}r^2}{1-kr^2}-R^2r^2(\mathrm{d}\upvartheta^2+\sin^2\upvartheta\mathrm{d}\upvarphi^2).
\end{equation}
On the one hand \textit{space}, defined by Cauchy surfaces containing material fluids at rest and spanned by dimensionless $r$, $\upvartheta$ and $\upvarphi$, has curvature constant $k$ equal to $1$, $0$ or $-1$. On the other \textit{time}, here the dimensionful cosmic time $t$, distinguishes those same surfaces and parametrises the evolution of the dimensionful scale factor $R$ along with derivative quantities such as the Hubble number $H$ and deceleration parameter $q$ 
\begin{equation}
  H=\partial_tR/R, \quad q=-R\partial_t^2R/(\partial_tR)^2.
  \label{handq}
\end{equation}
Einstein's general relativity (GR) predicts the geodesic trajectory of light, according to which recent measurements have been used to establish that at the present epoch the universe is expanding, accelerating and either spatially flat or very large
\begin{equation}
  H_0>0, \quad q_0<0, \quad |k|/H_0^2R_0^2\ll1.
\end{equation}
The cosmic concordance, or \textLambda CDM model \cite{2018arXiv180401318S}, aims to reconcile these observations with the rest of GR, whose contemporary Friedmann equations can be written as
\begin{subequations}
\begin{gather}
  \mathsf{h}^2=\omega_\text{r}+\omega_\text{m}+\omega_\Lambda+\omega_k,\label{Heqn} \\
  q_0\mathsf{h}^2=\omega_\text{r}+\tfrac{1}{2}\omega_\text{m}-\omega_\Lambda\label{qeqn}.
\end{gather}
\end{subequations}
In these equations the Hubble number (or today's Hubble \textit{constant}) is normalised to $\mathsf{h}$
\begin{equation}
  \mathsf{h}=H_0/\mathsf{H}, \quad \mathsf{H}=100\si{\km\per\second\per\mega\parsec},
  \label{littlehdef}
\end{equation}
while a material (non-gravitational) density $\rho_i$ gives rise to a contemporary dimensionless density according to
\begin{equation}
  \Omega_{i,0}=\kappa\rho_{i,0}/3H_0^2, \quad \omega_i=\Omega_{i,0}\mathsf{h}^2.
  \label{neldef}
\end{equation}
In particular, radiation is only partly accounted for by the photons of the CMB
\begin{equation}
  \omega_\text{r}=\left(1+\tfrac{7}{8}\left(\tfrac{4}{11}\right)^{4/3}N_{\text{eff}}\right)\omega_\gamma,
\end{equation}
with neutrinos making up the remaining relativistic degrees of freedom $N_\text{eff}=N_{\nu,\text{eff}}$.
Matter, or pressureless dust, can be partitioned into its baryonic and cold dark matter (CDM) fractions
\begin{equation}
  \omega_\text{m}=\omega_\text{b}+\omega_\text{c},
\end{equation}
and dark energy is assumed to emerge from a comsological constant $\Lambda$.
The \textit{deceleration equation }\eqref{qeqn} may be obtained from \eqref{Heqn} so long as the dependence of the various material energy-densities on $R$ --  their equations of state $w_i=p_i/\rho_i$ -- are known. In particular, these are
\begin{equation}
  w_\text{r}=1/3,\quad w_\text{m}=0, \quad w_k=-1/3, \quad w_\Lambda=-1.
  \label{eos}
\end{equation}
It is worth noting that the \textit{energy balance} equation \eqref{Heqn} may be understood heuristically as a dimensionless statement of zero net energy density, in the sense that the Einstein tensor provides a formal and covariant notion of gravitational energy in GR, although such a picture remains deeply dissatisfying (see \cite{2019JMP....60e2504B} and references therein). Accordingly, we may write
\begin{equation}
  \omega_\text{r}+\omega_\text{m}+\omega_\Lambda+\omega_H+\omega_k=0,
  \label{Heqn2}
\end{equation}
where the final two dimensionless densities are strictly gravitational in origin: the accepted quantity 
\begin{equation}
  \omega_k=-k/R_0^2\mathsf{H}^2,
\end{equation}
conveys the energy stored in curled-up Cauchy surfaces, while we define
\begin{equation}
  \omega_H=-\mathsf{h}^2,
  \label{silly}
\end{equation}
which might be thought of as the kinetic energy density of such surfaces as they expand or contract. Overall, \eqref{Heqn2} encodes a central tenet of modern cosmology: that $R$-evolution is fundamentally dependent on $k$.

Since its inception, many authors \cite{2016PDU....12...56B} have expressed concern with the \textLambda CDM model. In particular the required substances known as dark matter and dark energy remain unaccounted for, while the comparability of their densities at the present epoch is deemed so unlikely that it has become known as the \textit{cosmic coincidence problem} \cite{2014EPJC...74.3160V}. Similarly, the \textit{flatness problem} must be resolved by bolting on a non-gravitational inflationary mechanism at early times \cite{2014PhRvD..89f3505H}. While such long-standing objections stem from naturalness and Occam's razor, in recent years the prospect of observational inconsistencies with \textLambda CDM has become a reality. 
These possible inconsistencies appear at homogeneous scales in the form of the \textit{Hubble tension} \cite{2019NatAs...3..891V} and \textit{curvature tension} \cite{2019arXiv190809139H,2019arXiv191102087D}, and affect structure formation through the \textit{small scale crisis} \cite{2017ARA&A..55..343B}. The first of these is probably the most severe. At the far end of the cosmic distance ladder, major observational endeavours such as WMAP \cite{2004PhRvD..69j3501T} and most recently Planck \cite{2018arXiv180706209P} have caused a low value of $H_0$ or $\mathsf{h}$ to be inferred from the CMB. More local measurements using Cepheid-calibrated supernovae data (SH0ES) \cite{2016ApJ...826...56R}, the tip of the red giant branch (TRGB) \cite{2019arXiv190800993Y,2019ApJ...882...34F}, combined electromagnetic and gravitational observation of neutron star mergers \cite{2020arXiv200211355D}, or multiply lensed quasar systems (H0LiCOW) \cite{2019arXiv190704869W} indicate a somewhat higher value. Moreover, the situation has been exacerbated by each generation of experiments \cite{2019NatRP...2...10R}. By one current estimate \cite{2019ApJ...876...85R}, the $H_0$ discrepancy has placed \textLambda CDM in jeopardy to the tune of $4.4\sigma$.

In the present work, we will motivate a modified gravity theory, the effect of which on the background cosmology can be packaged into an augmentation of \textLambda CDM, involving the addition of a small extra component $\omega_\text{eff}$. The equation of state parameter $w_\text{eff}$ of this extra component `tracks' the dominant cosmic fluid in \eqref{eos}, such that
\begin{equation}
\begin{gathered}
  w_\text{r,eff}=1/3, \quad w_\text{m,eff}=(1-1/\sqrt{3})/2,\\
  w_{\Lambda,\text{eff}}=-1/\sqrt{3}.
  \label{proposal}
\end{gathered}
\end{equation}
Since $w_\text{r,eff}=w_\text{r}$, while $w_\text{m,eff}>w_\text{m}$ and $w_{\Lambda,\text{eff}}>w_\Lambda$, the extra component manifests an injection of \textit{dark radiation} in the early universe which redshifts away nontrivially at later times. In this sense, it can be cast as an extra relativistic species $N_{\text{eff}}=N_{\nu,\text{eff}}+\Delta N_{\text{dr,eff}}$. Similar models have recently become very popular \cite{Pandey,2018PhRvD..97l3504P,2018JCAP...09..025M,2007A&A...471...65E} as a means to alleviate the $H_0$ tension. Some of these are in conflict with the observational constraints from Big Bang nucleosynthesis (BBN) or even from the CMB itself (see e.g. \cite{2019JCAP...10..029S,2020arXiv200208942B,PhysRevD.101.035031,2020JCAP...01..004S,2013JCAP...09..013V,2018JCAP...09..025M}). Of greater concern is the reliance of many of these models on ad hoc physics.

In our case, the extra component picture is effective, since it emerges from a motivated modified gravity theory. Such alternatives to GR are themselves very popular, and may variously seek to cast early and late-time inflation as emergent gravitational phenomena, or conveniently resolve other tensions and crises in \textLambda CDM. 
A deeper motivation to modified gravity is the incompatibility of GR with quantum mechanics, and this provides further constraints on the theory. In particular GR is not perturbatively renormalisable, and modifications which fix this tend to do so at the expense of unitarity \cite{2016arXiv161008744B}.

Amongst the modified gravity theories, the gauge gravities have a heritage dating back to before the golden age of GR \cite{2010RPPh...73e6901N}, and are presently undergoing a renaissance due in part to the advent of computer algebra \cite{2019PhRvD..99f4001L,Lin2,2002IJMPD..11..747Y,1999IJMPD...8..459Y,maple}. Rather than the internal $\mathrm{SU}(3)\times\mathrm{SU}(2)\times\mathrm{U}(1)$ group of the standard model, these theories gauge the assumed external symmetry group of spacetime, where the specific gauge gravity depends on the group of choice. The diffeomorphism invariance of GR already encodes the gauged translational symmetry group $\mathbb{R}^{1,3}$ \cite{1984ucp..book.....W}. The least controversial extension ought to be such translations in combination with proper, orthochronous Lorentz rotations $\mathbb{R}^{1,3}\rtimes\mathrm{SO}^+(1,3)$, which constitute the Poincar\'e group $\mathrm{P}(1,3)$. This results in the Poincar\'e gauge theory (PGT), e.g. of Kibble \cite{1961JMP.....2..212K}, Utiyama \cite{PhysRev.101.1597} and Sciama \cite{RevModPhys.36.463}. Typical formulations of PGT split the metric into the square of a translational gauge field and introduce a rotational gauge field into the affine connection. This process introduces a geometric quality on the spacetime known as \textit{torsion}, which is distinct from \textit{curvature}. The spacetime is then said to be of Riemann-Cartan type. A special case of PGT known as \textit{teleparallelism}, in some sense antipodal to diffeomorphism gauge theories such as GR or $f(\mathcal{R})$ gravity, is reached by replacing curvature with torsion altogether -- in this case the flat but twisted spacetime is of Weitzenb\"ock type \cite{blagojevic2002gravitation}. 

An expanded choice of symmetry group is that of Weyl $\mathrm{W}(1,3)$. In this case, spacetime is symmetric under all elements of the extended conformal group excluding special conformal transformations. As an extension to PGT this adds Weyl rescalings to the list of symmetries which need to be gauged, and results in Weyl gauge theory (WGT) on Weyl-Cartan spacetime \cite{1975NCimB..28..127K}. It is not entirely clear how the rotational gauge field should respond to Weyl rescalings, and WGT was recently \textit{extended} (eWGT) \cite{lasenby-hobson-2016} by promoting this freedom to an internal gauge symmetry (the so-called \textit{torsion-scale gauge}). The relationship between PGT, WGT and eWGT is explained in detail in \cite{lasenby-hobson-2016}. In a world with discrete mass spectra, it is accepted that the scale gauge symmetry, if present, must be broken. In WGT this is usually done explicitly (e.g. by fixing to the Einstein -- sometimes called `unitary' \cite{2020arXiv200300664A} -- gauge), but it is possible to re-cast the equations of both WGT and eWGT in terms of scale-invariant variables which eliminate the scale gauge freedom and the need for explicit symmetry breaking. It is not yet clear that either method is preferable, or if they differ in a physical or merely philosophical sense. 

A similar question surrounds the role of geometry in these gauge gravities: it is perfectly feasible to eliminate any combination of curvature, torsion and scale as geometric qualities of the spacetime in favour of field strengths on a spacetime without these qualities, finally arriving at gauge gravity on Minkowski spacetime. This raises serious questions only when topology is considered important\footnote{For example a wormhole is difficult to cast in the Minkowski interpretation, as is the entire apparatus of Penrose diagrams.}. For our purposes, we find the Minkowski interpretation to be the simplest basis for comparison between competing gauge gravities.

As with diffeomorphism gauge theory, gauge gravities in general enjoy a large freedom in their Lagrangian structure. Each gauged spacetime symmetry introduces a new field strength, but may impose restrictions on the field strength invariants appearing in the Lagrangian. Stable PGTs may be powered by a gravitational sector constructed from invariants of two gauge field strengths, the curvature tensor $\mathcal{  R}_{abcd}$ and torsion tensor $\mathcal{  T}_{abc}$. Since the rather well accepted standard model which began this discussion relies exclusively on Yang-Mills gauge theories of internal symmetry groups, it is extremely tempting to consider \textit{quadratic} invariants of these tensors. Within gauge gravities, the dependence of $\mathcal{  R}_{abcd}$ and $\mathcal{  T}_{abc}$ on the $n$th derivatives of the gravitational gauge fields is not as clean-cut as in the standard model, and so it is also considered acceptable to include \textit{linear} invariants. The only linear invariant within PGT is the Ricci scalar $\mathcal{  R}$ which alone constitutes the minimal gauge gravity extension to GR known as Einstein-Cartan theory (ECT). We refer to PGTs and eWGTs including all possible quadratic and linear invariants as PGT\textsuperscript{q} and eWGT\textsuperscript{q}.
Within PGT\textsuperscript{q} it is possible to roughly halve the dimensionality of the parameter space by imposing parity invariance on the gravitational sector, resulting in PGT\textsuperscript{q+} and, analogously, eWGT\textsuperscript{q+}. This approach is commonly used in the literature, and constrains the theory in a natural manner. It must however be noted that a subset of authors (see e.g. \cite{2011PhRvD..83b4001B}) reject it on the grounds of poor physical motivation.

Applications of gauge gravity to cosmology began in the early 1970s and now constitute a large and established literature, with many authors progressing well beyond formalism to obtain analytical and numerical results. The earliest attempts narrowly focus on ECT, with the opening move being made by Kopczy\'nski \cite{1972PhLA...39..219K} who showed that the algebraic spin-torsion interaction could remove the singularity at the Big Bang. The modern notion of cosmological torsion in general, which we discuss in \cref{gravv}, was established by Tsamparlis \cite{1979PhLA...75...27T} before the end of the decade. Full PGT\textsuperscript{q+} was incorporated by Minkevich in 1980 \cite{1980PhLA...80..232M}, who identified a set of generalised cosmological Friedmann equations (GCFEs) which result from a \textit{single} parameter constraint on the PGT\textsuperscript{q+} action. Minkevich remains singularly prolific in this field, and the GCFEs have since been intensively studied in the context of singularity removal \cite{2003gr.qc....10060M,2007AcPPB..38...61M}, inflation \cite{2006CQGra..23.4237M} and dark energy \cite{2009PhLB..678..423M,2011arXiv1107.1566G}, see also \cite{2013JCAP...03..040M}. The GCFEs have also been analysed in the context of metric-affine gauge theory (MAGT) \cite{2000CQGra..17.3045M}. The first thorough (and widely cited) exposition on the cosmology of PGT\textsuperscript{q+} was undertaken four years later by Goenner and M\"uller-Hoissen \cite{1984CQGra...1..651G}, although their examination of the parameter space was by no means exhaustive. For a comprehensive review of the literature prior to 2004, see \cite{2005NewAR..49...59P}. In 2005 some of us were involved in an isolated study of pure Riemann-squared theory (RST) \cite{lasenby-doran-heineke-2005}. Within PGT\textsuperscript{q+}, RST is a minimal quadratic alternative to ECT known to accommodate at least a Schwarzschild-de Sitter vacuum solution, and although the cosmological model suffers from scale invariance (more specifically \textit{normal scale invariance}, NSI), it admits emergent inflationary behaviour. 

Superficially, these early classical endeavours may convey the impression that all emergent gravitational phenomena are available for free: questions raised by \textLambda CDM are simply absorbed into the fine-tuning of the ten PGT Lagrangian parameters.
In 2008 quantum feasability entered in a seminal paper \cite{2008PhRvD..78b3522S} by Shie, Nester and Yo (SNY), who observed that the $0^+$ and $0^-$ torsional modes of PGT are naturally suited to cosmological investigation. Their PGT\textsuperscript{q+} Lagrangian was constructed to target the $0^+$ mode, and as such their quadratic Riemann sector contains only $\mathcal{  R}^2$. In the same year Li, Sun and Xi performed a numerical study of the system \cite{2009PhRvD..79b7301L}. Chen, Ho, Nester, Wang and Yo later augmented their Lagrangian with the square pseudoscalar Riemann term in order to include the $0^-$ mode \cite{2009JCAP...10..027C}. Significant advances to the SNY Lagrangian were made in 2011 when Baekler, Hehl and Nester (BHN) included the parity-violating terms of PGT\textsuperscript{q} \cite{2011PhRvD..83b4001B}. The cosmological implications of all parity-violating \textit{shadow world} terms and parity-preserving \textit{world} terms were distilled by means of cosmologically harmless parameter constraints into their representative BHN Lagrangian. This work was still being explored by the same authors in 2015, see \cite{2011JPhCS.330a2005H,2011IJMPD..20.2125H,2011CQGra..28u5017B,2015arXiv151201202H}. Further work on the parity-preserving SNY Lagrangian was performed by Ao and Li in 2012 \cite{2011CQGra..28u5017B}. Most recently, Zhang and Xu (ZX) in \cite{2019arXiv190403545Z,2019arXiv190604340Z} have proposed a parameter constraint similar to that of Minkevich on PGT\textsuperscript{q+} which suggests a pleasing inflationary formalism. We note that the apparent trend toward quadratic Lagrangia is not universal, as ECT remains popular to this day \cite{2018IJMPD..2747020P,2019arXiv191108232M} as a simple way to import torsion, albeit algebraically bound to spin. Moreover, other authors have considered cosmological models with torsion which do not quite fit into the PGT\textsuperscript{q} category, such as $f(\mathcal{  R})$ and $\mathcal{  R}^n$ PGTs, see for example \cite{2019arXiv190604340Z}.

The theoretical development of eWGT was first introduced to the community in 2016, and from the outset it has been clear that structure of eWGT has more in common with PGT than WGT (for a recent incorporation of scale invariance to PGT\textsuperscript{q+}, see \cite{2020arXiv200300664A}). Indeed PGT\textsuperscript{q+} and eWGT\textsuperscript{q+} both sport ten Lagrangian parameters\footnote{For this reason, we will not attend to WGT cosmology.}. In the present work, which represents the first application of eWGT to cosmology, we aim to show that PGT\textsuperscript{q+} and eWGT\textsuperscript{q+} are cosmologically equivalent.

The remainder of this paper is structured as follows. In \cref{gaugetheories} we briefly explain the Minkowski interpretation of the two gauge gravities under consideration, PGT\textsuperscript{q+} and eWGT\textsuperscript{q+}, as in \cite{lasenby-hobson-2016}. In \cref{particles2w} we review the `cutting-edge' of PGT\textsuperscript{q+} quantum feasibility, as contained within our major references \cite{2019PhRvD..99f4001L,Lin2}. In \cref{cosmoans} we adapt the minisuperspace formalism to PGT\textsuperscript{q+} and eWGT\textsuperscript{q+} cosmology and set out a cosmological correspondence between the actions of the two theories. 

Our central results are confined to \cref{cosmo}. The generalised Friedmann equations, which are common to eWGT\textsuperscript{q+} and PGT\textsuperscript{q+}, are dissected in the context of quantum feasibility in \cref{map_section}, and the consequent $k$-screening in \cref{ksc_sec}. The new cosmology behind \eqref{proposal} is then developed in \cref{permitted}. Before concluding in \cref{conc}, we briefly discuss the application of Clifford algebra to general quadratic invariants in \cref{multi}. 
There follows a list of the spin projection operators (SPOs) used for \cite{2019PhRvD..99f4001L,Lin2} in \cref{spos}, a comparison to part of the literature mentioned above in \cref{chao}, and certain cumbersome functions in \cref{cc16}.

We provide a list of potentially nonstandard abbreviations in \cref{acro}. As far as possible we will adhere to the notation of \cite{lasenby-hobson-2016}. This entails the use of natural units, $c=\hbar=1$, in which energy has units $\si{\electronvolt}$ and the Einstein constant, $\kappa$, is used to account for dimensionality where necessary, though occasionally we revert to the reduced Planck mass, $M_\text{P}=\kappa^{-1/2}$. The signature is $\left( +,-,-,- \right)$.
\begin{table}
  \caption{\label{acro} Potentially nonstandard abbreviations.}
\begin{center}
\begin{tabularx}{\linewidth}{c|X}
\hline\hline
 PGT & Poincar\'e gauge theory \\
 MAGT & metric-affine gauge theory \\
 WGT & Weyl gauge theory \\
 eWGT & extended Weyl gauge theory \\
 PGT\textsuperscript{q}, eWGT\textsuperscript{q} & general PGTs and eWGTs with Lagrangia at most \textit{quadratic} in field strengths \\
 PGT\textsuperscript{q+}, eWGT\textsuperscript{q+} & general PGT\textsuperscript{q}s and eWGT\textsuperscript{q}s with \textit{parity-preserving} Lagrangia \\
 ECT & Einstein-Cartan theory \\
 RST & Riemann-squared theory \\
 GTG & gauge theory gravity \\
 STA & spacetime algebra \\
 SPO & spin-projection operator \\
 PCR & power-counting renormalisable \\
 NSI & normally scale-invariant \\
\hline\hline
\end{tabularx}
\end{center}
\end{table}
\section{Gauge theories}\label{gaugetheories}
\subsection{Symmetries, transformation laws and field strengths}\label{recap}
Gauge gravities may be cast (almost) without loss of generality in a manifold $\mathcal{M}$ with Minkowskian geometry. This Minkowski interpretation was pioneered by Kibble \cite{1961JMP.....2..212K} and later Lasenby \textit{et al} \cite{1998RSPTA.356..487L} and Blagojevi\'c \cite{blagojevic2002gravitation}, and used extensively in the initial proposal for eWGT \cite{lasenby-hobson-2016}. There is a potentially curvilinear coordinate system $\{x^\mu\}$ in this spacetime, with coordinates considered to be functions of the points of the manifold, and all fields written as functions of the coordinates. From the $\{x^\mu\}$ there is defined a basis of tangent vectors $\{\bm{e}_\mu\}$ and cotangent vectors $\{\bm{e}^\mu\}$ in the usual manner. The necessarily flat metric on $\mathcal{M}$, which is \textit{not} a gravitational gauge field, is then $\bm{e}_\mu\cdot\bm{e}_\nu=\gamma_{\mu\nu}$. The first gauge symmetry to consider is that of diffeomorphisms, though these are interpreted as passive general coordinate transformations (GCTs). Particularly, physical quantities should have zero \textit{total} (as supposed to \textit{form}) variations under GCTs. Taking new coordinates, $\{x'^\mu\}$, the covariance of a scalar matter field\footnote{The generic matter field $\varphi$ should not be confused with the azimuthal angle $\upvarphi$.} is expressed as
\begin{eqnarray}\label{passive_tran}
  \varphi'(x')=\varphi(x),
\end{eqnarray}
with the expected transformation of other quantities
\begin{eqnarray}\label{passive_trans}
  \bm{e}'^\mu=\frac{\partial x'^\mu}{\partial x^\nu}\bm{e}^\nu, \quad\bm{e}'_\mu=\frac{\partial x^\nu}{\partial x'^\mu}\bm{e}_\nu,\quad \partial'_\mu=\frac{\partial x^\nu}{\partial x'^\mu}\partial_\nu.
\end{eqnarray}
Independently of the coordinate basis, there exists an orthonormal Lorentz basis $\{\hat{\bm{e}}_a\}$ and dual basis $\{\hat{\bm{e}}^a\}$, such that $\hat{\bm{e}}_a\cdot\hat{\bm{e}}_b=\eta_{ab}$. While Greek indices transform under the Jacobian matrices of GCTs, Roman indices transform under local Lorentz rotations $\Lambda^a_{\ b}$. 
Indices are converted by means of the translational gauge fields (analogous to the \textit{tetrads} of the geometrical interpretation) $h_a^{\ \mu}$ and $b^a_{\ \mu}$, which themselves transform according to their indices\footnote{The gauge field $h_a^{\ \mu}$ and in particular its determinant $h$ should not be confused with the normalised Hubble constant, $\mathsf{h}=H_0/\mathsf{H}$.}, and which satisfy
\begin{equation}
  h_a^{\ \mu}b^a_{\ \nu}=\delta^{\mu}_{\nu}, \quad h_a^{\ \mu}b^c_{\ \mu}=\delta^{c}_{a}.
\end{equation}
The matter field should of course be generalised to some higher-spin representation of the Lorentz group. A spacetime derivative, covariantised with respect to both gauge freedoms, can then be defined as
\begin{equation}
  \mathcal{  D}_a\varphi= h_a^{\ \mu}\left( \partial_\mu+\tfrac{1}{2}A^{cd}_{\ \ \mu}\Sigma_{cd} \right)\varphi,
  \label{pgtcd}
\end{equation}
where $A^{cd}_{\ \ \mu}$ is the spin connection and the $\Sigma_{ab}$ are the Lorentz group generators of the spin-specific representation of $\varphi$. Note that in this general representation the associated indices are suppressed. By convention, calligraphic script is used to highlight components of tensors defined purely with respect to the Lorentz frames, while normal script is used for mixed or purely coordinate frame definitions\footnote{This is especially useful in the present work, as we can always refer to $\mathcal{R}_{abcd}$ instead of $R_{\alpha\beta\mu\nu}$, and thus avoid confusion with the dimensionful scale factor, $R$.}. Thus, we note the required transformation properties of the spin connection under a pure Lorentz rotation
\begin{equation}
  \mathcal{  A}'^{ab}_{\ \ \ c}=\Lambda^d_{\ c}\big( \Lambda^a_{\ e}\Lambda^b_{\ f}\mathcal{  A}^{ef}_{\ \ \ d}-\Lambda^{be}h_d^{\ \nu}\partial_\nu\Lambda^a_{\ e} \big).
\end{equation}
The field strength tensors of PGT are then defined in the Yang-Mills sense
\begin{equation}
  2\mathcal{  D}_{[c}\mathcal{  D}_{d]}\varphi=\big(\tfrac{1}{2}\mathcal{  R}^{ab}_{\ \ \ cd}\Sigma_{ab}-\mathcal{  T}^a_{\ \ cd}\mathcal{  D}_a\big)\varphi,
\end{equation}
where the Riemann (rotational) field strength tensor is
\begin{equation}
  \begin{aligned}
    \mathcal{  R}^{ab}_{\ \ cd}=&h_c^{\ \mu}h_d^{\ \nu}(\partial^{\vphantom{ab}}_{[\mu} A^{ab}_{\ \ \ \nu]}+A^a_{\ e[\mu}A^{eb}_{\ \ \ \nu]}),
  \label{moon}
\end{aligned}
\end{equation}
and the torsion (translational) field strength tensor is
\begin{equation}
  \mathcal{  T}^a_{\ \ bc}=-2b^a_{\ \mu}\mathcal{  D}_{[b}^{\phantom{\mu}}h_{c]}^{\ \mu}.
  \label{sun}
\end{equation}
Under local Weyl transformations, the various PGT quantities are expected to transform as
\begin{equation}
  \varphi'=e^{w\rho}\varphi, \quad {h'}_a^{\ \mu}=e^{-\rho}h_a^{\ \mu}, \quad A'^{ab}_{\ \ \ \mu}=A^{ab}_{\ \ \ \mu},
  \label{wresc}
\end{equation}
where $w$ is the Weyl weight of the matter field. To arrive at WGT, the covariant derivative \eqref{pgtcd} must then be augmented with an extra Weyl gauge field. In eWGT, the spin connection obeys a more general transformation law
\begin{equation}
  A'^{ab}_{\ \ \ \mu}=A^{ab}_{\ \ \ \mu}-2\theta \eta^{c[a}b^{b]}_{\ \mu}h_c^{\ \nu}\partial_\nu\rho.
\end{equation}
The dimensionless parameter\footnote{The parameter $\theta$ should not be confused with the polar angle $\upvartheta$.} $\theta\in[0,1]$ is introduced to \textit{extend} the \textit{normal} transformation law of \eqref{wresc} to the \textit{special} alternative, including admixtures between the two in its range\footnote{Note that although the special transformation is defined as $\theta=1$, the apparatus of eWGT also functions outside the range $\theta\in[0,1]$.}.
The induced transformation of the PGT torsion contraction, $\mathcal{  T}_a=\mathcal{  T}^b_{\ \ ab}$, combined with another $\theta$-dependent transformation law for the Weyl gauge field
\begin{equation}
  T'_\mu=T_\mu+3(1-\theta)\partial_\mu\rho, \quad V'_\mu=V_\mu+\theta\partial_\mu\rho,
\end{equation}
allows a suitable eWGT covariant derivative to then be constructed
\begin{equation}
  \mathcal{  D}^\dagger_a\varphi= h_a^{\ \mu}\big( \partial_\mu+\tfrac{1}{2}{A^\dagger}^{cd}_{\ \ \mu}\Sigma_{cd} -wV_{\mu}-\tfrac{1}{3}wT_\mu\big)\varphi.
  \label{ewgtcd}
\end{equation}
In general, eWGT quantities are distinguished from PGT counterparts by an obelisk superscript: the eWGT spin connection is
\begin{equation}
  {\mathcal{  A}^\dagger}^{ab}_{\ \ \ c}=\mathcal{  A}^{ab}_{\ \ \ c}+2\mathcal{  V}^{[a}\delta^{b]}_c.
\end{equation}
By generalising \eqref{pgtcd} to \eqref{ewgtcd}, the translational and rotational gauge field strengths are themselves redefined, and the extra gauge symmetry introduces its own field strength tensor
\begin{equation}
  2\mathcal{  D}^\dagger_{[c}\mathcal{  D}^\dagger_{d]}\varphi=\big(\tfrac{1}{2}{ \mathcal{  R}^\dagger }^{ab}_{\ \ \ cd}\Sigma_{ab}-w{\mathcal{  H}^\dagger}_{cd}-{ \mathcal{  T}^\dagger }^a_{\ \ cd}\mathcal{  D}_a\big)\varphi.
\end{equation}
In particular the eWGT Riemann tensor differs from \eqref{moon} according to
\begin{equation}
  \begin{aligned}
    {\mathcal{  R}^\dagger}^{ab}_{\ \ \ cd}=&\mathcal{  R}^{ab}_{\ \ \ cd}+2\delta^{[b}_{d}(\mathcal{  D}_c+\mathcal{  V}_c)\mathcal{  V}^{a]}_{\phantom{d}}\\
    &-2\delta^{[b}_{c}(\mathcal{  D}_d+\mathcal{  V}_d)\mathcal{  V}^{a]}_{\phantom{c}}-2\mathcal{  V}^e\mathcal{  V}_e\delta^{[a}_{c\phantom{d}}\delta^{b]}_{d}\\
    &+2\mathcal{  V}^{[a}_{\phantom{d}}\mathcal{  T}^{b]}_{\ \ cd},
\end{aligned}
\end{equation}
while the eWGT torsion differs from \eqref{sun} according to
\begin{equation}
  {\mathcal{  T}^\dagger}^a_{\ \ bc}=\mathcal{  T}^a_{\ \ bc}+\tfrac{2}{3}\delta^a_{[b}\mathcal{  T}^{\phantom{a}}_{c]},
\end{equation}
and has the property that \textit{all} of its contractions vanish.
We will not give the precise form of the field strength ${\mathcal{  H}^\dagger}_{ab}$ associated with Weyl rescalings, since it is not used in the eWGT\textsuperscript{q+} actions which follow on the grounds of potential instablility.
\subsection{Restricted actions}\label{ras}
The PGT\textsuperscript{q+} Lagrangian density should be linear in gauge-invariant quantities with dimensions of energy density, $\si{\electronvolt^{4}}$. Displacement gauge invariance naturally demands that these quantities be tensor densities of rank zero, while parity invariance further eliminates pseudoscalar densities. We are therefore interested in scalars, which we can always convert to densities by combination with the factor $h^{-1}=1/\det(h_a^{\ \mu})$. Within the gravitational sector, we are free to use invariants of the field strengths up to second order. The only such first order term is that of Einstein and Hilbert, which we write as\footnote{Note that in \cite{lasenby-hobson-2016} the notation $\alpha_0=a$ is used, which we will require for the dimensionless scale factor, $a=R/R_0$.}
\begin{equation}
  L_{\mathcal{  R}}=-\tfrac{1}{2}\alpha_0\mathcal{  R},
\end{equation}
where $\alpha_0$ is a dimensionless parameter of the theory. Likewise, there are six such parameters in the quadratic Riemann sector
\begin{equation}
  \begin{aligned} 
  L_{{\mathcal{R}}^2}=&\alpha_1{\mathcal{R}}^2+\alpha_2{\mathcal{R}}_{ab}{\mathcal{R}}^{ab}+\alpha_3{\mathcal{R}}_{ab}{\mathcal{R}}^{ba}\\
  &+\alpha_4{\mathcal{R}}_{abcd}{\mathcal{R}}^{abcd}+\alpha_5{\mathcal{R}}_{abcd}{\mathcal{R}}^{acbd}\\
  &+\alpha_6{\mathcal{R}}_{abcd}{\mathcal{R}}^{cdab},
  \end{aligned}
  \label{inic}
\end{equation}
and three more in the quadratic torsion sector
\begin{equation}
  \begin{aligned}
    L_{{\mathcal{T}}^2}=\beta_1{\mathcal{T}}_{abc}{\mathcal{T}}^{abc}+\beta_2{\mathcal{T}}_{abc}{\mathcal{T}}^{bac}+\beta_3{\mathcal{T}}_a{\mathcal{T}}^a.
  \end{aligned}
  \label{inoc}
\end{equation}
We also reserve the freedom at this stage to introduce an ad hoc cosmological constant, $\Lambda\sim\si{\electronvolt^2}$. Anticipating various mechanisms which may give rise to an effective cosmological constant through the introduction of new dynamical fields, $\Lambda$ will not be re-cast as a dimensionless theory parameter, and will enter into the Lagrangian as
\begin{equation}
  L_\Lambda=-\Lambda.
\end{equation}
Finally, the various matter fields will couple to the gravitational gauge fields within their own Lagrangian densities: we will denote the resulting scalar simply as $L_\text{m}$. The general PGT\textsuperscript{q+} action thus has ten dimensionless parameters, and by introducing Einstein's constant to compensate for dimensionality we may write it as
\begin{equation}\label{pgtaction}
  \begin{aligned}
    S_\text{T}=\int\mathrm{d}^4xh^{-1}\big[&L_{\mathcal{  R}^2}\\
    &+\kappa^{-1}\left(L_{\mathcal{  T}^2}+L_{\mathcal{  R}}+L_\Lambda\right)+L_\text{m}\big].
  \end{aligned}
\end{equation}
The situation for eWGT\textsuperscript{q+} differs through the structure of the eWGT torsion tensor and the imposition of Weyl gauge invariance. The forms of $L_{{\mathcal{  R}^\dagger}}$ and $L_{{\mathcal{  R}^\dagger}^2}$ are identical to those of $L_{\mathcal{  R}}$ and $L_{\mathcal{  R}^2}$: one needs simply to replace the PGT Riemann tensor with its eWGT counterpart
\begin{equation}
  L_{\mathcal{  R}^\dagger}=-\tfrac{1}{2}\alpha_0\mathcal{  R}^\dagger,
\end{equation}
and likewise for the quadratic Riemann sector.
The quadratic torsion sector in eWGT\textsuperscript{q+} contains only \textit{two} degrees of freedom, because the eWGT torsion has identically vanishing contraction
\begin{equation}
  \begin{aligned}
    L_{{\mathcal{T}^\dagger}^2}=\beta_1{\mathcal{T}^\dagger}_{abc}{\mathcal{T}^\dagger}^{abc}+\beta_2{\mathcal{T}^\dagger}_{abc}{\mathcal{T}^\dagger}^{bac}.
  \end{aligned}
\end{equation}
The quadratic torsion and linear Riemann sectors cannot be directly admitted to the Lagrangian because their Weyl weight is too low. This can be fixed by multiplication with a \textit{compensator} field of dimension $\si{\electronvolt}$ and weight $w=1$
\begin{equation}
  \phi'=e^\rho\phi.
\end{equation}
The generally dynamical nature of the compensator field demands the addition of an extra Lagrangian contribution, which we write as a sum of kinetic and potential terms
\begin{equation}\label{lagphi}
  L_\phi=\tfrac{1}{2}\nu\mathcal{D}^\dagger_a\phi{\mathcal{D}^\dagger}^a\phi-\lambda\phi^4.
\end{equation}
The constraint on the Weyl weight of Lagrangian densities means that the second term in \eqref{lagphi} already functions as a suitably general cosmological constant, therefore $\nu$ is the only new dimensionless theory parameter. A final possibility is a term quadratic in the Weyl gauge field strength
\begin{equation}
  L_{{\mathcal{  H}^\dagger}^2}=\tfrac{1}{2}\xi{\mathcal{  H}^\dagger}_{ab}{\mathcal{  H}^\dagger}^{ab},
  \label{H2term}
\end{equation}
though in the present work we will take $\xi=0$ as the field strength is incompatible with the strong cosmological principle. Moreover, $\mathcal{  H}^\dagger_{ab}$ has the unusual property of containing second derivatives of the $h_a^{\ \mu}$ gauge field: such a structure \textit{might} be expected to introduce an Ostrogradsky instability to the equations of motion\footnote{We note however that there is some reason to believe \cite{lasenby-hobson-2016} that such problems, when caused by \eqref{H2term}, may be self-resolving in practice.}. This may be compared to candidate terms in the PGT Lagrangian, quadratic in the first derivatives of the PGT torsion: these are traditionally excluded on similar grounds. 
The matter coupling will in general differ between eWGT and PGT, so we denote the matter Lagrangian by $L_{\text{m}}^\dagger$ and write the total action as
\begin{equation}\label{ewgtaction}
  \begin{aligned}
    S_\text{T}=\int\mathrm{d}^4xh^{-1}\big[&L_{ {\mathcal{  R}^\dagger}^2}+\phi^2\left(L_{ {\mathcal{  T}^\dagger}^2}+L_{\mathcal{  R}^\dagger}\right)+L_\phi+L_{\text{m}}^\dagger\big].
  \end{aligned}
\end{equation}

Note that while eWGT incorporates scale invariance by guaranteeing homogeneous transformation of the covariant derivative $\mathcal{  D}_a^\dagger$, some choices of PGT action are naturally scale invariant despite the inhomogeneous transformation of $\mathcal{  D}_a$. In the context of PGT\textsuperscript{q+}, this holds for \textit{normally} scale invariant $L_\text{m}$ in combination with
\begin{equation}
  L_{\mathcal{  R}}=L_{\mathcal{  T}^2}=0,
  \label{<+label+>}
\end{equation}
or the theory parameter constraint
\begin{equation}
  \alpha_0=\beta_1=\beta_2=\beta_3=0.
  \label{NSI}
\end{equation}
This imposes severe restrictions on both the gravitational sector, which is confined to the quadratic Riemann sector, and the matter content, which is confined to radiation. We refer to such PGT\textsuperscript{q+}s as \textit{normally scale invariant} (NSI). 

In \cite{lasenby-hobson-2016} it is noted that more general NSI versions of PGT\textsuperscript{q+} can be formed by allowing for the compensator $\phi$ field in PGT to make up for weights in both gravitational and matter sectors, as with eWGT. So long as no term proportional to $\mathcal{  D}_a\phi\mathcal{  D}^a\phi$ is is added to the matter sector, the constraints \eqref{NSI} on the gravitational sector can then be relaxed because the only remaining concern is the inhomogeneous transformation of $\mathcal{  T}^a_{\ \ bc}$. This can be eliminated (up to a total derivative) by a specific restriction on the $\{\beta_i\}$
\begin{equation}
  2\beta_1+\beta_2+3\beta_3=0.
  \label{NSI_phi}
\end{equation}
In what follows, as a matter of convenience, we will confine the $\phi$ field to eWGT.

We see therefore that the PGT\textsuperscript{q+} and eWGT\textsuperscript{q+} both contain ten freedoms at the level of the theory, and possibly an eleventh freedom in the form of the cosmological constant. There is some subtlety regarding the true freedom of the quadratic Riemann sector in both cases, because of the Gauss-Bonnet identity, which states that the quantity
\begin{equation}\label{gaussbonnet}
  \mathcal{  G}= \mathcal{  R}^2-4\mathcal{  R}_{ab}\mathcal{  R}^{ba}+\mathcal{  R}_{abcd}\mathcal{  R}^{cdab},
\end{equation}
is a total derivative in $n\leq 4$ dimensions, as is the analogous quantity in eWGT. This allows us to set one of $\alpha_1$, $\alpha_3$ or $\alpha_6$ to zero without loss of generality. Since the invariance of physical results under a Gauss-Bonnet variation is a useful test, we will not make any such reduction for the purpose of simplifying calculations and instead maintain all six quadratic Riemann parameters as far as possible. 

Of greater relevance to the present work is the reparametrisation freedom under linear combinations: the $\{\alpha_i\}$, $\{\beta_i\}$ and $\nu$ are conveniently chosen to agree with the canonical form of tensor components. Unfortunately, this formulation does little to convey the effects of symmetry properties of the field strength tensors on the quadratic invariants. The symmetries of the Riemann tensor are of fundamental importance when comparing these torsionful theories to more traditional metrical alternatives, and with this in mind we will work with the following reparametrisation
\begin{equation}
  \begin{gathered}
    \check{\alpha}_0=\alpha_0, \quad \check{\alpha}_1=\alpha_1,  \quad \check{\alpha}_2=\alpha_2, \quad \check{\alpha}_3=\alpha_3,\\
    \check{\alpha}_4=2\alpha_4+\alpha_5, \quad \check{\alpha}_5=\alpha_5, \quad \check{\alpha}_6=2\alpha_6,\\
    \check{\beta}_1=-2\beta_1-\beta_2, \quad \check{\beta}_2=\beta_2, \quad \check{\beta}_3=\beta_3.
  \end{gathered}
  \label{newparams}
\end{equation}
These parameters drop out of a new scheme for expressing quadratic invariants, which we set out in \cref{multi}. Note that as with $\beta_3$, the term parametrised by $\check\beta_3$ vanishes identically in eWGT\textsuperscript{q+}.
\section{Ghosts, tachyons and loops}\label{particles2w}
The perturbative QFT of PGT\textsuperscript{q+} begins with the linearisation
\begin{equation}
\begin{gathered}
  h_a^{\ \mu}=\delta_a^\mu+f_a^{\ \mu},\quad b^a_{\ \mu}=\delta^a_\mu-f^a_{\ \mu}+\mathcal{  O}\left( f^2 \right),\\
  A^{ab}_{\ \ \mu}=\mathcal{  O}\left( f \right).
\end{gathered}
\end{equation}
The perturbative gravitational gauge fields with which we work are then
\begin{equation}
  \mathfrak{s}_{ab}=f_{(a}^{\ \mu}\eta^{\phantom{\mu}}_{b)\mu},\quad \mathfrak{a}_{ab}=f_{[a}^{\ \mu}\eta^{\phantom{\mu}}_{b]\mu},\quad \mathcal{  A}_{abc}=\delta_c^{\mu}A_{ab\mu},
  \label{dyfel}
\end{equation}
i.e. two four-tensor fields of rank two and one of rank three. Upon canonical quantization, in composition with states of definite momentum or position, the four-tensor content of these fields will be distributed amongst states of definite spin-parity $J^P$. The $J^P$ spectrum of any field is generally set by the rank $n$ of the four-tensor, which is a tensor product of $n$ four-vectors. Under a spacetime rotation $\Lambda^a_{\ b}$ confined to a spatial rotation orthogonal to some timelike vector $k^a$ the timelike part of the four-vector transforms as a $0^+$ state and the spacelike part as a $1^-$ state. A rank-two four-tensor such as $f_{ab}$ thus transforms as a state under the following equivalent representations of $\mathrm{SO}(3)$
\begin{equation}
  \begin{aligned}
(\mathbf{D}&(0^+)\oplus\mathbf{D}(1^-))\otimes(\mathbf{D}(0^+)\oplus\mathbf{D}(1^-))\\
&\simeq(\mathbf{D}(0^+)\otimes\mathbf{D}(0^+))\oplus(\mathbf{D}(0^+)\otimes\mathbf{D}(1^-))\\
&\phantom{\simeq}\oplus(\mathbf{D}(1^-)\otimes\mathbf{D}(0^+))\oplus(\mathbf{D}(1^-)\otimes\mathbf{D}(1^-))\\
&\simeq\mathbf{D}(0^+)\oplus\mathbf{D}(0^+)\oplus\mathbf{D}(1^-)\oplus\mathbf{D}(1^-)\\
&\phantom{\simeq}\oplus\mathbf{D}(1^+)\oplus\mathbf{D}(2^+),
\end{aligned}
\end{equation}
indicating that the tensor is a direct sum of two $0^+$, two $1^-$, one $1^+$ and one $2^+$ states. An analogous calculation reveals that a general rank-three four-tensor is a direct sum of four $0^+$, one $0^-$, three $1^+$, six $1^-$, three $2^+$, one $2^-$ and one $3^+$ states. By adding the multiplicities of the states $2J+1$ for either field one recovers the $4^2$ or $4^3$ tensor degrees of freedom, illustrating the completeness of the $J^P$ decomposition. 

In practice, the fields defined in \eqref{dyfel} contain \textit{a priori} symmetries which reduce their $J^P$ content.
Thus the $21$ $J^P$ sectors of $f_{ab}$ are neatly partitioned among the symmetric and antisymmetric parts $\mathfrak{s}_{ab}$ and $\mathfrak{a}_{ab}$. This procedure was historically applied to the symmetric perturbation of metrical gravity in order to classify $J^P$ graviton states. In PGT, the antisymmetric part of $f_{ab}$ introduces a $1^-$ and additional $1^+$ sector to the theory, though both $\mathfrak{s}_{ab}$ and $\mathfrak{a}_{ab}$ excitations are always considered \textit{gravitons}. The assumed antisymmetry of the spin connection $\mathcal{A}_{[ab]c}=\mathcal{A}_{abc}$ eliminates three $0^+$, one $1^+$, four $1^-$ and two $2^+$ sectors along with the curious $3^+$ sector -- excitations of the $\mathcal{  A}^a_{\ \ bc}$ field are sometimes called \textit{tordions}. In a general therefore, the gravitational particles of PGT remain maximally spin-$2$. 

It is worth noting that the distinction between symmetric and antisymmetric gravitons is rather artificial, as is the distinction between gravitons and tordions. This is because in many cases the various fields are related by gauge transformations or the excitations are coupled. The various $J^P$ components of all fields may be extracted by means of well-established spin projection operators (SPOs). In the case of the field $\mathcal{  A}_{abc}$, these generically take the form $\mathcal{  P}_{ij}(J^P)$, where the three Roman indices are suppressed and $i$ and $j$ label independent sectors with the same $J^P$. In particular, the diagonal elements $i=j$ form a complete set over all $J^P$ sectors in $\mathcal{  A}_{abc}$, and $i\neq j$ is only possible within the $1^-$ and $1^+$ sectors, since the direct sum contains \textit{two} independent representations of these $J^P$. In the remainder of this work, we will be working at the level of the torsion rather than the spin-connection. Within the linearised regime set out above, $\mathcal{  T}^a_{\ \ bc}$ and $\mathcal{  A}^{ab}_{\ \ \ c}$ are two sides of the same coin and related by the contortion
\begin{equation}
  \mathcal{  T}_{abc}=\mathcal{  N}_{abc}^{\ \ \ ijk}\mathcal{  A}_{ijk},\quad \mathcal{  N}_{abc}^{\ \ \ ijk}=2\delta^j_a\delta^i_{[c}\delta^k_{b]}.
\end{equation}
Thus all freedoms in the spin-connection are inherited by the torsion. It is natural that the $J^P$ sectors of one field map onto the other, indeed generally we find
\begin{equation}
 \mathcal{  N}_{abc}^{\ \ \ ijk}\mathcal{  P}_{nn}(J^P)_{ijk}^{\ \ \ def}\mathcal{  A}_{def}=\mathcal{  P}_{nn}(J^P)_{abc}^{\ \ \ ijk}\mathcal{  T}_{kji}.
  \label{nomix}
\end{equation}
Some nuance is however required in the case of the pseudovector tordion triplet, since $\mathcal{  N}$ does not commute with $\mathcal{  P}_{ij}(1^+)$. The correct mixing in this case is given by the off-diagonal SPOs
\begin{equation}
  \begin{aligned}
\mathcal{  N}_{abc}^{\ \ \ ijk}&\mathcal{  P}_{11}(1^+)_{ijk}^{\ \ \ def}\mathcal{  A}_{def}=\\
&\left(\mathcal{  P}_{22}(1^+)_{abc}^{\ \ \ ijk}-\tfrac{1}{\sqrt{2}}\mathcal{  P}_{12}(1^+)_{abc}^{\ \ \ ijk}\right)\mathcal{  T}_{kji},\\
\mathcal{  N}_{abc}^{\ \ \ ijk}&\mathcal{  P}_{22}(1^+)_{ijk}^{\ \ \ def}\mathcal{  A}_{def}=\\
&\left(\mathcal{  P}_{11}(1^+)_{abc}^{\ \ \ ijk}+\tfrac{1}{\sqrt{2}}\mathcal{  P}_{12}(1^+)_{abc}^{\ \ \ ijk}\right)\mathcal{  T}_{kji}.
  \end{aligned}
  \label{psmixing}
\end{equation}
With \eqref{nomix} and \eqref{psmixing} in mind, it is therefore possible to consider $J^P$ tordions as well-defined excitations of the torsion and/or the spin connection, though the latter is more conventional from the perspective of quantisation. A full list of the diagonal SPOs of the $\mathcal{  A}_{abc}$ field is given in \cref{spos}.

The theory parameters employed in \cite{2019PhRvD..99f4001L,Lin2} differ from those in \cite{lasenby-hobson-2016} chiefly through mixing of the linear Riemann and quadratic torsion sectors\footnote{Note that in \cite{2019PhRvD..99f4001L} the Gauss-Bonnet identity is used to eliminate $\check\alpha_1$, which we resurrect through $r_6$, and the notation $l=\lambda$ is used, which we will require for the effective cosmological constant in eWGT, $\kappa^{-1}\Lambda=\lambda\phi_0^4$.}
\begin{equation}
  \begin{gathered}
    r_1=\check\alpha_4-\tfrac{1}{2}\check\alpha_5, \quad r_2=\check\alpha_4-2\check\alpha_5,\\
    r_3=\tfrac{1}{2}\check\alpha_4-\tfrac{1}{2}\check\alpha_5-\tfrac{1}{2}\check\alpha_6,\\
    r_4=\tfrac{1}{2}\check\alpha_2+\tfrac{1}{2}\check\alpha_3, \quad r_5=\tfrac{1}{2}\check\alpha_2-\tfrac{1}{2}\check\alpha_3, \quad r_6=\check\alpha_1,\\
    \kappa t_1=-\check\beta_1-\tfrac{1}{2}\check\alpha_0, \quad \kappa t_2=-2\check\beta_1-6\check\beta_2+\tfrac{1}{2}\check\alpha_0,\\
    \kappa t_3=-\tfrac{1}{2}\check\beta_1+\tfrac{3}{2}\check\beta_3+\tfrac{1}{2}\check\alpha_0,\quad \kappa l=\tfrac{1}{2}\check\alpha_0.
  \end{gathered}
  \label{ycp}
\end{equation}
In terms of these parameters, \cite{2019PhRvD..99f4001L,Lin2} analyse the viability of the free-field theory from the perspective of the physical propagator. Also known as the saturated propagator, this quantity can be obtained when the SPO decomposition of the free-field action is expressible in terms of \textit{invertible} matrices which quadratically combine the $\mathfrak{s}_{ab}$, $\mathfrak{a}_{ab}$ and $\mathcal{  A}_{abc}$ fields within each $J^P$ sector. As might be expected, there exist certain \textit{critical cases} for which some of these matrices become singular. Each such case is defined by certain equations which are linear in the parameters of \eqref{ycp}, and represents one or more emergent gauge symmetry in the linearised theory that must be eliminated before proceeding. Beyond such gauge symmetries, further critical cases alter the factorised form of the matrix determinants, which encode the bare mass spectrum of each $J^P$ sector. 
In \cite{2019PhRvD..99f4001L}, the 1918 such critical cases of PGT\textsuperscript{q+} were exhaustively determined. A systematic survey of these theories identified the 450 for which unitarity can be achieved through additional inequality constraints on the parameters of \eqref{ycp}. This requires the elimination of ghost modes by fixing a positive propagator residue about the relevant pole, and tachyonic particles by fixing a positive square of the relevant bare mass. Any of these critical cases can be discarded if a power counting shows that the superficial degree of divergence in a diagram scales with the number of loops. In \cite{2019PhRvD..99f4001L}, such an analysis was restricted to cases in which the propagator was diagonal not only in the $J^P$ sectors, but also in the fields themselves. This yielded 10 cases which were power-counting renormalisable (PCR). 

Although the PCR condition is thought to be necessary for full renormalisability, it raises ambiguities when applied to PGT\textsuperscript{q+}. Firstly, there may be two or three gauge choices which eliminate the symmetries of a critical case, of which not all are PCR. Secondly, a mode with unsatisfactory high-energy behaviour may yet be non-propagating, and thus inconsequential. Such modes tend to arise precisely when the propagator is non-diagonal in the fields, in particular when the $1^+$ and $1^-$ sectors of $\mathcal{  A}_{abc}$ are mixed. Of the 450 unitary cases, a further 48 were found in \cite{Lin2} which can be considered PCR according to these extended criteria. In the present work, we exclude from all 58 theories only those for which the divergence of non-propagating modes is most egregious\footnote{While this is probably a conservative move, it is foremost a matter of convenience.}, going as $k^2$ rather than $k^{-2}$. This leaves us with 33 critical cases, which include all of the original 10 in \cite{2019PhRvD..99f4001L}. These are listed in \cref{table}. Note that while the methods in \cite{2019PhRvD..99f4001L,Lin2} can identify the \textit{definite} $J^P$ sectors of propagating massive modes, it can only identify the \textit{possible} $J^P$ sectors of propagating massless modes, and their \textit{definite} degrees of freedom. In the present work, we will adhere to the numbering of critical cases used in \cite{Lin2}, in which the select 33 cases we consider range from \criticalcase{1} to \criticalcase{41}. We also use the convention of \cite{Lin2} in which cases previously discovered in \cite{2019PhRvD..99f4001L} are listed with their original numbering in a superscript, such as \criticalcase{9}, \criticalcase{10}, \criticalcase{11} and \criticalcase{13}, which are the only four cases with gauge-invariant PCR.
\DeclareRobustCommand{\particle}[1]{%
  \begingroup\normalfont
  \includegraphics[height=\fontcharht\font`\B]{#1}%
  \endgroup
}

\begin {table*}[htp]
  \caption{\label{table}The select 33 of the unitary, PCR critical cases of PGT\textsuperscript{q+}, according to parameter constraints and particle content. The given numbers are as in \cite{Lin2}, with the original numbers in \cite{2019PhRvD..99f4001L} denoted by an asterisk where applicable. The criticality equalities include an implicit $r_6=0$. The particle content of each $J^P$ sector is as follows. Possible massless excitations of $\mathcal{A  }_{abc}$, $\mathfrak{s}_{ab}$ and $\mathfrak{a}_{ab}$ are respectively `\particle{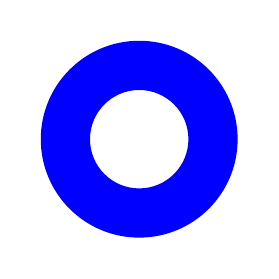}', `\particle{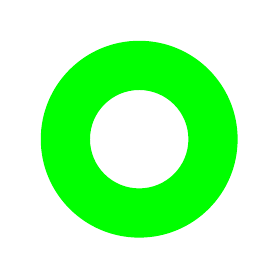}' and `\particle{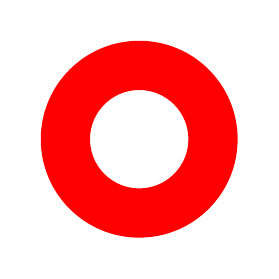}'. Definite massive excitations are `\particle{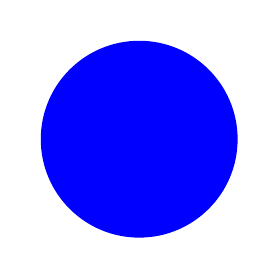}', `\particle{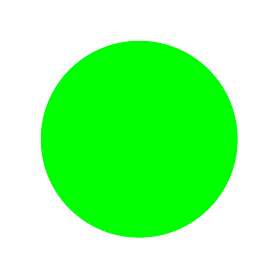}' and `\particle{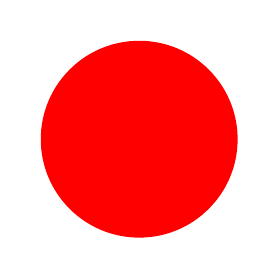}'. Possible massless excitations may have a different field character in a different gauge, e.g. `\particle{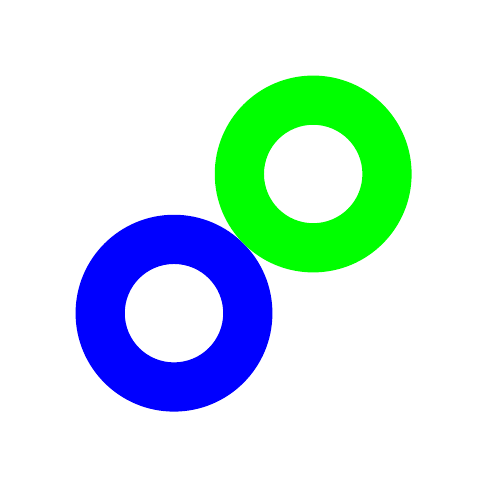}', or be of uncertain field character in one or more such gauge, e.g. `\particle{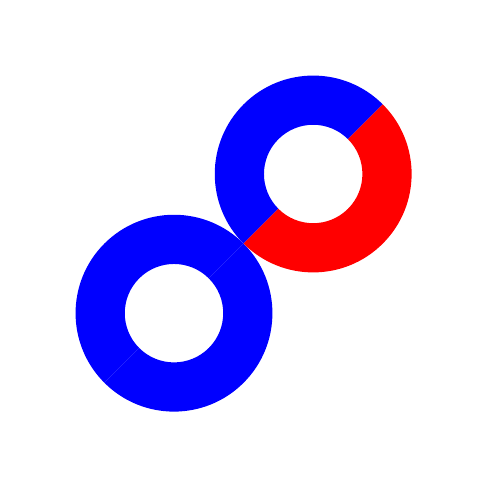}'. While the $J^P$ character of propagating massless excitations remains ambiguous, there are always two, if any, massless degrees of freedom.}
  \includegraphics[width=\linewidth]{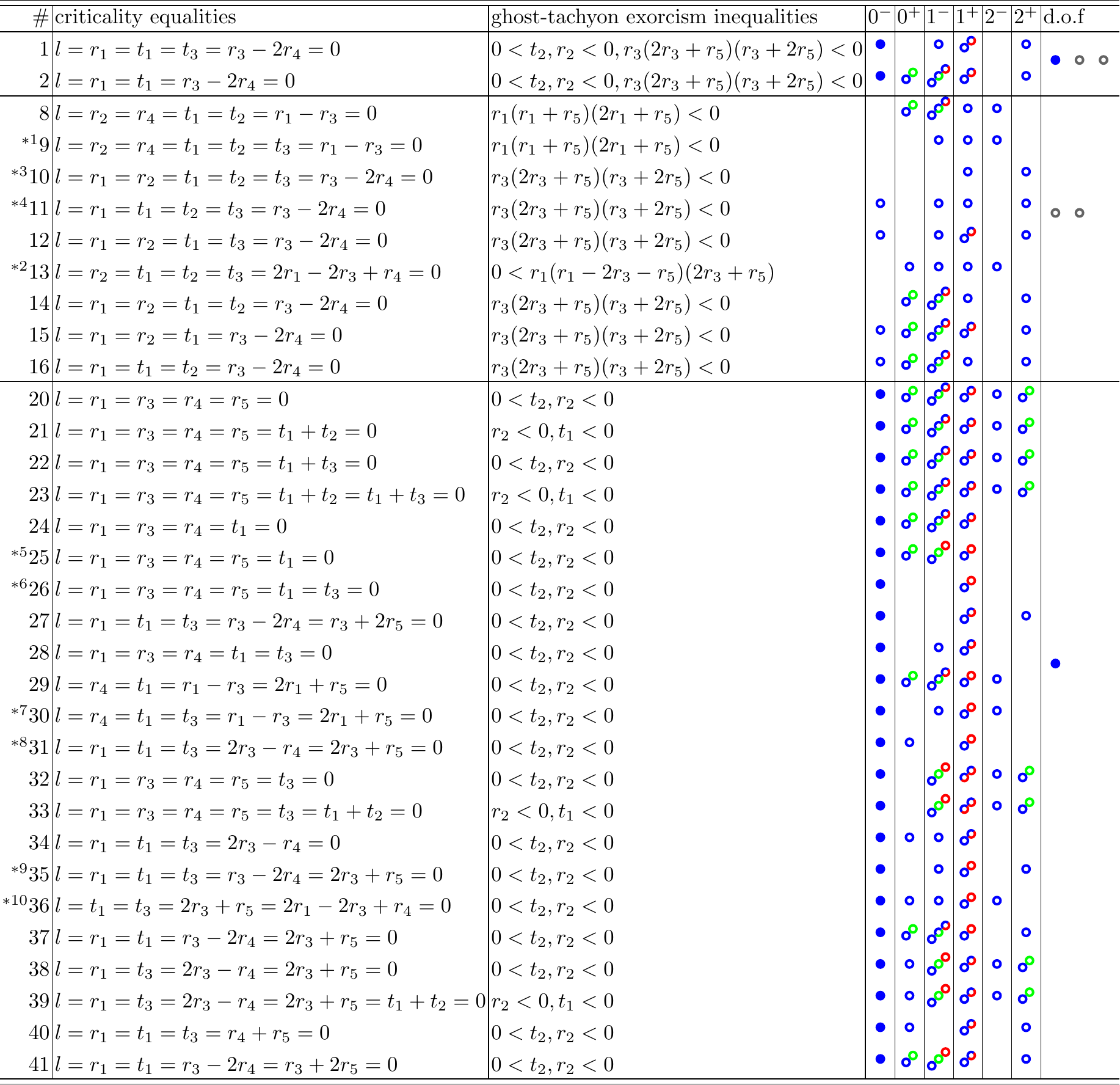}
\end{table*}
\section{The cosmological ansatz}\label{cosmoans}
\subsection{Lessons out of superspace}
The equations of motion of a field theory are usually obtained using the Lagrangian, or less commonly Hamiltonian, formalism. In the theories of (potentially) high-spin fields such as those of gravity considered here, this process is typically lengthy and necessarily results in tensor equations. Once the gravitational field equations are to hand, it is most convenient either to solve the fields for a desirable source, or vice versa. In cases where the solutions are known to be highly symmetric, a suitable ansatz for both sources and fields may be substituted and these solved simultaneously: this is often done in cases where the strong cosmological principle applies. It is worth noting that an alternative `intrinsic' method of solution has been developed for the special formulation of ECT known simply as gauge theory gravity (GTG) \cite{1998RSPTA.356..487L}. Whilst the bulk of what follows was first obtained using similar formulations of PGT\textsuperscript{q+} and eWGT\textsuperscript{q+}, we do not use the intrinsic method and include only a sample of the relevant Clifford algebra in \cref{multi}. Our main results are translated back into the passive tensor formulation set out above. 

In obtaining the field equations, we take a short cut by substituting the source and field ansatz into the action directly, and taking variations with respect to the remaining free parameters. It should be stressed that this method is \textit{not} always justifiable, as variable reduction and variational differentiation are generally non-commuting operations. Nor is it entirely without precedent. In the quantum cosmology of GR, similar methods are frequently employed as part of the minisuperspace approximation \cite{2013PhRvD..88h3518R}. Moreover, the approach has been shown to hold true in GR for all Bianchi A class cosmological models \cite{Ashtekar_1991} and similar methods are even employed for PGT\textsuperscript{q+} in \cite{2009JCAP...10..027C}. Special care must be taken, so that the field ansatz preserves some notion of the ADM lapse and shift freedoms, and that the source ansatz comes pre-packaged with the expected conservation laws \cite{faroni-2009,brown-1993}. In this way, we can avoid intermediate tensor expressions, arriving at an unorthodox but useful statement of the general cosmological equations. These are given in \cref{tor1,tor2,S,R}.
\subsection{Gravitational fields}\label{gravv}
The first task is to find the most general ansatz for each of the four gauge fields $h_a^{\ \mu}$, $b^a_{\ \mu}$, $A^{ab}_{\ \ \mu}$ and $V^\mu$ consistent with the strong cosmological constraints of spatial homogeneity and isotropy. These constraints do not apply directly to the gauge fields, but to the observable quantities derived from them. It is convenient to adopt spherical polar coordinates $\{x^\mu\}=\{t,r,\upvartheta,\upvarphi\}$ where the only dimensionful coordinate is $t$. This fixes the diffeomorphism gauge via the basis vectors $\{\boldsymbol{e}_\mu\}$ and covectors $\{\boldsymbol{e}^\mu\}$. 

By orthogonality, the normalised counterparts of these eight quantities provide a natural choice of Lorentz rotation gauge, $\{\hat{\boldsymbol{e}}_a\}$ and $\{\hat{\boldsymbol{e}}^a\}$, should we choose to fix it. An interval which suitably generalises \eqref{frw} is then
\begin{equation}\label{interval}
  \begin{aligned}
  \mathrm{d}s^2=S^2\bigg[&\mathrm{d}t^2-\frac{R^2\mathrm{d}r^2}{1-kr^2}\\
  &-R^2r^2(\mathrm{d}\upvartheta^2+\sin^2\upvartheta\mathrm{d}\upvarphi^2)\bigg],
\end{aligned}
\end{equation}
where $S=S(t)$ is a dimensionless conformal factor which establishes the length scale of the theory, $R=R(t)$ is the dimensionful relative scale factor while the constant $k\in\{0,\pm 1\}$ dictates the curvature of Cauchy surfaces. Note that setting $S=1$ corresponds to the Friedmann diffeomorphism gauge, in which $R$ becomes the usual scale factor of the universe. The interval \eqref{interval} determines the components $b^a_{\ \mu}$ only up to the rotation gauge, which we leave arbitrary. The diffeomorphism gauge fields are then fixed to
\begin{equation}
  \begin{gathered}
    b^a_{\ t}=S(\hat{\boldsymbol{e}}_t)^a, \quad b^a_{\ r}=\frac{SR}{\sqrt{1-kr^2}}(\hat{\boldsymbol{e}}_r)^a,\\
    b^a_{\ \upvartheta}=rSR(\hat{\boldsymbol{e}}_\upvartheta)^a, \quad b^a_{\ \upvarphi}=rSR(\hat{\boldsymbol{e}}_\upvarphi)^a,
  \end{gathered}
  \label{bs}
\end{equation}
up to a choice of sign. In practice, we will work exclusively with the inverse fields, which we define by $h_a^{\ \mu}=\eta_{ab}g^{\mu\nu}b^b_{\ \nu}$. 

Whilst $h_a^{\ \mu}$ has thus been determined by a cosmological $g_{\mu\nu}$, $A^{ab}_{\ \ \mu}$ must be determined by a cosmological $\mathcal{T}^a_{\ \ bc}$. The unique form adopted by the torsion tensor under the restrictions of homogeneity and isotropy may be written down immediately
\begin{equation}
  \mathcal{  T}^a_{\ \ bc}=(\hat{\boldsymbol{e}}_t)^d\left(\tfrac{2}{3}U\delta^a_{[c}\eta^{\phantom{a}}_{db]}-Q\epsilon^a_{\ dbc}\right),
  \label{torsionobv}
\end{equation}
where the fields $U=U(t)$ and $Q=Q(t)$ have units of $\si{\electronvolt}$ and are observable quantities which may be easily extracted through the quadratic invariants
\begin{equation}
  \begin{gathered}
    \mathcal{  T}^a\mathcal{  T}_{\ a}=U^2, \quad \mathcal{  T}_{\ abc}\mathcal{  T}^{bac}=\tfrac{1}{3}U^2+6Q^2,\\
  \mathcal{  T}_{abc}\mathcal{  T}^{abc}=\tfrac{2}{3}U^2-6Q^2.
  \label{degen}
\end{gathered}
\end{equation}
This form was first rigorously identified by Tsamparlis \cite{1979PhLA...75...27T}, and has been used by both Boehmer and Bronowski \cite{2006gr.qc.....1089B} and Brechet, Hobson and Lasenby \cite{2008CQGra..25x5016B} in the study of cosmologies filled with Weyssenhoff fluids. One may arrive at \eqref{torsionobv} by noting that, under the strong cosmological principle, the spacetime contains six global Killing vector fields $\{\mathcal{  K}^a\}$, each tangent to the local Cauchy surface. Furthermore, cosmic fluids share a global, normalised velocity field $u^a$, to which the Cauchy surfaces are orthogonal $u^a\mathcal{  K}_a=0$. We can use this to define the intrinsic metric on the Cauchy surfaces, which is also a projection tensor with vanishing Lie derivative
\begin{equation}\label{projector}
  s_{ab}=\eta_{ab}-u_au_b, \quad \mathcal{  L}_\mathcal{  K}s_{ab}=0,
\end{equation}
along with the projection of any tensor, $\mathcal{  F}^{a_1\dots a_i}_{\ \ \ \ \ \  c_1 \dots c_j}$, and its projected covariant derivative
  \begin{align}
  \hat{\mathcal{  F}}^{a_1\dots a_i}_{\ \ \ \ \ \  c_1 \dots c_j}&=s^{a_1}_{\ a'_1}\dots s^{c'_j}_{\ c_j}\mathcal{  F}^{a'_1\dots a'_i}_{\ \ \ \ \ \  c'_1 \dots c'_j},\\
  \hat{\mathcal{  D}}_e\mathcal{  F}^{a_1\dots a_i}_{\ \ \ \ \ \  c_1 \dots c_j}&=s^n_{\ e}s^{a_1}_{\ a'_1}\dots  s^{c'_j}_{\ c_j}\mathcal{  D}_n\mathcal{  F}^{a'_1\dots a'_i}_{\ \ \ \ \ \  c'_1 \dots c'_j}.
\end{align}
Our fundamental requirement is that $\mathcal{  L}_\mathcal{  K}\mathcal{  T}^a_{\ \ bc}=0$, but by \eqref{projector} we must have $\mathcal{  L}_\mathcal{  K}\hat{\mathcal{  T}}^a_{\ \ bc}=0$ also. Examining this, we find
\begin{equation}
\begin{aligned}
    \mathcal{  K}^d&\hat{\mathcal{  D}}_d\hat{\mathcal{  T}}^a_{\ \ bc}=\hat{\mathcal{  T}}^d_{\ \ bc}\hat{\mathcal{  D}}_d\mathcal{  K}^a-\left(\hat{\mathcal{  T}}^a_{\ \ dc}\hat{\mathcal{  D}}_b+\hat{\mathcal{  T}}^a_{\ \ bd}\hat{\mathcal{  D}}_c\right)\mathcal{  K}^d\\
    &=\Big(s^{ea}\hat{\mathcal{  T}}^d_{\ \ bc}+s^e_{\ b}\hat{\mathcal{  T}}^{ad}_{\ \ \ c}+s^e_{\ c}\hat{\mathcal{  T}}^{a\ d}_{\ \ b} \Big)\hat{\mathcal{  D}}_{[d}\mathcal{  K}_{e]}.
\end{aligned}
\label{masterkilling}
\end{equation}
There is freedom in the choice of the $\mathcal{  K}^a$ to set to zero either side of \eqref{masterkilling}. Doing so on the RHS enforces spatial homogeneity, so that the components $\mathcal{  T}^a_{\ \ bc}$ are functions only of the coordinate $t$. On the LHS, we enforce isotropy, so that
\begin{equation}
  s^{[e}_{\ \ a}s^{n]}_{\ \ r}\hat{\mathcal{  T}}^r_{\ \ bc}+s^{[e}_{\ \ b}s^{n]}_{\ \ r}\hat{\mathcal{  T}}^{\ r}_{a \ c}+s^{[e}_{\ \ c}s^{n]}_{\ \ r}\hat{\mathcal{  T}}^{\ \ r}_{ab}=0.
  \label{comm}
\end{equation}
From examination of \eqref{comm} we then arrive at the following pair of projected component constraints
\begin{equation}
  \hat{\mathcal{  T}}^a_{\ \ ba}=0, \quad \hat{\mathcal{  T}}_{\ abc}=\hat{\mathcal{  T}}_{\ [abc]},
\end{equation}
and by inspection we see that these admit only the form set out in \eqref{torsionobv}. 

The fields $U$ and $Q$ are sometimes referred to as the \textit{torsion contraction} and \textit{torsion protraction} respectively -- the reference to the protraction will be explained in \cref{multi}. Furthermore, it is easy to show that the strong cosmological principle has done nothing more than pick the $0^-$ and $0^+$ sectors out of the general torsion tensor, since setting $k^a=(\boldsymbol{e}_t)^a$, we find without loss of generality
\begin{equation}
  \begin{aligned}
    \mathcal{  P}_{11}(0^+)^{\ \ aij}_{bc\ \ k}\mathcal{  T}^k_{\ \ ij}&=\tfrac{2}{3}(\hat{\boldsymbol{e}}_t)^dU\delta^a_{[c}\eta^{\phantom{a}}_{db]},\\
    \mathcal{  P}_{11}(0^-)^{\ \ aij}_{bc\ \ k}\mathcal{  T}^k_{\ \ ij}&=-(\hat{\boldsymbol{e}}_t)^dQ\epsilon^a_{\ dbc}.
\end{aligned}
\end{equation}
In this manner, the quantities $U$ and $Q$ then encode the freedoms in the scalar and pseudoscalar tordion singlets.
From \eqref{degen} we see right away that there is some degeneracy among the dimensionless theory parameters $\{\beta_i\}$ under cosmological conditions. This behaviour is to be expected, and is even more pronounced in the quadratic Riemann sector: we will make extensive use of it in \cref{cosmo}. 

For the purposes of the ansatz, we take the torsion tensor to have the form
\begin{equation}\label{torsion}
  \mathcal{  T}^a_{\ \ bc}=\frac{2}{SR}(\hat{\boldsymbol{e}}_t)^d\bigg[\left( X+\frac{\partial_t(SR)}{S} \right)\delta^a_{[c}\eta^{\phantom{a}}_{db]}-\frac{Y}{2}\epsilon^a_{\ dbc}\bigg].
\end{equation}
The dimensionless fields $X=X(t)$ and $Y=Y(t)$ now inherit the two degrees of freedom in $U$ and $Q$. The form of the first term in \eqref{torsion} is designed to absorb those Ricci rotation coefficients containing $\partial_t S$ and $\partial_t R$, and the rotational gauge fields which generate this torsion are
\begin{equation}
\begin{aligned}
    A^{ab}_{\ \ r}=&\frac{1}{\sqrt{1-kr^2}}(\hat{\boldsymbol{e}}_t)^c(\hat{\boldsymbol{e}}_r)^d\left(2X\delta^a_{[d}\delta^b_{c]}+Y\epsilon^{ab}_{\ \ cd}\right),\\
    A^{ab}_{\ \ \upvartheta}=&2(\hat{\boldsymbol{e}}_\upvartheta)^c\bigg[\frac{1}{r}\left( 1-\sqrt{1-kr^2}\right)(\hat{\boldsymbol{e}}_r)^{d}\\
    &+X(\hat{\boldsymbol{e}}_t)^d\bigg]\delta^a_{[c}\delta^b_{d]}+Y(\hat{\boldsymbol{e}}_t)^c(\hat{\boldsymbol{e}}_\upvartheta)^d\epsilon^{ab}_{\ \ cd},\\
    A^{ab}_{\ \ \upvarphi}=&2(\hat{\boldsymbol{e}}_\upvarphi)^c\bigg[\frac{1}{r}\left( 1-\sqrt{1-kr^2}\right)(\hat{\boldsymbol{e}}_r)^{d}\\
    &+X(\hat{\boldsymbol{e}}_t)^d\bigg]\delta^a_{[c}\delta^b_{d]}+Y(\hat{\boldsymbol{e}}_t)^c(\hat{\boldsymbol{e}}_\upvarphi)^d\epsilon^{ab}_{\ \ cd}.
\end{aligned}
\end{equation}

The equations of motion are therefore to be obtained through variation with respect to $R$, $S$, $X$ and $Y$, yet the cosmological equations are ideally expressed in terms of observable quantities. Clearly $S$ is not observable, because after variation we would like to adopt the Friedmann gauge by globally setting $S=1$. Having done this, we also note that $R$ is not generally a quantity with good physical motivation, since when $k=0$ it may be chosen arbitrarily. With this in mind, we prefer to substitute for $R$, $X$ and $Y$ in terms of the Hubble number and deceleration parameter defined in \eqref{handq} and physical torsion fields once the Friedmann gauge has been adopted
\begin{equation}
  U=\frac{3}{R}\left( X+\partial_t R \right), \quad Q=\frac{Y}{R}.
  \label{physicalgrav}
\end{equation}
Having established the gravitational field ansatz in PGT, the extension to eWGT is quite straightforward. The compensator, $\phi$, naturally satisfies the strong cosmological principle as a scalar field, $\phi=\phi(t)$. The obvious choice for the Weyl gauge field is then to define a dimensionless $V=V(t)$ such that
\begin{equation}
  \mathcal{  V}^a=\frac{V}{SR}(\hat{\boldsymbol{e}}_t)^a.
\end{equation}
\subsection{Gravitational sources}\label{outlay}
From a mathematical perspective we will consider four distinct sources in our models, though three of these may correspond to a variety of physical matter fields. Firstly the curvature constant $k$ is deeply embedded in the gravitational rather than matter sector of the action, yet as we discussed in \cref{intro}, it has become acceptable to view it as a source term in the cosmological equations. Dark energy, or vacuum energy is included via the cosmological constant $\Lambda$ in PGT\textsuperscript{q+} and parameter $\lambda$ in eWGT\textsuperscript{q+}, and is already a valid cosmological source having both homogeneity and isotropy. Directly observable baryonic matter and dark matter are modelled by \textit{dust}, while photons and neutrinos are modelled by \textit{radiation}. In making these approximations we forfeit any effects arising from the spin content of the real sources, but avoid the complexities of constructing Weyssenhoff fluids\footnote{Note that if the Dirac Lagrangian is rendered scale-invariant by means of the compensator $\phi$, the resulting matter stress-energy tensor resembles that of a perfect fluid \cite{1973RSPSA.333..403D,1990GReGr..22..289M,1992ApJ...391..429M}.}. 

In establishing the form of $L_\text{m}$ and $L_{\text{m}}^\dagger$, we adopt the techniques set out in \cite{brown-1993,bertolami-lobo-paramos-2008}, taking the Lagrangian densities to be the negative on-shell energy densities of the fluids,
\begin{equation}
\begin{aligned}
  L_\text{m}&=-\rho_\text{m}-\rho_\text{r}=-\kappa^{-\frac{1}{2}}\frac{\varrho_\text{m}}{S^3R^3}-\frac{\varrho_\text{r}}{S^4R^4},\\
  L_{\text{m}}^\dagger&=-\kappa^{\frac{1}{2}}\phi\rho_\text{m}-\rho_\text{r}=-\phi\frac{\varrho_\text{m}}{S^3R^3}-\frac{\varrho_\text{r}}{S^4R^4},
\end{aligned}
\end{equation}
where $\rho_\text{m}=\rho_\text{m}(t)$ and $\rho_\text{r}=\rho_\text{r}(t)$ have dimension $\si{\electronvolt^4}$ and $\varrho_\text{r}$ and $\varrho_\text{m}$ are dimensionless constants. As with the gravitational variables, we will prefer to express the matter content in the cosmological equations in terms of observable quantities. The constants $\Lambda$ and $k$ along with the densities $\rho_\text{m}$ and $\rho_\text{r}$ are already perfectly acceptable from this perspective, but we will make use of the popular \textit{dimensionless densities} as they are defined in the Friedmann gauge,
\begin{equation}
  \begin{gathered}
    \Omega_k=-\frac{k}{R^2H^2}, \quad \Omega_\Lambda=\frac{\Lambda}{3H^2},\\
    \Omega_\text{m}=\frac{\kappa\rho_\text{m}}{3H^2}, \quad \Omega_\text{r}=\frac{\kappa\rho_\text{r}}{3H^2}.
  \end{gathered}
  \label{physmat}
\end{equation}
These quantities are well suited to the analysis that follows in \cref{cosmo}, but differ from the contemporary densities in \cref{intro}, which are typically used in the field of cosmological inference, through the normalisation of $H$ according to \eqref{neldef}.
\section{General cosmologies}\label{cosmo}
\subsection{A demonstration: Einstein-Cartan theory}
The equations of motion are to be obtained by considering PGT\textsuperscript{q+} actions of the form
\begin{equation}
  \tilde{S}_\text{T}=\int \mathrm{d}t\tilde{\mathcal{L}}_T(X(t),Y(t),S(t),R(t)),
\end{equation}
and eWGT\textsuperscript{q+} actions of the form
\begin{equation}
  \tilde{S}_\text{T}=\int \mathrm{d}t\tilde{\mathcal{L}}_T(X(t),Y(t),S(t),R(t),\phi(t),V(t)).
\end{equation}
To check the efficacy of our approach, we will obtain the Friedmann equations from the minimal gravitational gauge theory in which the $\{\check{\alpha}_i\}$ and $\{\check{\beta}_i\}$ are all set to zero except for $\check{\alpha}_0$: this is ECT. The action $S_\text{T}$ in \eqref{pgtaction} is the integral of the dimensionless \textit{reduced} action, $\tilde{S}_\text{T}$, over the Cauchy-surface
\begin{equation}
\begin{aligned}
  \tilde{S}_\text{T}=-\int\mathrm{d}t\big[&3\check{\alpha_0}\kappa^{-1}S^2R\left( R\partial_t X+Y^2/4-X^2-k \right)\\
  &+\kappa^{-1}\Lambda S^4R^3+\kappa^{-\frac{1}{2}}\varrho_\text{m}S+\varrho_\text{r}/R\big].
  \label{redactecpgt}
\end{aligned}
\end{equation}
There are four dynamical fields: two for curvature, $R$ and $S$, and two for torsion, $X$ and $Y$. It is with respect to these quantities, rather than their physical counterparts, that we must take variations. Once we set $S=1$, the equations of motion for $X$ and $Y$ are 
\begin{subequations}
  \begin{align}
    \left(\delta \tilde{\mathcal{ L}}_\text{T}/\delta X\right)_{\text{F}}&\propto R\left( \partial_t R+X \right),\label{ectorsp}\\
    \left(\delta \tilde{\mathcal{ L}}_\text{T}/\delta Y\right)_{\text{F}}&\propto RY,
  \label{ectors}
  \end{align}
\end{subequations}
which immediately confirms that cosmic torsion is prohibited in an Einstein-Cartan universe filled with the simplistic source fluids considered here, or $U=Q=0$. The curvature equations for $R$ and $S$ are
\begin{subequations}
  \begin{align}
    \left(\delta \tilde{\mathcal{ L}}_\text{T}/\delta R\right)_{\text{F}}&\propto3\check{\alpha}_0R^2\left( 2R\partial_t X-X^2+Y^2/4-k\right)\nonumber\\
    &\phantom{\propto}+3R^4\Lambda-\kappa\varrho_\text{r},\label{eccurvp}\\
    \left(\delta \tilde{\mathcal{ L}}_\text{T}/\delta S\right)_{\text{F}}&\propto6\check{\alpha}_0R\left(R\partial_t X-X^2+Y^2/4-k\right)\nonumber\\
    &\phantom{\propto}+4R^3\Lambda-\kappa^{\frac{1}{2}}\varrho_\text{m}.
  \label{eccurv}
  \end{align}
\end{subequations}

The four \cref{ectorsp,ectors,eccurvp,eccurv} may then be re-arranged in terms of the preferred variables to give the cosmic equations of motion
\begin{equation}
  \begin{gathered}
    \check{\alpha}_0=\Omega_\text{m}+\Omega_\text{r}+\Omega_\Lambda+\Omega_k,\\
    \check{\alpha}_0q=\tfrac{1}{2}\Omega_\text{m}+\Omega_\text{r}-\Omega_\Lambda.
\end{gathered}
\end{equation}
The Friedmann equations are recovered when we choose $\check{\alpha}_0=1$, thus making the connection to ECT. 

The reduced action in eWGT naturally takes a very similar form to \eqref{redactecpgt}
\begin{equation}
\begin{aligned}
  \tilde{S}_\text{T}=-\int\mathrm{d}t\big[&3\check{\alpha}_0\phi^2S^2R\big( R\partial_t (X+V)+Y^2/4\\
      &-(X+V)^2-k \big)+\lambda\phi^4S^4R^3\\
    &+\varrho_\text{m}\phi S+\varrho_\text{r}/R\big],
  \label{redactedpgt}
\end{aligned}
\end{equation}
the important difference being the appearance of the $\phi$ field, which always appears in the combination $\phi S$, and the $V$ field, which appears in the combination $X+V$. These are perfectly general features of cosmological eWGT\textsuperscript{q+}: the extra gauge fields are degenerate with two of the original four in PGT\textsuperscript{q+}
\begin{equation}
  \phi\leftrightharpoons S, \quad V\leftrightharpoons X.
  \label{slip}
\end{equation}
The degeneracy \eqref{slip} clearly indicates that we will have no more independent equations of motion in eWGT\textsuperscript{q+} than in PGT\textsuperscript{q+}, but the fixing of the Friedmann gauge in the former case remains to be defined. In particular, $V$ can be absorbed directly into $X$ since both fields are dimensionless. Finally, if the fixing of $S=1$ is carried over to eWGT\textsuperscript{q+}, we find the appropriate Einstein gauge $\phi=\phi_0=\kappa^{-1/2}$ completes the correspondence. Note that in this case, the freedom in $\Lambda$ is truly inherited by the dimensionless $\lambda$ rather than $\phi$.
\subsection{The cosmic theory parameters}\label{map_section}
We would now like to consider the general actions of PGT\textsuperscript{q+} and eWGT\textsuperscript{q+}, \eqref{pgtaction} and \eqref{ewgtaction}. The parameter degeneracy among the torsion variables identified in \eqref{degen} extends throughout the gravitational sector, allowing us to express the equations of motion minimally in terms of parameter combinations which uniquely affect the cosmology. 
It is expedient to use vector notation to discuss theories, for example any PGT\textsuperscript{q+} may be written in terms of its theory parameters as
\begin{equation}
    \boldsymbol{x}=\sum_{i=0}^6\check{\alpha}_i\check{\boldsymbol{\alpha}}_i+\sum_{i=1}^3\check{\beta}_i\check{\boldsymbol{\beta}}_i,
\end{equation}
such that the vectors on the RHS form an orthonormal set, and any theory parameter may be extracted by projecting with the relevant vector, e.g. $\check{\alpha}_1=\check{\boldsymbol{\alpha}}_1\cdot\boldsymbol{x}$. The form of \eqref{gaussbonnet} then suggests that (at the classical level) any theory is unchanged under a transformation in the Gauss-Bonnet sense
\begin{equation}
  \boldsymbol{x}\to\boldsymbol{x}+\check{\alpha}_{\text{GB}}\boldsymbol{L}, \quad \boldsymbol{L}=\check{\boldsymbol{\alpha}}_1-4\check{\boldsymbol{\alpha}}_3+2\check{\boldsymbol{\alpha}}_6.
\end{equation}

The quadratic Riemann sector thus has a five-dimensional parameter space in general. When we demand homogeneity and isotropy as with cosmology, we might reasonably expect this number to be reduced. To identify the reduced degrees of freedom we should turn to the equations of motion. Doing so, we find the cosmological conditions eliminate a further two degrees of freedom from the quadratic Riemann sector. Let us define two coordinates
\begin{equation}
  {\chi}_1=\tfrac{3}{2}\check{{\alpha}}_1+\tfrac{1}{4}\check{{\alpha}}_3-\tfrac{1}{4}\check{{\alpha}}_6, \quad {\chi}_2=\tfrac{3}{2}\check{{\alpha}}_1+\tfrac{1}{2}\check{{\alpha}}_3+\tfrac{1}{4}\check{{\alpha}}_6,
\end{equation}
which are oblivious to the Gauss-Bonnet content of the theory:
\begin{equation}
  \boldsymbol{\chi}_2\cdot\boldsymbol{L}=\boldsymbol{\chi}_1\cdot\boldsymbol{L}=0.
\end{equation}
The cosmologically meaningful coordinates of the quadratic Riemann sector are then equally oblivious, as we might expect, and are given by
\begin{equation}
  \begin{gathered}
    \sigma_1=\chi_1+\tfrac{1}{4}\check\alpha_2+\tfrac{1}{4}\check\alpha_5, \quad \sigma_2=\chi_2+\tfrac{1}{2}\check\alpha_2+\tfrac{3}{4}\check\alpha_4-\check\alpha_5,\\
    \sigma_3=\chi_2+\tfrac{1}{2}\check\alpha_2+\tfrac{1}{4}\check\alpha_4.
  \end{gathered}
\end{equation}
We have already seen that the three $\{\check\beta_i\}$ of PGT\textsuperscript{q+} must reduce to two cosmic theory parameters for PGT torsion. Denoting these by $\{\upsilon_i\}$ we find
\begin{equation}
  \upsilon_1=\check\beta_1+3\check\beta_2, \quad \upsilon_2=3\check\beta_3-\check\beta_1.
\end{equation}
In eWGT\textsuperscript{q+} there \textit{is} no $\check\beta_3$, but we find that its r\^ole is filled by $\nu$, so that
\begin{equation}
  \upsilon_1=\check\beta_1+3\check\beta_2, \quad \upsilon_2=-\nu/6.
\end{equation}

We therefore find that the ten theory parameters of PGT\textsuperscript{q+} and eWGT\textsuperscript{q+} reduce to five cosmic theory parameters. The freedoms of the quadratic Riemann sector are reduced from six to three, and those of the torsion and compensator sectors are reduced from three to two.
\subsection{$k$-screening}\label{ksc_sec}
Having defined the Lagrangian parameters relevant to cosmology, we are now in a position to express the equations of motion in a form valid simultaneously for \textit{both} gauge theories. As before, these constitute a coupled system of four equations. For brevity, we write these in terms of dimensionless conformal time
\begin{equation}
  \mathrm{d}\tau=\mathrm{d}t/R,
  \label{dim_conf}
\end{equation}
and the dynamical variables introduced above, with the Friedmann gauge fixed:
\begin{widetext}
  \begin{subequations}
\begin{align}
  \left(\delta \tilde{\mathcal{ L}}_\text{T}/\delta X\right)_{\text{F}}&\propto\left( \upsilon_2+\check{\alpha}_0 \right)R\left(RX+\partial_\tau R\right)-8\kappa\sigma_3\partial^2_\tau X-4\kappa\sigma_1Y\partial_\tau Y-4\kappa X\left( \sigma_2 Y^2-4\sigma_3\left( X^2+k\right) \right),\label{tor1}\\
  \left(\delta \tilde{\mathcal{  L}}_\text{T}/\delta Y\right)_{\text{F}}&\propto\left( 4\upsilon_1-\check{\alpha}_0 \right)R^2Y-4\kappa\left( \sigma_3-\sigma_2 \right)\partial^2_\tau Y+16\kappa\sigma_1Y\partial_\tau X+4\kappa Y\left(\sigma_3Y^2-4\kappa\left( \sigma_2X^2+\sigma_3k \right)\right),\label{tor2}\\
  \left(\delta \tilde{\mathcal{  L}}_\text{T}/\delta S\right)_{\text{F}}&\propto12\upsilon_2\partial^2_\tau R+12\left( \upsilon_2+\check{\alpha}_0 \right)R\left(\partial_\tau X-X^2  \right)-3\left( 4\upsilon_1-\check{\alpha}_0 \right)RY^2-12\check{\alpha}_0kR+2\kappa^{\frac{1}{2}}\varrho_\text{m}+8\Lambda R^3,\label{S}\\
  \left(\delta \tilde{\mathcal{  L}}_\text{T}/\delta R\right)_{\text{F}}&\propto12\upsilon_2\left(2R\partial^2_\tau R-(\partial_\tau R)^2\right)+12\left( \upsilon_2+\check{\alpha}_0 \right)R^2\left( 2\partial_\tau X-X^2 \right)-3\left( 4\upsilon_1-\check{\alpha}_0 \right)R^2Y^2-12\check{\alpha}_0kR^2\nonumber\\
&\phantom{\propto}+6\kappa\sigma_3\left( 16X^2\left( X^2+2k \right)+Y^2\left( Y^2-8k \right)  +16k^2-2(\partial_\tau Y)^2-16(\partial_\tau X)^2 \right)\nonumber\\
&\phantom{\propto}+12\kappa\sigma_2\left((\partial_\tau Y)^2-2X^2Y^2\right)-4\kappa\varrho_\text{r}+12\Lambda R^4\label{R}.
\end{align}
\end{subequations}
\end{widetext}
A cursory examination of this system reveals a degree of similarity between the torsion equations \eqref{tor1} and \eqref{tor2} which we will mention again in \cref{chao}, along with the parameters $\sigma_2$ and $\sigma_3$, and $\upsilon_1$ and $\upsilon_2$. The single linear Riemann parameter, $\check{\alpha}_0$, has an entirely different effect to the quadratic Riemann parameters $\sigma_1$, $\sigma_2$ and $\sigma_3$, and while it mostly combines with $\upsilon_1$ or $\upsilon_2$, it couples uniquely with $k$ in \eqref{S}. This gives us some insight into the cosmological overlap between Einstein-Hilbert and Yang-Mills gravities: the latter are not expected to conventionally interact with the bulk curvature of space. 

In fact, a pure Yang-Mills theory
\begin{equation}\label{yangmill}
  \check{\alpha}_0=0,
\end{equation}
may be `screened' from this curvature altogether, since the \textit{single} parameter constraint
\begin{equation}
  \sigma_3=0,
  \label{ksc}
\end{equation}
promptly eliminates $k$ from the entire system. In the context of our opening remarks regarding $\omega_k$ in \cref{intro}, this is a superficially disastrous choice of theory, in which the global geometry of space is decoupled from the dynamics. On the other hand, \eqref{ksc} is a tempting starting point for the study of PGT\textsuperscript{q+} and eWGT\textsuperscript{q+} cosmologies, since it eliminates many other unattractive derivative terms from the system, and does so with a very high degree of naturalness.

\subsection{Cosmological normal scale invariance}

In our narrow $\phi$-free definition of PGT\textsuperscript{q+}, the NSI condition on the gravitational sector \eqref{NSI} clearly imposes
\begin{equation}
  \check{\alpha}_0=\upsilon_1=\upsilon_2=0. 
  \label{cNSI}
\end{equation}
The effect of \eqref{cNSI} on \cref{tor1,tor2,S,R} is profound, as it sets
\begin{equation}
  \Omega_\text{m}=\Omega_{\Lambda}=0,
  \label{<+label+>}
\end{equation}
in all relevant solutions. We use this to write such theories off as \textit{cosmologically} NSI. It should be noted that the cosmological NSI condition \eqref{cNSI} is slightly less restrictive than \eqref{NSI}. It is also interesting to note that if $\phi$ were minimally included in PGT\textsuperscript{q+} (i.e. without any term proportional to $\mathcal{  D}_a\phi\mathcal{  D}^a\phi$), from \eqref{slip}, the condition \eqref{NSI_phi} would reduce to
\begin{equation}
  \upsilon_2=0,
  \label{u20}
\end{equation}
without any such loss of generality.

\begin{figure*}[t!]
  \includegraphics[width=\linewidth]{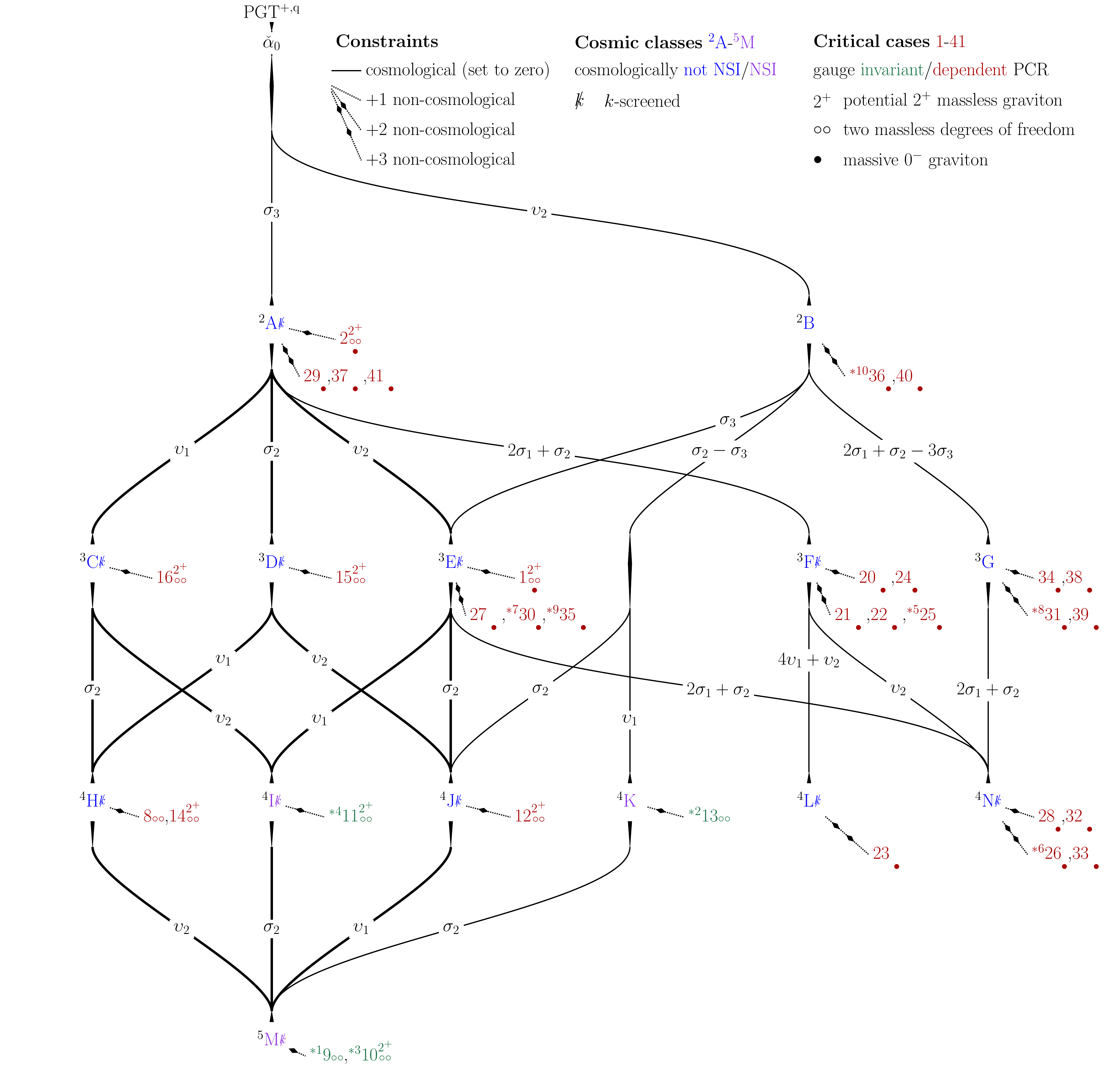}
  \caption{\label{map_plot}The select 33 unitary, PCR critical cases of PGT\textsuperscript{q+} identified in \cite{2019PhRvD..99f4001L,Lin2} and listed in \cref{table} span \numbercosmicclass{} cosmic classes. Note that the traditional Einstein-Hilbert term is the first to be excluded, $\check{\alpha}_0=0$. Desirable critical cases admit the possibility of a massless $2^+$ graviton, i.e. \criticalcase{15}, \criticalcase{16}, \criticalcase{14}, \criticalcase{12}, \criticalcase{11} and \criticalcase{10}. We cannot exclude \criticalcase{2} and \criticalcase{1} on the basis of their additional massive $0^-$ gravitons. Superficially, cosmic classes are excluded by cosmological NSI, which arises when $\check{\alpha}_0=\upsilon_1=\upsilon_2=0$. By these criteria the only truly desirable cosmologies are clearly of \cosmicclass{2}, \cosmicclass{16}, \cosmicclass{15}, \cosmicclass{1}, \cosmicclass{14} or \cosmicclass{12}, and this restricts us to two faces of the cube at the far left of the diagram. All such cosmologies are $k$-screened, with $\check\alpha_0=\sigma_3=0$.}
\end{figure*}

The select 33 critical cases of PGT\textsuperscript{q+} listed in \cref{table} may now be categorised into \numbercosmicclass{} \textit{cosmic classes} according to the effects of their defining parameter constraints on the general PGT\textsuperscript{q+} cosmology. This is illustrated in \cref{map_plot}. Independent cosmic classes are labelled by letters, with a superscript denoting the minimum number of constraints that must be applied to the `root' PGT\textsuperscript{q+} Lagrangian \eqref{pgtaction} to obtain them, e.g. \cosmicclass{2}, \cosmicclass{23} etc. Note that no critical case is completely determined by its cosmic class, in that there are always two or three non-cosmological constraints in the critical case definition which do not appear to affect \cref{tor1,tor2,S,R}.
\subsection{Motivated Cosmologies}\label{permitted}
A glaring feature of \cref{map_plot} is that all critical cases begin with the Yang-Mills constraint, \eqref{yangmill}. Beyond this, the $k$-screening condition, \eqref{ksc}, defines the most general vertex, \cosmicclass{2}, of the cube containing all critical cases with possible $2^+$ massless gravitons.
\subsubsection{\cosmicclass{16}: Einstein freezing}\label{newroot}
To gain some traction, we will not start with \cosmicclass{2}, but enforce a third trivial constraint on the torsion
\begin{equation}
  \upsilon_1=0.
  \label{tor1cons}
\end{equation}
\cosmicclass{16} is the most general comsology defined by these three constraints. 

A useful property common to \cosmicclass{16} and some of its children is that \eqref{tor1} allows us to eliminate $U$ from the system immediately,
\begin{equation}
  U=\frac{12\kappa Q\left( \left( \sigma_2-\sigma_1 \right)QH-\sigma_1\partial_tQ \right)}{4\kappa\sigma_2Q^2-\upsilon_2}.
  \label{Heqn2m}
\end{equation}
An energy balance equation may then be constructed by linear combination of \eqref{S} and \eqref{R}
\begin{equation}
  \Omega_\text{r}+\Omega_\text{m}+\Omega_\Lambda+\Omega_\Psi+\Omega_\Phi=0,
  \label{density}
\end{equation}
differing from \eqref{Heqn2} in the dependence of modified gravitational dimensionless energy densities $\Omega_\Psi$ and $\Omega_\Phi$, on the torsion. These are given in \cref{cc16}, and are rational functions\footnote{Note also that there is considerable freedom between these densities, if they are constrained only by \eqref{oh} and \eqref{ok}, and that the notation is designed with \cref{JJ} and \cosmicclass{14} and \cosmicclass{11} in mind.} of the form
\begin{subequations}
  \begin{gather}
    \Omega_\Phi=\Omega_\Phi\big(\kappa^{\frac{1}{2}}Q\big|\sigma_1,\sigma_2,\upsilon_2\big),\label{oh}\\
    \Omega_\Psi=\Omega_\Psi\big(\kappa^{\frac{1}{2}}\partial_tQH^{-1},\kappa^{\frac{1}{2}}Q\big|\sigma_1,\sigma_2,\upsilon_2  \big).\label{ok}
\end{gather}
\end{subequations}
This dependence may in principle be eliminated in favour of $H$ by means of the remaining torsion equation \eqref{tor2} which takes the form
\begin{equation}
  f_1\frac{\partial^2_tQ}{Q}+f_2\frac{\left(\partial_tQ\right)^2}{Q^2}+f_3\frac{\partial_tQ}{Q}H+f_4\partial_tH+f_5H^2=0,
  \label{aux}
\end{equation}
where the various coefficients are again confined to \cref{cc16} for the sake of brevity, and are also rational functions of the form
\begin{equation}
  f_i=f_i\big(\kappa^{\frac{1}{2}}Q\big|\sigma_1,\sigma_2,\upsilon_2\big).
\end{equation}

The coupled second order system of \eqref{density} and \eqref{aux} is generally challenging to solve, but despite the doubtful nature of the constraints \eqref{yangmill} and \eqref{ksc}, we are not disappointed if we look for the kind of curvature evolution suggested by GR. Since \cosmicclass{16} is fundamentally $k$-screened, it is logical to consider analogies with traditional $k=0$ solutions -- as discussed in \cref{intro}, these are in contemporary focus anyway. The evolution of $R$ in GR is often broken down into regimes where a particular cosmic fluid is dominant. For the material sources under consideration, \eqref{Heqn} and \eqref{qeqn} demand that $R$ then approach a power-law in $t$, depending on the dominant equation-of-state parameter $w_i$ in \eqref{eos}
\begin{equation}
    H_\text{m}=2/3t, \quad H_\text{r}=1/2t, \quad H_\Lambda=\sqrt{\Lambda/3}.
\label{powerlaws}
\end{equation}
Remarkably, \cosmicclass{16} can mimic this behaviour. We require only that the modified gravitational densities be constant when a fluid of particular $w_i$ is dominant
\begin{equation}
  \Omega_\Phi+\Omega_\Psi=-1/g_i,
\end{equation}
at which point \eqref{density} will then coincide with \eqref{Heqn2} up to a modified Einstein constant
\begin{equation}
  \breve{\kappa}=g_i\kappa.
\end{equation}

Examination of \eqref{oh} and \eqref{ok} suggests that this can be achieved by constant $Q=Q_i$, which in turn greatly simplifies \eqref{aux} to a form which, for $H=H_i$ as in \eqref{powerlaws}, \textit{remains consistent} for as long as pure fluid dominance holds. We may thus hypothesise that a universe of \cosmicclass{16} will routinely `freeze out' into epochs of traditional flat GR behaviour. In this case the full complexity of the modified cosmological equations is confined to turnover epochs, and otherwise manifest in the specific value of the constant torsion $Q_i$ and modified Einstein constant $\breve\kappa_i$ during pure fluid dominance.

\begin{figure*}[t!]
  \includegraphics[width=\linewidth]{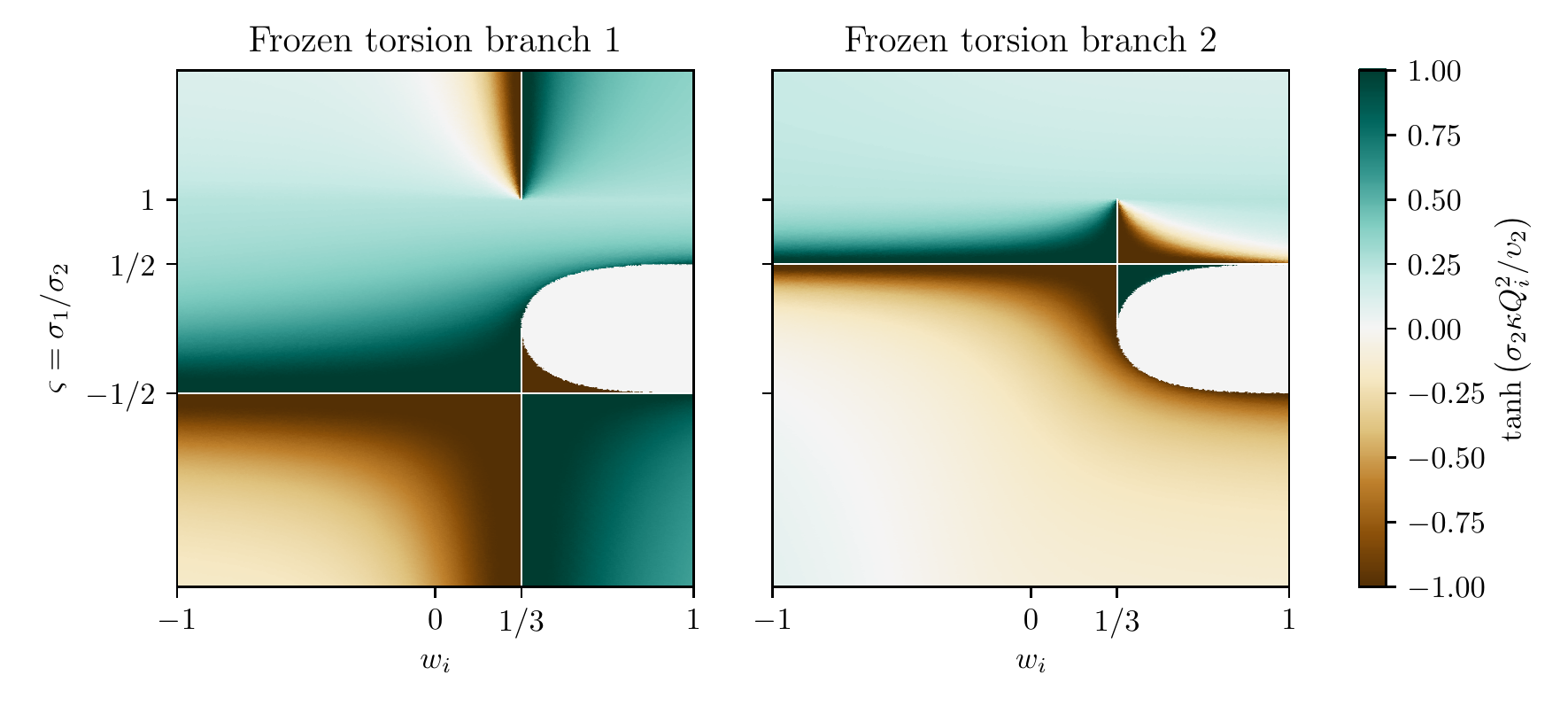}
  \caption{\label{frozen_plot}Within \cosmicclass{16}, the root system \eqref{frozen_eq} of constant torsion, $Q_i$, frozen out by a dominant cosmic fluid with equation of state parameter $w_i$, depends on the ratio of cosmic theory parameters, $\sigma_1/\sigma_2$. Freezing at a real torsion value generally appears possible for inflationary fluids ranging from dark energy, $w_\Lambda=-1$, through to curvature, $w_k=-1/3$ (although curvature cannot be re-imagined as a source in $k$-screened theories), and so on to matter, $w_\text{m}=0$. Radiation, at $w_\text{r}=1/3$, clearly occupies a privileged position in the overall theory, while extension to `stiff matter' with $w_\text{s}=1$ may be impossible over a range of $\varsigma=\sigma_1/\sigma_2$. Of particular interest is the case $\varsigma=1$, which corresponds to \cosmicclass{null}, and for which $\kappa Q_i^2\equiv\kappa Q_{\text{cor}}^2=\upsilon_2/4\sigma_1$ across all fluids except for radiation, which requires special treatment.}
\end{figure*}

The potential for this behaviour is worth some general investigation within \cosmicclass{16}, whose Lagrangian freedoms are partially parametrised by the ratio 
\begin{equation}
\varsigma=\sigma_1/\sigma_2.
\label{vsdef}
\end{equation}
Setting $Q=Q_i$ under a dominant cosmic fluid with equation of state parameter $w=w_i$, the remaining torsion equation \eqref{aux} may be naively solved for $Q_i$ by setting $H_i=2/3(1+w_i)t$, which yields the following
\begin{equation}
  \begin{aligned}
    (4\sigma_2/\upsilon_2)(&12\varsigma^2w_i-4\varsigma^2-3w_i+1)\kappa Q_i^2=\\
&6w_i\varsigma^2+2\varsigma^2+6w_i\varsigma-6\varsigma-3w_i+1\\
&\pm 2\big[9\varsigma^4w_i^2+6\varsigma^4w_i-18\varsigma^3w_i^2+\varsigma^4\\
  &\hphantom{\pm 2\big[}-12\varsigma^3w_i+9\varsigma^2w_i^2-2\varsigma^3+3\varsigma^2\\
    &\hphantom{\pm 2\big[}+12\varsigma w_i-4\varsigma-6w_i+2\big]^{1/2}.
  \label{frozen_eq}
\end{aligned}
\end{equation}
The somewhat complementary branches of this root system are illustrated in \cref{frozen_plot}. Superficially, this suggests that Einstein freezing can occur across many instances of \cosmicclass{16} for a variety of source fluids. Note however that radiation with $w_\text{r}=1/3$ appears to occupy a special place in \cosmicclass{16}.

Numerically, it proves easy to induce such emergent flat GR behaviour, and this is best demonstrated by means of a series expansion out of the classical radiation-dominated Big Bang. When propagating the cosmological equations of motion, a convenient choice of dimensionless time similar to \eqref{dim_conf} is given by normalising with the contemporary Hubble number
\begin{equation}
  \mathrm{d}\tilde\tau=R_0H_0\mathrm{d}t/R.
\end{equation}
When combined with the dimensionless scale factor
\begin{equation}
  a=R/R_0,
\end{equation}
this has the advantage that the Friedmann equations of GR, \eqref{Heqn} and \eqref{qeqn}, in the flat case become
\begin{subequations}
\begin{gather}
  (\partial_{\tilde\tau}a)^2=\Omega_{\text{r},0}+\Omega_{\text{m},0}a+\Omega_{\Lambda,0}a^4, \\
  (\partial_{\tilde\tau}a)^2-a\partial^2_{\tilde\tau}a=\Omega_{\text{r},0}+\tfrac{1}{2}\Omega_{\text{m},0}a-\Omega_{\Lambda,0}a^4,
\end{gather}
\end{subequations}
i.e. a form where the contemporary dimensionless densities are the only free parameters. It is then easy to obtain the following power series for GR out of radiation dominance
\begin{equation}
  a=\sqrt {{ {\Omega_{\text{r},0}}}}{\tilde\tau}+{\frac {{ {\Omega_{\text{m},0}}}}{4}}\,{{\tilde\tau}}^{2}+{\frac {{ {\Omega_{\Lambda,0}}}}{10}{{ {\Omega_{\text{r},0}}}}^{{\frac{3}{2}}}}{{\tilde\tau}}^{5}+\mathcal{O}(\tilde\tau^6).

  \label{GR_series}
\end{equation}
Applying this approach to \cosmicclass{16} results in a power seies for $a$ and separate series for $Q$ and $U$. These are all rather cumbersome, but can be used to integrate the modified cosmological equations as follows. Assuming \eqref{Heqn2m} remains valid, we can propagate the coupled second-order system in $Q$ and $R$ formed from the modified deceleration equation (the linear combination of \eqref{S} and \eqref{R} orthogonal to \eqref{density}), and \eqref{aux}, using \eqref{density} as a constraint.
The resulting evolution of the comoving Hubble horizon $H_0/aH$ is plotted against the scale factor $a$ in \cref{horizon_plot}, over a range of $\varsigma$.
Note that in \cref{horizon_plot}, the initial conditions are tweaked to agree with the flat GR model as far as possible. This involves, for every instance of \cosmicclass{16} defined by $\varsigma$, adapting $\upsilon_2$ so that $\breve{\kappa}=\kappa$. We see that for $\varsigma$ of order unity, the radiation, matter and dark energy dominated regimes familiar from flat GR are cleanly picked out.
The freezing of torsion by radiation, matter and dark energy is also apparent for some values of $\varsigma$ in \cref{horizon_plot}.
\subsubsection{\cosmicclass{null}: dark radiation}\label{dt}
\begin{figure*}[t!]
  \includegraphics[width=\linewidth]{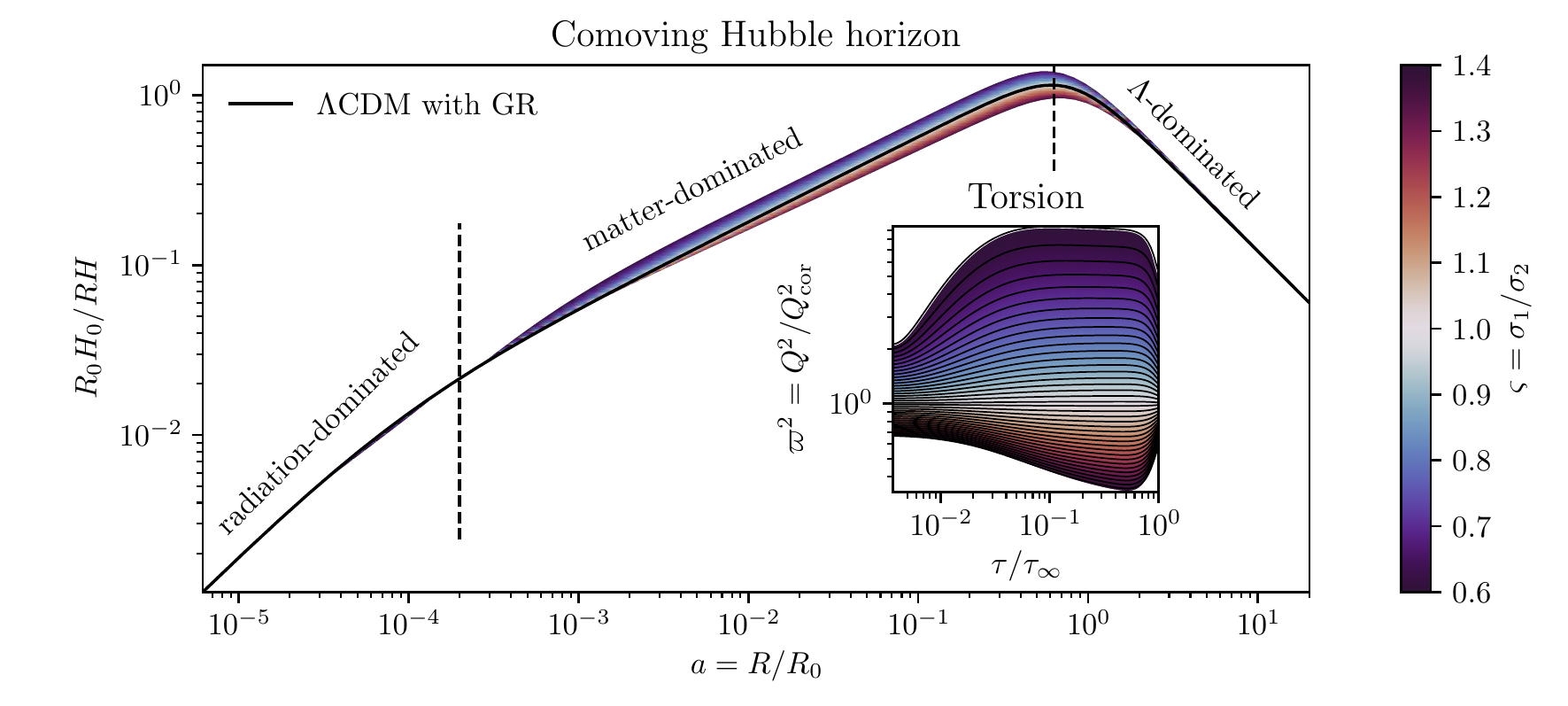}
  \caption{\label{horizon_plot} \textit{Main}: The cosmological equations of \cosmicclass{16} are propagated from $z\approx \num{1.63e5}$ ($12$ $e$-folds) using the corresponding primordial density parameters of flat GR (based on $\Omega_{\text{r},0}=\num{2.47e-5}$, $\Omega_{\text{m},0}=0.3089-\Omega_{\text{r},0}/2$ and $\Omega_{\Lambda,0}=0.6911-\Omega_{\text{r},0}/2$ with neutrinos neglected), with the GR evolution also shown. At this initial radiation-dominated epoch, $\breve{\kappa}=\kappa$ is fixed with $\upsilon_2=4\sigma_1/(\sigma_2-4\sigma_1)$. \textit{Inset}: The $Q$ torsion remains finite for the whole evolution, and may be plotted up to the Future Conformal Boundary at $\tau_\infty$. For general $\varsigma$, each epoch of equality triggers a smooth transition to a new torsion value, the intermediate $Q_\text{m}$ plateau is visible for $\varsigma<1$. Arbitrarily close agreement with GR is seen as $\varsigma=\sigma_1/\sigma_2\to 1$, which corresponds to \cosmicclass{null}. In this case, the correspondence solution keeps the torsion fixed throughout at $Q\equiv Q_{\text{cor}}$, or $\varpi=Q/Q_{\text{cor}}\equiv 1$.}
\end{figure*}

From the analysis in \cref{horizon_plot,frozen_plot} of the variable $\varsigma$ which parameterises \cosmicclass{16}, we see that an algebraically natural choice of theory defined by the additional constraint
\begin{equation}
  \sigma_1-\sigma_2=0,
  \label{b_constraint}
\end{equation}
or $\varsigma=1$, is especially significant. We will refer to \cosmicclass{16} in combination with \eqref{b_constraint} as \cosmicclass{null}. Since it is not defined by any critical case, \cosmicclass{null} does not appear in the map of cosmologies in \cref{map_plot} -- note however that \cosmicclass{null} and \criticalcase{16} remain compatible. 

To see the significance of \eqref{b_constraint}, first note from \cref{horizon_plot} that \cosmicclass{null} is defined by precisely the value $\varsigma=1$ that imitates the expansion of flat GR cosmology, when propergated from the same initial conditions. In this case the $Q_i$ and $g_i$ all coincide at the same `correspondence values' across the three $w_i$ of radiation, matter and dark energy
\begin{equation}
  \kappa Q_i^2\equiv\kappa Q_\text{cor}^2=\upsilon_2/4\sigma_1,\quad g_i\equiv g_\text{cor}= -4/3\upsilon_2,
  \label{b_torsion}
\end{equation}

and moreover do not deviate from these values during turnover epochs\footnote{It is important to note that the particular form of the equations of motion \eqref{density}, \eqref{aux} and particularly \eqref{Heqn2m} only allow for this solution if a careful limit is taken.}. 
In order to recover the correct sign of the modified Einstein constant, we will need 
\begin{equation}
  \upsilon_2<0,
  \label{u2_ineq}
\end{equation}
and likewise for real torsion
\begin{equation}
  \sigma_1=\sigma_2<0.
  \label{s2_ineq}
\end{equation}
Confirmation of this behaviour can be seen in \cref{frozen_plot}, since $\varsigma=1$ is actually a contour in both branches of the frozen torsion value, except at the intersection with $w_i=1/3$. Moreover, we see that $\varsigma=1$ is one of the special cases of \cosmicclass{16} for which frozen torsion cannot escape the vertical radiation asymptote simply by switching branches. We refer to the solution \eqref{b_torsion} to \cosmicclass{null}, in which flat GR evolution is naturally recovered, as the \textit{correspondence solution}.

While very encouraging in itself, in the absence of any measurement of $Q_0$ today and pinning $g_\text{cor}=1$ to recover $\breve{\kappa}\equiv\kappa$, the correspondence solution introduces no new parameters to cosmology: we thus seek to relax it. To do so, we will turn back to the series expansion out of the radiation-dominated Big Bang. It proves useful to define the dimensionless deviation from the correspondence torsion as
\begin{equation}
  \varpi=Q/Q_\text{cor}.
\end{equation}
Guided by \cref{frozen_plot}, closer examination of the intersection of $w_i=1/3$ with $\varsigma=1$ reveals something interesting: the spectrum of possible $Q_\text{r}$ or $\varpi_\text{r}$ is in fact continuous here, introducing a free parameter. 
If therefore, we do not need to fix $\varpi_\text{r}=1$ at the singularity, the general power series for the scale factor in \cosmicclass{null} is
\begin{widetext}
  \begin{align}
    a=&{\frac {{g_{\text{cor}}}}{{ {\varpi_{\text{r}}}}}\sqrt {{ {\Omega_{\text{r},0}}}}}{\tilde\tau}+{\frac {{ {\Omega_{\text{m},0}}}\, \left( 3\,{{ {\varpi_{\text{r}}}}}^{2}+1 \right) {{g_{\text{cor}}}}^{2}}{16\,{{ {\varpi_{\text{r}}}}}^{2}}}{{\tilde\tau}}^{2}+{\frac {5\,{{ {\Omega_{\text{m},0}}}}^{2}{{g_{\text{cor}}}}^{3} \left( {{ {\varpi_{\text{r}}}}}^{2}-1 \right) }{512\,{{ {\varpi_{\text{r}}}}}^{3}}{\frac {1}{\sqrt {{ {\Omega_{\text{r},0}}}}}}}{{\tilde\tau}}^{3}\nonumber\\&+{\frac {{{ {\Omega_{\text{m},0}}}}^{3} \left( 27\,{{ {\varpi_{\text{r}}}}}^{2}-121 \right) {{g_{\text{cor}}}}^{4} \left( {{ {\varpi_{\text{r}}}}}^{2}-1 \right) }{49152\,{{ {\varpi_{\text{r}}}}}^{4}{ {\Omega_{\text{r},0}}}}}{{\tilde\tau}}^{4}\nonumber\\&+{\frac { \left( -441\,{{ {\varpi_{\text{r}}}}}^{4}{{ {\Omega_{\text{m},0}}}}^{4}+98304\,{{ {\varpi_{\text{r}}}}}^{2}{ {\Omega_{\Lambda,0}}}\,{{ {\Omega_{\text{r},0}}}}^{3}+1421\,{{ {\varpi_{\text{r}}}}}^{2}{{ {\Omega_{\text{m},0}}}}^{4}+32768\,{ {\Omega_{\Lambda,0}}}\,{{ {\Omega_{\text{r},0}}}}^{3}-980\,{{ {\Omega_{\text{m},0}}}}^{4} \right) {{g_{\text{cor}}}}^{5}}{1310720\,{{ {\varpi_{\text{r}}}}}^{5}}{{ {\Omega_{\text{r},0}}}}^{-{\frac{3}{2}}}}{{\tilde\tau}}^{5}\nonumber\\&+\mathcal{O}(\tilde\tau^6),
\label{a_series}
  \end{align}
\end{widetext}
and by comparing \eqref{GR_series} to \eqref{a_series} we see that the two series can be made to coincide by setting $\varpi_\text{r}=1$. Doing so guarantees the other half of the correspondence solution -- the constancy of $\varpi\equiv 1$ throughout the evolution -- which can be seen by examining the \cosmicclass{null} power series for $\varpi$
\begin{widetext}
  \begin{align}
    \varpi=&{ {\varpi_{\text{r}}}}+{\frac {3\,{ {\Omega_{\text{m},0}}}\,{g_{\text{cor}}} \left( {{ {\varpi_{\text{r}}}}}^{2}-1 \right) }{16}{\frac {1}{\sqrt {{ {\Omega_{\text{r},0}}}}}}}{\tilde\tau}+{\frac {{{ {\Omega_{\text{m},0}}}}^{2}{{g_{\text{cor}}}}^{2} \left( 18\,{{ {\varpi_{\text{r}}}}}^{2}+13 \right)  \left( {{ {\varpi_{\text{r}}}}}^{2}-1 \right) }{512\,{ {\Omega_{\text{r},0}}}\,{ {\varpi_{\text{r}}}}}}{{\tilde\tau}}^{2}\nonumber\\&+{\frac {{{ {\Omega_{\text{m},0}}}}^{3}{{g_{\text{cor}}}}^{3} \left( 324\,{{ {\varpi_{\text{r}}}}}^{4}+279\,{{ {\varpi_{\text{r}}}}}^{2}+299 \right)  \left( {{ {\varpi_{\text{r}}}}}^{2}-1 \right) }{49152\,{{ {\varpi_{\text{r}}}}}^{2}}{{ {\Omega_{\text{r},0}}}}^{-{\frac{3}{2}}}}{{\tilde\tau}}^{3}\nonumber\\&-{\frac {{{g_{\text{cor}}}}^{4} \left( -1620\,{{ {\Omega_{\text{m},0}}}}^{4}{{ {\varpi_{\text{r}}}}}^{6}-1620\,{{ {\varpi_{\text{r}}}}}^{4}{{ {\Omega_{\text{m},0}}}}^{4}-1462\,{{ {\varpi_{\text{r}}}}}^{2}{{ {\Omega_{\text{m},0}}}}^{4}+98304\,{ {\Omega_{\Lambda,0}}}\,{{ {\Omega_{\text{r},0}}}}^{3}-2327\,{{ {\Omega_{\text{m},0}}}}^{4} \right)  \left( {{ {\varpi_{\text{r}}}}}^{2}-1 \right) }{1310720\,{{ {\Omega_{\text{r},0}}}}^{2}{{ {\varpi_{\text{r}}}}}^{3}}}{{\tilde\tau}}^{4}\nonumber\\&+\mathcal{O}(\tilde\tau^5).
\label{q_series}
  \end{align}
\end{widetext}

\begin{figure*}[t!]
  \includegraphics[width=\linewidth]{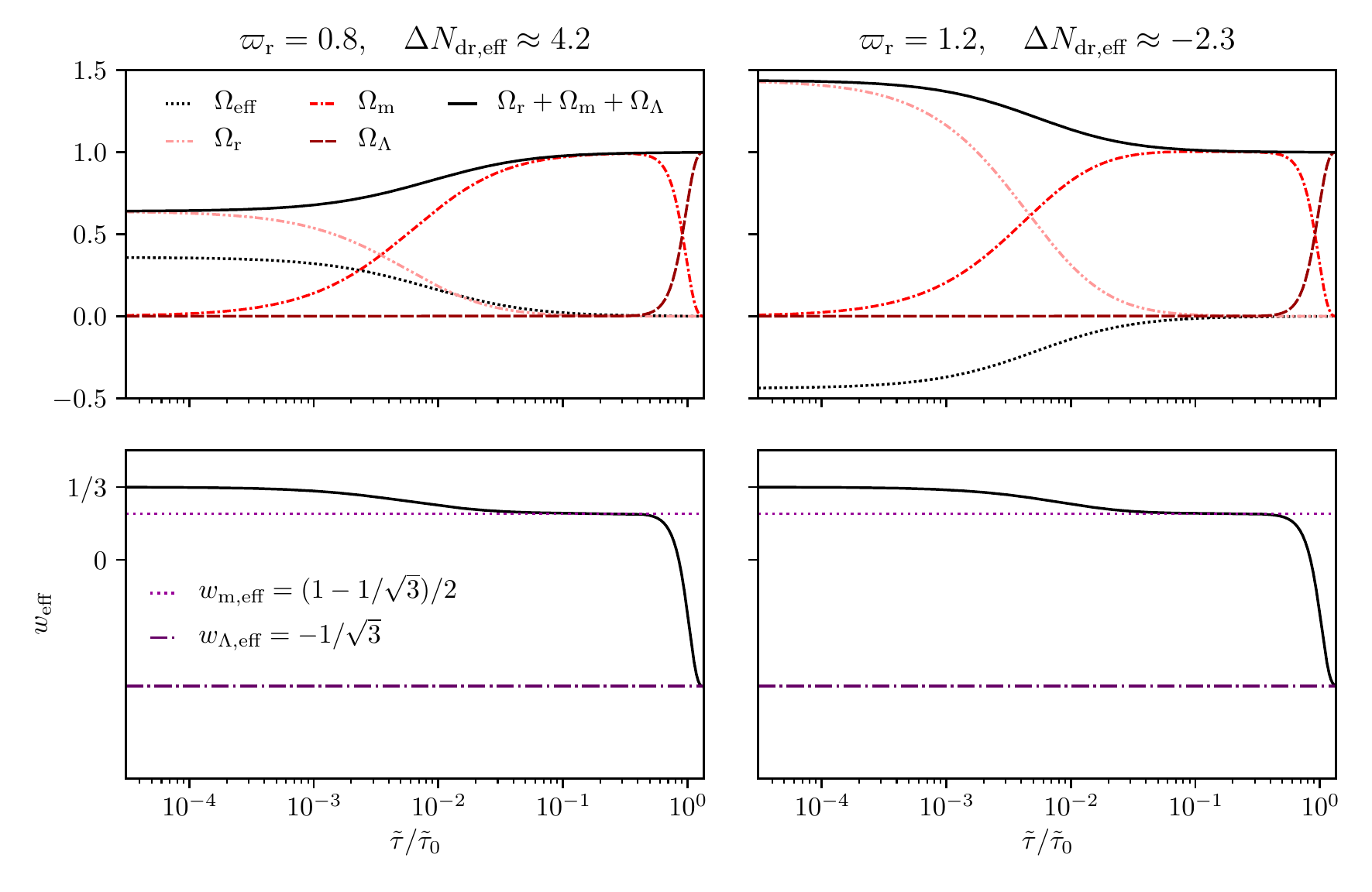}
  \caption{\label{density_plot}Within \cosmicclass{null}, density parameters and effective equation of state parameter for a Big Bang with positive and negative dark radiation fractions. The effective equation of state cleanly picks out the frozen regimes in \eqref{pert_f}, and consequently the dark sector redshifts away more slowly than radiation after the first turnover. Note that dark radiation with positive energy has a tendency to advance the epoch of equality.}
\end{figure*}

This translates into precisely the relaxation of the correspondence solution we had sought. Rather than interpreting the effect of arbitrary $Q_\text{r}$ through a time-varying renormalisation of the Einstein constant $\breve{\kappa}$, it is useful to cast it as a gravitational \textit{extra component} which must be added to the bare (physical) matter in \eqref{Heqn} to account for the actual curvature evolution. This we will now do, and take the opportunity to combine the analysis with a crude stability check of the correspondence solution itself. To this end, we perturb the cosmological equations around the correspondence solution of some pure bare matter $w_i$, taking the origin of $\tilde\tau$ to be either the Big Bang as exited to the right or Future Conformal Boundary as approached from the left
\begin{equation}
  \sgn(3w_i+1)=\sgn(\tilde\tau).
\end{equation}
The perturbation of the correspondence curvature evolution is supposedly generated by a perturbation from correspondence torsion, or taking $\varepsilon$ to be some small parameter
\begin{equation}
  \begin{aligned}
    \varpi&=1+\varepsilon\delta \varpi+\mathcal{O}(\varepsilon^2),\\ a&=\left(\tfrac{3w_i+1}{2}\tilde\tau\right)^{\frac{2}{3w_i+1}}+\varepsilon\delta a+\mathcal{O}(\varepsilon^2).
  \label{pert_q}
\end{aligned}
\end{equation}
For the bare fluids anticipated here, we find that to first perturbative order the deviation from correspondence torsion typically decays away as a power law in normalised conformal time $\tilde\tau$ away from the Big Bang or towards the Future Conformal Boundary
\begin{equation}
  \delta \varpi=
  \begin{cases}
    \left(c_1\tilde\tau^{-1}+c_2\right)^2  & w_i=1/3 \\
    \left(c_1\tilde\tau^{-\frac{3+\sqrt{3}}{2}}+c_2\tilde\tau^{-\frac{3-\sqrt{3}}{2}}\right)^2  & w_i=0 \\
    \left(c_1\tilde\tau^{\frac{3+\sqrt{3}}{2}}+c_2\tilde\tau^{\frac{3-\sqrt{3}}{2}}\right)^2  & w_i=-1. \\
  \end{cases}
  \label{pert_a}
\end{equation}
We take this to confirm the stability of the correspondence solution under pure fluid dominance.
The obvious exception is the arbitrary constant torsion deviation under bare \textit{radiation} dominance. This was of course anticipated as part of the relaxation procedure, and it need not be perturbative at all.
The solutions \eqref{pert_a} and \eqref{pert_q} can now be used to account for the extra components to which they correspond
\begin{equation}
  \left( \partial_{\tilde\tau}a \right)^2-a^{1-3w_i}=\varepsilon \kappa a^4\delta\rho/3H_0^2+\mathcal{O}(\varepsilon^2),
\end{equation}
which take the following forms\footnote{The precise dependece of $c_3$ and $c_4$ on $c_1$ and $c_2$ is suppressed for brevity.}
\begin{equation}
  a^4\delta\rho=
  \begin{cases}
    c_3+c_4a^{-2}  & w_i=1/3 \\
    c_3a^{-\frac{1+\sqrt{3}}{2}}+c_4a^{-\frac{1-\sqrt{3}}{2}}  & w_i=0 \\
    c_3a^{1+\sqrt{3}}+c_4a^{1-\sqrt{3}} & w_i=-1. \\
  \end{cases}
  \label{pert_f}
\end{equation}
Note that \eqref{pert_f} is consistent with \eqref{pert_a} in that a \textit{decaying} deviation from correspondence torsion is manifest as a strictly sub-dominant extra component. After a while, the extra component may be approximated by the contribution from the slowest-decaying torsion mode, and we see that it quietly redshifts away under the dominant bare matter in all cases but bare radiation. For this reason, we anticipate an arbitrary co-dominant dark radiation component to accompany bare radiation until the epoch of equality, a small amount of hot dark matter with $w_{\text{m,eff}}\approx 0.211$ to accompany bare matter and, after the contemporary turnover, a miniscule amount of non-phantom dark energy with $w_{\Lambda\text{,eff}}\approx -0.577$ to accompany bare dark energy. These values, which we introduced in \eqref{proposal} in \cref{intro}, can readily be obtained from \eqref{pert_f}.

\begin{figure*}[t!]
  \includegraphics[width=\linewidth]{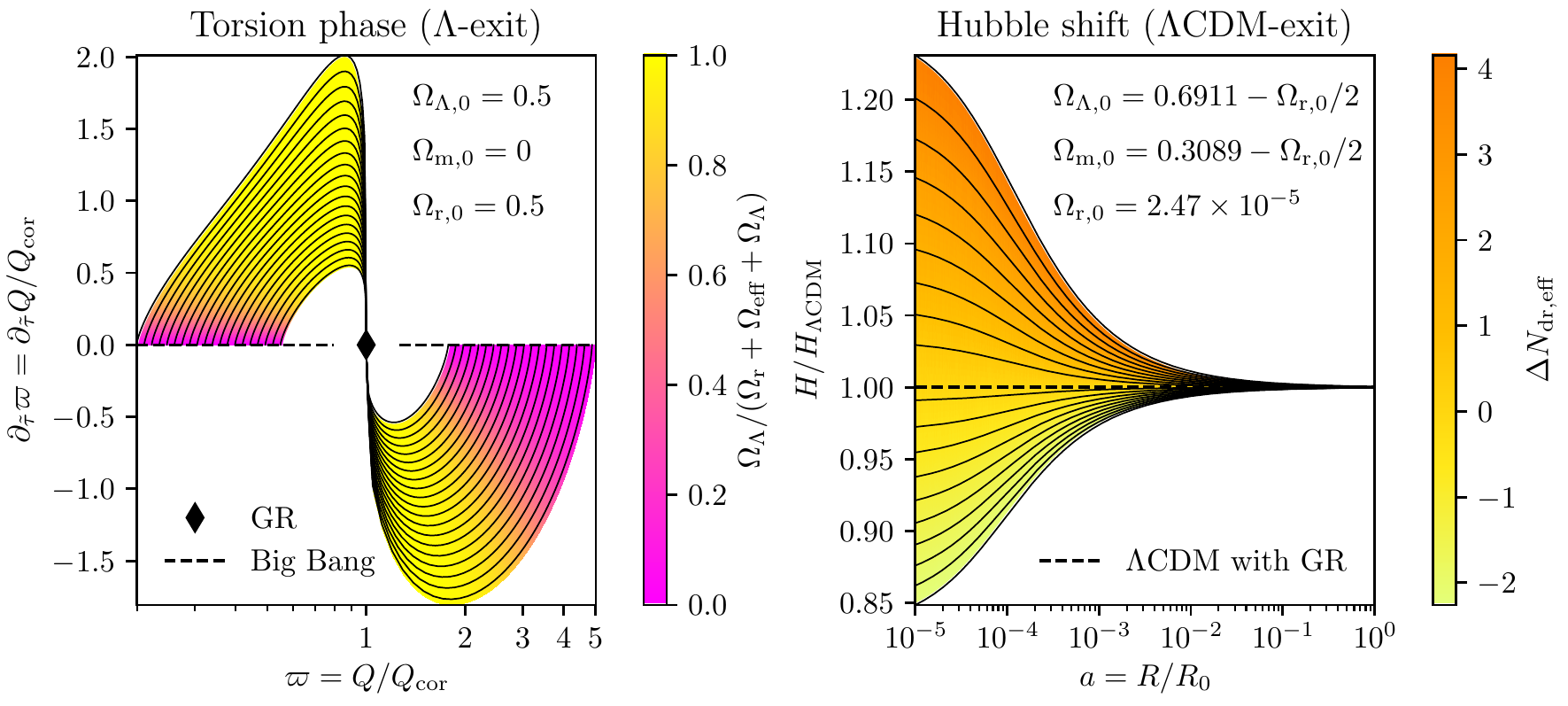}
  \caption{\label{attractor_plot}Reliable emergence of Einsteinian cosmology from \cosmicclass{null}. \textit{Left}: The correspondence solution attracts the torsion to the value $Q_\text{cor}$, here illustrated in the phase space of $Q$ for a toy model with no matter. Torsion is at rest at the Big Bang when the only sources are dark energy and radiation as shown here -- though if matter is present it begins to decay immediately, and propagates off a parabola in phase space; \textit{right}: compared to \textLambda CDM, initial dark radiation allows one-parameter tuning of the expansion rate during radiation-dominance. Compared to the equivalent plot in \cite{Pandey} we are allowed both increased and decreased early expansion because our extra component is effective, furthermore the effect is heavily suppressed at modern times.}
\end{figure*}

Numerical investigation suggests that this version of events is surprisingly robust, in that large positive or negative dark radiation fractions in the early universe are typically eliminated by the first turnover they encounter. The analytic predictions for the effective equation of state parameter are borne out in \cref{density_plot}. The ability of the theory to recover \textLambda CDM evolution at late times over a wide range of $\varpi_\text{r}$ is especially striking in toy universes without bare matter, as illustrated in \cref{attractor_plot}: the correspondence solution superficially resembles a damped harmonic attractor out of initial dark radiation dominance\footnote{We will not attempt to prove that the critical solution is actually an attractor state, but rather suffice with the stability properties mentioned here.}.

In the broadest terms, we can understand the arbitrary-$\varpi_\text{r}$ solution to \cosmicclass{null} as a positive or negative dark radiation component in the early universe. A crude translation into the nomenclature of \textLambda CDM mentioned in \cref{intro} is simply to absorb this dark radiation into the effective post-standard model relativistic degrees of freedom $\Delta N_\text{dr,eff}$ as follows
\begin{equation}
  \Delta N_\text{dr,eff}=\left( \varpi_\text{r}^{-2}-1 \right)\left( \tfrac{8}{7}\left( \tfrac{11}{4}\right)^{4/3}+N_{\nu,\text{eff}} \right).
  \label{neff}
\end{equation}
This heuristic formula is the basis of the $\Delta N_\text{dr,eff}$ values referenced in \cref{density_plot} and \cref{attractor_plot}, given the Planck 2018 estimate of $N_{\nu,\text{eff}}=2.99\pm0.17$ \cite{2018arXiv180706209P}. This estimate may fall foul of circularity arguments due to the GR interpretation of the Planck data, and direct \cite{2018PhR...754....1P} $\Delta N_{\nu,\text{eff}}$ estimations based on Big Bang nucleosynthesis (BBN) may be more appropriate. Finally we emphasise that the dark radiation approximation \textit{remains} an approximation: since the general arbitrary-$\varpi_\text{r}$ solution predicts a complicated dark sector with a dynamical equation of state.

\subsubsection{Other special cases of \cosmicclass{16}}\label{JJ}
We initially proposed that \cosmicclass{16} defined by \cref{yangmill,ksc,tor1cons}, be refined to \cosmicclass{null} by the final constraint \eqref{b_constraint} in order to satisfy correspondence with flat GR. It is also worthwhile investigating alternative constraints appearing in \cref{map_plot} which definitely alter the particle content of the theory -- ideally for the better.
\paragraph*{\cosmicclass{14}: $k$-screened dynamically open.}
An additional constraint
\begin{equation}
  \sigma_2=0,
  \label{s20}
\end{equation}
focuses \cosmicclass{16} onto \cosmicclass{14}. This is the cosmic class of \criticalcase{14} which may admit massless $2^+$ gravitons as with \criticalcase{16}, and also of \criticalcase{8}, though the massless graviton in this case is not expected to be $2^+$. Furthermore, \eqref{s20} appears to have as profound a `taming' effect on \cosmicclass{16} as the constraint \eqref{b_constraint} does. Since our analysis in \eqref{frozen_eq} cannot be recycled to show this without a certain amount of difficulty, we will begin again from first principles. The cosmic implications of the quadratic Riemann sector in \cosmicclass{14} are characterised by the single parameter $\sigma_1$, and those of the quadratic torsion by $\upsilon_2$. The latter generally maintains the broken cosmological NSI, allowing for matter as a cosmic fluid. The cosmological equations of motion are significantly simplified by defining two fields from the observable torsion quantities, $\Phi$ and $\Psi$ of dimension $\si{\electronvolt}$
\begin{equation}
  \Psi=\frac{\upsilon_2U}{4\sqrt{3}\sigma_1\kappa Q^2}-\frac{\sqrt{3}\partial_tQ}{Q}, \quad \Phi=\Psi-\frac{U}{\sqrt{3}}.
  \label{feldef}
\end{equation}
The density balance equation now adopts a form very similar to the first Friedmann equation, \eqref{Heqn} or \eqref{Heqn2}
\begin{equation}
  \Omega_\text{r}+\Omega_\text{m}+\Omega_\Lambda+\Omega_\Phi+\Omega_\Psi=0,
  \label{HeqnJ}
\end{equation}
where the dimensionless densities of the torsion fields are entirely analogous to that of the cosmological constant, in that $\rho_\Phi=-\kappa^{-1}\Phi^2$ and $\rho_\Psi=\kappa^{-1}\Psi^2$ are incorporated as
\begin{equation}
  \Omega_\Phi=\frac{\upsilon_2\kappa\rho_\Phi}{3H^2},\quad \Omega_\Psi=\frac{\upsilon_2\kappa\rho_\Psi}{3H^2}.
  \label{dimdens}
\end{equation}
This re-labelling becomes meaningful when we apply it to the torsion equations \eqref{tor1} and \eqref{tor2}, which, if $Q\neq 0$, become respectively
\begin{subequations}
  \begin{gather}
    \Psi=\sqrt{3}H\label{timpro},\\
  \partial_t\Phi+H\Phi=0\label{tempro}.
\end{gather}
\end{subequations}
These immediately allow us to express \eqref{HeqnJ} purely in terms of $R$, $H$ and various constants, thus encoding the curvature-evolution. Specifically, we have from \eqref{tempro}
\begin{equation}
  \Phi=\chi/R,
\end{equation}
where $\chi$ is a constant of integration, so that the density equation reduces to
\begin{equation}
  \Omega_\text{r}+\Omega_\text{m}+\Omega_\Lambda-\frac{\upsilon_2\chi^2}{3H^2R^2}=-\upsilon_2.
  \label{fundamental}
\end{equation}
Given the \textit{same} inequality constraint \eqref{u2_ineq} that was so vital to the root theory when constrained by \eqref{b_constraint}, and accepting that for $\Phi$ be an observable quantity we must have $\chi^2\geq 0$, we have again uncovered emergent GR evolution, but now with a strictly negative \textit{effective} $k$. This, in a theory that is fundamentally $k$-screened, results in dynamically \textit{open} but geometrically \textit{arbitrary} cosmology.
It remains only to examine the evolution of the observable torsion quantities $U$ and $Q$. We find
\begin{subequations}
  \begin{gather}
    U=3H-\frac{\sqrt{3}\chi}{R},\label{Usol}\\
    \kappa Q^2=\frac{\upsilon_2}{4\sigma_1}-\frac{\upsilon_2\chi}{2\sqrt{3}\sigma_1R^2}\int\mathrm{d}t R.\label{Qsol}
\end{gather}
\end{subequations}
Once more this is not entirely dissimilar to the torsion evolution in classes \cosmicclass{16} and \cosmicclass{null}: on the approach to the radiation dominated Big Bang the unobservable torsion $U$ diverges, while the observable $Q$ converges.

\paragraph*{\cosmicclass{11}: power-law inflation.}
Yet another alternative constraint to \eqref{b_constraint} is \eqref{u20}: this acts on the torsion rather than curvature sector -- eliminating the former entirely.
This constraint defines \cosmicclass{11}, of \criticalcase{11} which again contains a propagating massless, potentially $2^+$ graviton and also has gauge-invariant PCR. An undesirable and damning side effect of \eqref{u20} is of course the introduction of cosmological NSI. Nonetheless, we repeat the procedure used for \cosmicclass{14} by redefining \eqref{feldef} as
\begin{equation}
\begin{gathered}
  \Psi=\frac{1}{\sigma_2-\sigma_1}\left( \frac{\sigma_2U}{\sqrt{3}}+\frac{\sigma_1\sqrt{3}\partial_t Q}{Q} \right),\\
  \Phi=\frac{\sigma_2\kappa^{\frac{1}{2}}}{\sigma_2-\sigma_1}\left( \frac{QU}{\sqrt{3}}+\sqrt{3}\partial_tQ \right).
  \label{newss}
\end{gathered}
\end{equation}
This time, the $\Psi$-field does not appear in the density balance equation, and the only possible source fluid is naturally NSI radiation
\begin{equation}
  \Omega_\text{r}+\Omega_\Phi=0.
  \label{final_density}
\end{equation}
The coupling constant is also redefined according to
\begin{equation}
  \Omega_{\Phi}=\frac{(4\sigma_1^2-\sigma_2^2)\kappa\rho_{\Phi}}{3\sigma_2H^2}.
\end{equation}
From \cosmicclass{14} we find \eqref{timpro} is remains valid, while \eqref{tempro} is slightly modified to
\begin{equation}
  \partial_t\Phi+2H\Phi=0\label{tempro2}.
\end{equation}
This immediately translates to another effective radiation component which renders \eqref{final_density} useless. The curvature evolution is thus determined by the remaining torsion equations, which may be solved to give the following
\begin{equation}
  U=0, \quad \frac{\partial_t Q}{Q}=\frac{\sigma_1-\sigma_2}{(\sigma_1+\sigma_2)t}, \quad H=\frac{\sigma_2}{(\sigma_1+\sigma_2)t},
\end{equation}
implying a potentially inflationary expansion, according to a power-law (see also \cite{1988PhLB..207...31Y}) which depends on the theory parameters.
\paragraph*{\cosmicclass{10}}
The final combination of \eqref{s20} with \eqref{u20} results in \cosmicclass{10}. While \criticalcase{10} (unlike \criticalcase{9}) again may contain a massless $2^+$ graviton and has the gauge-invariant PCR property, the cosmology is even more impoverished than \cosmicclass{11}, and we will not discuss it further. We will stop short of generalising the $\Phi$-$\Psi$ formalism in reverse to \cosmicclass{16} or repeating the analysis of \eqref{frozen_eq} with conformally transformed $\varsigma$ so as to better accommodate \cosmicclass{14}. This concludes the summary of the child theories of \cosmicclass{16}.

\subsubsection{\cosmicclass{1}: cyclic cosmologies}\label{sibling}
In focussing on \cosmicclass{16} in \cref{newroot} and its child cosmologies in \cref{dt,JJ}, we have neglected the parent \cosmicclass{2} and siblings \cosmicclass{1} and \cosmicclass{15}. The particle content of \criticalcase{15} of \cosmicclass{15} is similar to that of \criticalcase{16} of \cosmicclass{16}, with a potential massless $2^+$ graviton. Indeed, \cosmicclass{15} and its child \cosmicclass{12} are good candidates for further investigation. In this section, we will very briefly focus on \cosmicclass{1}, which instead has a similar particle content to the parent cosmology, \cosmicclass{2}. Both classes are richly populated by critical cases with massive $0^-$ gravitons, though \criticalcase{1} in \cosmicclass{1} may additionally contain a massless $2^+$ graviton. 

In particular, we will retain the fundamentals of a $k$-screened Yang-Mills theory with \eqref{yangmill} and \eqref{ksc}, but instead of \eqref{tor1cons} we will enforce \eqref{u20}. To highlight the emergent inflationary effects we will set $\Lambda=0$, admitting radiation and matter only. As a $k$-screened theory, the formula \eqref{Heqn2m} still allows us to solve for $U$ in terms of $Q$ and $H$. The usual energy balance equations are no longer especially insightful, and so we work again at the level of the dynamical variables. Curiously \eqref{S} allows $Q$ to be expressed purely in terms of the matter content
\begin{equation}
  Q^2=H^2\Omega_\text{m}/2\upsilon_1.
  \label{Q_dust}
\end{equation}
By substituting \eqref{Q_dust} and \eqref{Heqn2m} into \eqref{R} we then obtain the following solution
\begin{equation}
  a=c_1\left(\cosh(c_2t) -1 \right),
  \label{hyperbolic}
\end{equation}
where the amplitude depends on the ratio of radiation to matter, and the characteristic time on the cosmic theory parameters
\begin{equation}
  c_1=\Omega_{\text{r},0}/\Omega_{\text{m},0}, \quad \kappa c_2^2=\sigma_2\upsilon_1/(\sigma_2^2-4\sigma_1^2).
  \label{chartime}
\end{equation}
Thus, through a suitable choice of the theory parameters we may obtain either cyclic universes in which the Big Crunch and Big Bang are periodic, or perpetual exponential inflation to the Future Conformal Boundary.
\subsubsection{Unitarity inequalities}
Our analysis of each cosmic class relies only on the equalities that define promising critical cases. We now wish to combine this analysis with the accompanying unitarity inequalities in \cref{table}, so as to further constrain each theory. Of greatest concern is \cosmicclass{16}. Naturally not all of the inequalities are expressible purely in terms of the cosmic theory parameters, and we find the relevant inequality constraint on \criticalcase{16} reduces to
\begin{equation}
  (3r_5+2\sigma_1+\sigma_2)(3r_5+8\sigma_1+4\sigma_2)(2\sigma_1+\sigma_2)<0,
  \label{ineq_master}
\end{equation}
from which $r_5$ cannot be eliminated in favor of $\sigma_1$, $\sigma_2$ or $\upsilon_2$. This means that the unitarity of \criticalcase{16} does not constrain the cosmological picture of \cosmicclass{16} discussed in \cref{newroot}, or the cosmology of \cosmicclass{null} discussed in \cref{dt}.

Of the other child cosmologies of \cosmicclass{16} examined in \cref{JJ}, we find that unitarity of \criticalcase{11} of the cosmologically NSI \cosmicclass{11} also requires \eqref{ineq_master}. On the other hand, the quite promising \cosmicclass{14} requires
\begin{equation}
  \sigma_1(3r_5+2\sigma_1)(3r_5+8\sigma_1)<0,
  \label{ineq_mistress}
\end{equation}
for the unitarity of \criticalcase{14} -- once more $r_5$ cannot be expressed in terms of $\sigma_1$ or $\upsilon_2$. The other cosmologically NSI \cosmicclass{10} also requires \eqref{ineq_mistress} for the unitarity of \criticalcase{10}. Although not considered in the present work, we note that the intriguing \criticalcase{15} and \criticalcase{12} respectively of \cosmicclass{15} and \cosmicclass{12} also require \eqref{ineq_mistress}.

In fact, the unitarity inequalities only begin to impinge on the cosmology when massive $0^-$ gravitons are present. In \cref{sibling} we touched on \cosmicclass{1}. The relevant \criticalcase{1} which may contain a $2^+$ graviton also requires \eqref{ineq_master}, and two additional inequalities
\begin{equation}
  \sigma_2<0, \quad \upsilon_1<0.
  \label{<+label+>}
\end{equation}
Although these explicitly affect the cosmic theory parameters remaining to \cosmicclass{1}, they do not fully constrain the characteristic time \eqref{chartime} of the hyperbolic solution we consider in \eqref{hyperbolic}. We will not examine \criticalcase{27}, \criticalcase{30} or \criticalcase{35} of \cosmicclass{1}, since they do not contain massless particles.

\section{An alternative formalism}\label{multi}
\subsection{The spacetime algebra}
Having completed the physical picture, we will apply an alternative formulation of the relevant gravitational gauge theories to the quadratic invariants. We refer to the apparatus of geometric algebra used in gauge theory gravity (GTG)\footnote{Note that the name GTG may be confusing, since it is locally equivalent to ECT: the mathematical formulation of the theories is however quite different.}, which is to be contrasted with the ubiquitous tensor formalism employed above -- though both depict gravitational gauge fields on an unobservable Minkowskian background. As noted in \cref{intro}, this is already a potential source of difficulty (if not controversy) if one wishes to extend to aspects of GR such as wormholes. A superficial difference arises in the treatment of diffeomorphisms, which are \textit{actively} interpreted in GTG. The defining feature however is the extensive use of Clifford algebras, more specifically the \textit{spacetime algebra} (STA). A comprehensive introduction to the STA is provided by Hestenes and Sobczyk \cite{1985AmJPh..53..510H}, and also in \cite{doran-lasenby}. GTG itself is adequately explained in \cite{doran-lasenby,1998RSPTA.356..487L}, and we have also offered a brief introduction in \cite{2019JMP....60e2504B}. We will not recapitulate the gauge theory structure of \cref{recap} here, but follow \cite{lewis,ewgt_conformal} in applying the STA in a targeted manner to the field strength tensors and their quadratic invariants.

The elements of the STA, known as \textit{multivectors}, may be constructed from the Lorentz basis of vectors $\{\hat{\mathsf{e}}_a\}$ and dual basis, $\{\hat{\mathsf{e}}^a\}$. The \textit{geometric product} is represented by a simple juxtaposition of quantities; it is associative and distributative over addition, but not commutative. A geometric product between two vectors can be expanded into symmetric (interior) and antisymmetric (exterior) products
\begin{equation}
  \hat{\mathsf{e}}_a\hat{\mathsf{e}}_b=\hat{\mathsf{e}}_a\cdot\hat{\mathsf{e}}_b+\hat{\mathsf{e}}_a\wedge\hat{\mathsf{e}}_b.
  \label{gp}
\end{equation}
The first term on the RHS of \eqref{gp} is a scalar, and the basis vectors are Lorentzian in the sense that
\begin{equation}
  \hat{\mathsf{e}}_a\cdot\hat{\mathsf{e}}_b=\eta_{ab}, \quad \hat{\mathsf{e}}^a\cdot\hat{\mathsf{e}}^b=\eta^{ab}, \quad \hat{\mathsf{e}}^a\cdot\hat{\mathsf{e}}_b=\delta^a_b.
\end{equation}
The second term on the RHS of \eqref{gp} is a \textit{bivector}, and the antisymmetry allows six such quantities to be defined from the $\{\hat{\mathsf{e}}_a\}$. In the same manner, four \textit{trivectors} can be constructed, along with the unique \textit{pseudoscalar}
\begin{equation}
  I=\hat{\mathsf{e}}_0\wedge\hat{\mathsf{e}}_1\wedge\hat{\mathsf{e}}_2\wedge\hat{\mathsf{e}}_3.
\end{equation}
Repeating the procedure with the $\{\hat{\mathsf{e}}^a\}$ generates the same quantities, modulo sign differences. This defines the five \textit{grades} of the spacetime algebra. The PGT curvature tensor is represented by a bivector-valued linear function of its bivector argument, with the usual components recovered as scalars via the appropriate interior product
\begin{equation}
  \mathcal{  R}_{abcd}=(\hat{\mathsf{e}}_a\wedge\hat{\mathsf{e}}_b)\cdot\mathcal{  R}(\hat{\mathsf{e}}_d\wedge\hat{\mathsf{e}}_c).
  \label{riecomp}
\end{equation}
Note the unfortunate reversal in the last two indices. Equivalently, the PGT torsion is a bivector-valued linear function of its vector argument
\begin{equation}
  \mathcal{  T}^a_{\ \ bc}=(\hat{\mathsf{e}}_b\wedge\hat{\mathsf{e}}_c)\cdot\mathcal{  T}(\hat{\mathsf{e}}^a).
  \label{torcomp}
\end{equation}
A major advantage of the geometric algebra formulation is that it renders such components unnecessary for formal calculations, since the $\{\hat{\mathsf{e}}_a\}$ and $\{\hat{\mathsf{e}}^a\}$ may be replaced by arbitrary constant vectors, denoted similarly by lower-case Roman letters, e.g.\footnote{Note that this notation, which unavoidably clashes with the dimensionless scale factor, is confined to \cref{multi}.} $a$, $b$, $c$, and multivector derivatives with respect to them, $\partial_a$, $\partial_b$, $\partial_c$. These have the desired properties in common with the usual basis and dual basis
\begin{equation}
  \partial_a\cdot a=4, \quad \partial_a\wedge a=0.
\end{equation}
Thus, in a notation which makes no reference to any Lorentz basis, we can define the vector-valued Ricci tensor, torsion contraction and Ricci scalar
\begin{equation}
\begin{gathered}
  \mathcal{  R}(a)=\partial_b\cdot\mathcal{  R}(b\wedge a),\quad \mathcal{  R}=\partial_a\cdot\mathcal{  R}(a),\\ 
  \mathcal{  T}=\partial_a\cdot\mathcal{  T}(a).
\end{gathered}
\end{equation}

It should by this point be clear that the formalism is advantageous for identifying tensor symmetries. In particular, the essential symmetries
\begin{equation}
  \mathcal{  R}_{(ab)cd}=\mathcal{  R}_{ab(cd)}=\mathcal{  T}^a_{\ \ (bc)}=0,
\end{equation}
follow immediately from \eqref{riecomp} and \eqref{torcomp}. Less obvious are those symmetries of the Riemann and Ricci tensors which emerge in the metrical limit of vanishing torsion. To discuss these, we define the \textit{adjoint} functions,
\begin{equation}
  \begin{gathered}
  (a\wedge b)\cdot\mathcal{  R}(c\wedge d)=\bar{\mathcal{  R}}(a\wedge b)\cdot(c\wedge d),\\
  a\cdot\mathcal{  R}(b)=\bar{\mathcal{  R}}(a)\cdot b,
\end{gathered}
\end{equation}
which are distinguishable from the functions themselves only when torsion is present. Without torsion, the overbars can be removed and by inserting the Lorentz basis we can easily recover
\begin{equation}
  \mathcal{  R}_{abcd}=\mathcal{  R}_{cdab}, \quad \mathcal{  R}_{ab}=\mathcal{  R}_{ba}.
\end{equation}

As was illustrated in \cref{ras}, the eWGT counterparts of the PGT field strength tensors have a very similar structure, though the torsion contraction vanishes by construction, $\partial_a\cdot\mathcal{  T}^\dagger(a)=0$. Although we will only apply this formalism to the Lagrangian structure of the gauge theories, we note that it has many other advantages. For example, once the direction associated with cosmic time is known $\hat{\mathsf{e}}_t$, construction of the most general isotropic torsion bivector equivalent to \eqref{torsionobv} follows straightforwardly
\begin{equation}
  \mathcal{  T}(a)=(\tfrac{1}{3}U+QI)(a\wedge\hat{\mathsf{e}}_t).
\end{equation}
\subsection{Quadratic invariants}
A natural reshuffling of the gravitational action is possible within the STA. The usual arrangement of quadratic invariants such as \eqref{inic} and \eqref{inoc} are obtained by asking for all unique contraction permutations between squared tensors. Alternatively, we can ask for all unique geometric quantities formed from the same tensor, and square them. 

Applied to the quadratic Riemann sector, most of the terms in either decomposition are identical, for example
\begin{equation}
  \begin{aligned}
    \mathcal{  R}_{abcd}\mathcal{  R}^{abcd}&= 2\mathcal{  R}(c\wedge d)\cdot\mathcal{  R}(\partial_d\wedge\partial_c),\\
    \mathcal{  R}_{abcd}\mathcal{  R}^{cdab}&= 2\bar{\mathcal{  R}}(c\wedge d)\cdot\mathcal{  R}(\partial_d\wedge\partial_c),\\
  \end{aligned}
\end{equation}
with analogous formulae in the quadratic Ricci sector. The only theory parameter that requires much thought in its conversion is $\alpha_5$. Tellingly this is the only quadratic invariant that does not is not generated by a clean symmetry operation on its Riemann tensor factors
\begin{equation}
  \mathcal{  R}_{abcd}\mathcal{  R}^{acbd}= ((b\cdot \bar{\mathcal{  R}}(d\wedge c))\cdot\left(\partial_c\cdot\mathcal{  R}(\partial_d\wedge\partial_b)\right).
  \label{ght}
\end{equation}
The RHS of \eqref{ght} does not conform to the principle of the new decomposition, but can itself be further decomposed using
\begin{equation}
  \begin{aligned}
  \left( \partial_b\wedge\mathcal{  R}(b\wedge d) \right)\cdot&\left( c\wedge\mathcal{  R}(\partial_c\wedge\partial_d) \right)=\\
  &\left( c\cdot\mathcal{  R}(b\wedge d) \right)\cdot\left( \partial_b\cdot\mathcal{  R}(\partial_c\wedge\partial_d) \right)\\
  &-\mathcal{  R}(d\wedge c)\cdot\mathcal{  R}(\partial_c\wedge\partial_d).
  \label{sillion}
  \end{aligned}
\end{equation}
This results in the following decomposition of the quadratic Riemann sector
\begin{equation}
  \begin{aligned}
    L_{{\mathcal{R}}^2}=&\check{\alpha}_1{\mathcal{R}}^2+\check{\alpha}_2\mathcal{R}(\partial_b)\cdot\mathcal{R}(b)+\check{\alpha}_3\bar{\mathcal{R}}(\partial_b)\cdot\mathcal{R}(b)\\
    &+\check{\alpha}_4\mathcal{R}(\partial_b\wedge\partial_c)\cdot\mathcal{R}(c\wedge b)\\
    &+\check{\alpha}_5\left( \partial_b\wedge\mathcal{R}\left( b\wedge d \right) \right)\cdot\left( c\wedge\mathcal{R}\left( \partial_c\wedge\partial_d \right) \right)\\
    &+\check{\alpha}_6\bar{\mathcal{R}}(\partial_b\wedge\partial_c)\cdot\mathcal{R}(c\wedge b), 
    \label{lillion2}
  \end{aligned}
\end{equation}
while the same methodology decomposes the quadratic torsion sector as follows
\begin{equation}
  \begin{aligned}
    L_{{\mathcal{T}}^2}=&\check{\beta}_1\mathcal{T}(\partial_b)\cdot\mathcal{T}(b)\\
    &+\check{\beta}_2\left( \partial_a\wedge\mathcal{T}(a) \right)\cdot\left( \partial_b\wedge\mathcal{T}(b) \right)+\check{\beta}_3{\mathcal{T}}^2.
    \label{lillion}
  \end{aligned}
\end{equation}
The decompositions in \eqref{lillion2} and \eqref{lillion} are the origin of the theory parameters, \eqref{newparams}. Note that the first term on the RHS of \eqref{sillion} and the second term on the RHS of \eqref{lillion} are the squares of the Riemann and torsion \textit{protractions} which were mentioned in \cref{gravv}.
\subsection{Conformal gravity vs $k$-screened gravity}
The Weyl tensor is defined as 
\begin{equation}
  \mathcal{  W}(a\wedge b)=\mathcal{  R}(a\wedge b)-\mathcal{  S}(a)\owedge b,
\end{equation}
where the Schouten tensor is defined in terms of the Ricci tensor and scalar as
\begin{equation}
  \mathcal{ S }(a)=\tfrac{1}{2}\left( \mathcal{  R}(a)-\tfrac{1}{6}\mathcal{  R}a \right),
\end{equation}
and in geometric algebra the Kulkarni-Nomizu product of two tensors (as usual represented by linear functions on vectors) is
\begin{equation}
  \mathcal{  A}(a)\owedge\mathcal{  B}(b)=\mathcal{  A}(a)\wedge\mathcal{  B}(b)-\mathcal{  B}(a)\wedge\mathcal{  A}(b).
\end{equation}
This allows us to translate the Weyl tensor directly into the Riemann and Ricci as follows\footnote{Note this is a standard result in tensor notation also.}
\begin{equation}
  \begin{aligned}
  \mathcal{  W}(a\wedge b)=&\mathcal{  R}(a\wedge b)-\tfrac{1}{2}\left( \mathcal{  R}(a)\wedge b+a\wedge \mathcal{  R}(b) \right)\\
  &+\tfrac{1}{6}a\wedge b\mathcal{  R}.
\end{aligned}
\end{equation}
It is also easy to find the adjoint Weyl tensor in the presence of torsion
\begin{equation}
  \begin{aligned}
  \bar{\mathcal{  W}}(a\wedge b)=&\bar{\mathcal{  R}}(a\wedge b)-\tfrac{1}{2}\left( \bar{\mathcal{  R}}(a)\wedge b+a\wedge \bar{\mathcal{  R}}(b) \right)\\
  &+\tfrac{1}{6}a\wedge b\mathcal{  R}.
\end{aligned}
\end{equation}
While is not possible, by invoking torsion, to resurrect the contractions of the Weyl tensor or its adjoint
\begin{equation}
\partial_a\cdot\mathcal{  W}(a\wedge b)=\partial_a\cdot\bar{\mathcal{  W}}(a\wedge b)=0,
\end{equation}
we do find that the Weyl protraction no longer vanishes in general
\begin{equation}
  \partial_a\wedge\mathcal{  W}(a\wedge b)=\partial_a\wedge\mathcal{  R}(a\wedge b)-\tfrac{1}{2}\partial_a\wedge\mathcal{  R}(a)\wedge b.
\end{equation}

By combining these results and by analogy with the six quadratic curvature invariants, we find three obvious candidates for the quadratic invariants of the Weyl
\begin{widetext}
\begin{subequations}
  \begin{align}
  \mathcal{W}(\partial_b\wedge\partial_a)\cdot\mathcal{W}(a\wedge b)=&\mathcal{R}(\partial_b\wedge\partial_a)\cdot\mathcal{R}(a\wedge b)-\mathcal{R}(\partial_a)\cdot\mathcal{R}(a)+\tfrac{1}{6}\mathcal{R}^2,\\
  \bar{\mathcal{W}}(\partial_b\wedge\partial_a)\cdot\mathcal{W}(a\wedge b)=&\bar{\mathcal{R}}(\partial_b\wedge\partial_a)\cdot\mathcal{R}(a\wedge b)-\bar{\mathcal{R}}(\partial_a)\cdot\mathcal{R}(a)+\tfrac{1}{6}\mathcal{R}^2,\\
  \left( \partial_a\wedge\mathcal{W}(a\wedge b) \right)\cdot\left( c\wedge\mathcal{W}(\partial_c\wedge\partial_b) \right)=&\left( \partial_a\wedge\mathcal{R}(a\wedge b) \right)\cdot\left( c\wedge\mathcal{R}(\partial_c\wedge\partial_b) \right)+\tfrac{1}{2}\left( \mathcal{R}(\partial_a)\cdot\mathcal{R}(a)-\bar{\mathcal{R}}(\partial_a)\cdot\mathcal{R}(a) \right).
\end{align}
\end{subequations}
\end{widetext}
This motivates three further theory parameters for the quadratic Weyl sector
\begin{equation}
  \begin{gathered}
  \check{{\mu}}_1=\tfrac{1}{6}\check{{\alpha}}_1-\check{{\alpha}}_2+\check{{\alpha}}_4, \quad
  \check{{\mu}}_2=\tfrac{1}{6}\check{{\alpha}}_1-\check{{\alpha}}_3+\check{{\alpha}}_6,\\
  \check{{\mu}}_3=\tfrac{1}{2}\check{{\alpha}}_2-\tfrac{1}{2}\check{{\alpha}}_3+\check{{\alpha}}_5.
\end{gathered}
\end{equation}
It is then easy to see that the $k$-screening condition \eqref{ksc} is indeed compatible with \textit{any} generalisation of conformal gravity theory to nonzero torsion, since
\begin{equation}
  \check{\boldsymbol{\mu}}_1\cdot\boldsymbol{\sigma}_3=\check{\boldsymbol{\mu}}_2\cdot\boldsymbol{\sigma}_3=\check{\boldsymbol{\mu}}_3\cdot\boldsymbol{\sigma}_3=0,
\end{equation}
moreover we may relate some of the more specific cosmologies (e.g. \cosmicclass{14} defined by \eqref{s20}) mentioned in \cref{permitted} to the quadratic Weyl sector as follows:
\begin{equation}
  \check{\boldsymbol{\mu}}_1\cdot\boldsymbol{\sigma}_1=\check{\boldsymbol{\mu}}_2\cdot\boldsymbol{\sigma}_2=0.
\end{equation}

We finally note that the parameter space of the quadratic Weyl sector is \textit{three} dimensional, whilst that of the quadratic Riemann sector is \textit{five} dimensional as discussed in \cref{map_section}. It should therefore be possible to construct a fourth theory which is simultaneously $k$-screened and independent of the quadratic Weyl sector. \section{Conclusions}\label{conc}
Had the standard model of particle physics predated general relativity, we might be left wondering at the classical successes of the Einstein-Hilbert action. In fact the order was reversed, and the standard model of cosmology has instead cemented it. In this final section we will summarise the combined classical and quantum aspects of the Yang-Mills actions considered here.

We should not lose sight of the gauge theories that underlie these actions. In the short term, these results will principally be of relevance to PGT, but the classical equivalence of PGT\textsuperscript{q+} and eWGT\textsuperscript{q+} cosmologies should save considerable time as the latter field develops. Moreover, we are hopeful that it may be generalised to other simple spacetimes, such as pp-waves, anisotropic Bianchi models and axisymmetric sources.

The guiding results of \cite{2019PhRvD..99f4001L,Lin2} should themselves be thought of as preliminary, as the analysis only considers the linearised theory of PGT\textsuperscript{q+}. Moreover, we do not necessarily expect them to extend to eWGT\textsuperscript{q+} at any level of approximation. We note that work is now well underway \cite{Lin3} to perform a similar systematic search for unitary PCR instances of WGT\textsuperscript{q+} with the ultimate aim of a full eWGT\textsuperscript{q+} survey.
Next, the additional gauge symmetries which define the various critical cases have not themselves been studied, and there is no guarantee that they survive in the nonlinear theory. Of greater concern is the question of renormalisability, as the power-counting formalism is very much a \textit{first step} in its determination.
The need for a nonlinear quantum feasibility analysis is thus obvious. One possible method is the Hamiltonian analysis \cite{1999IJMPD...8..459Y,2002IJMPD..11..747Y}, which was used to eliminate certain of Sezgin and Nieuwenhuizen's theories \cite{1980PhRvD..21.3269S} on the grounds of constraint bifurcation and field activation.

Within PGT\textsuperscript{q+}, we grouped 33 of the 58 new critical cases into \numbercosmicclass{} cosmic classes. Most of these classes are $k$-screened, in the sense that the evolution of the universe is decoupled from the spatial curvature. We stress that this does not equate to an assertion that $k=0$, but rather that the flat, open or closed nature of the geometry does not affect the expansion rate or torsion evolution. This includes \cosmicclass{16} and its special case, \cosmicclass{14}, which contain the very promising \criticalcase{16} and \criticalcase{14}. Despite $k$-screening, these classes can be understood to mimic the cosmology of GR, powered `under the hood' by involved curvature-torsion interactions. In \cosmicclass{16}, flat GR cosmology emerges through `Einstein freezing', when a pure fluid with equation of state parameter $w_i$ becomes dominant, up to a $w_i$-specific renormalisation of the Einstein constant that depends on a parameter of the theory $\varsigma$. Such a renormalisation is better understood in terms of an extra-component model, in which context it could be exploited for various purposes, such as dark energy enhancement -- this is of course objectionable on the grounds of fine-tuning. 
To eliminate $\varsigma$ \textit{naturally} we may either change the quantum theory to the \criticalcase{14} of \cosmicclass{14}, or pick an instance of \cosmicclass{16} that appeals on classical and algebraic grounds without contradicting \criticalcase{16}, such as \cosmicclass{null}. \cosmicclass{14} requires $\varsigma\to\infty$ in our (short-sighted) choice of notation, but remains a promising theory in that the Friedmann equations emerge \textit{exactly} along with an effective $k\leq 0$. 
\cosmicclass{null} simply sets $\varsigma=1$, but again a `correspondence solution' can be found in which $k=0$. 

In thus avoiding fine-tuning, we have in some sense returned to flat GR on square one. Remarkably however, the special significance of radiation in \cosmicclass{16} gives rise to an extra torsion freedom at the radiation-dominated Big Bang in \cosmicclass{null}, and this allows the complexity of the theory to shine through. 
In the extra-component picture, this is manifest as a dark `tracker matter' fraction, whose equation of state reflects that of the dominant cosmic fluid. Post-equality, this matter is always subdominant, and its principal effect is that of dark radiation in the early universe. 

We have been driving at a popular proposal in the resolution of the $H_0$ discrepancy, which is worth some explanation. Generally, the expansion history of the universe must be tweaked so as to revise the CMB-inferred value of $H_0$ and $\mathsf{h}$ \textit{upwards}, towards less history-sensitive measurements (e.g. the SH0ES program or HOLiCOW project). The CMB data can be roughly characterised by two quantities \cite{2018JCAP...09..025M,2016JCAP...10..019B,2007A&A...471...65E,1999MNRAS.304...75E}, the \textit{shift parameter} $\mathscr{R}$ and multipole position $l_\text{a}$ of the first CMB peak
\begin{equation}
  \mathscr{  R}=\mathsf{H}\sqrt{\omega_\text{m}}D_\text{A}(z_{\text{rec}}), \quad l_\text{a}=\pi\frac{D_\text{A}(z_{\text{rec}})}{r_\text{s}}.
  \label{shifts}
\end{equation}
These quantities rely on the comoving angular diameter distance to recombination (as a proxy for CMB decoupling), $D_\text{A}$ at $z_\text{rec}$, and sound horizon $r_\text{s}$ at that same epoch $t_\text{rec}$. Both length scales are highly model dependent. Expressions for $D_\text{A}$ which hold for general $k$ illustrate its sensitivity to the expansion history
\begin{equation}
  \begin{aligned}
    D_\text{A}(z_{\text{rec}})&=(1+z_{\text{rec}})d_\text{A}(z_{\text{rec}})\\
  &=\frac{\sin\left( \sqrt{-\Omega_{k,0}}\int_0^{z_\text{rec}}\frac{H_0\mathrm{d}z}{H} \right)}{H_0\sqrt{-\Omega_{k,0}}}\\
  &=\frac{\sinh\left( \sqrt{\omega_{k}}\int_0^{z_\text{rec}}\frac{\mathsf{H}\mathrm{d}z}{H} \right)}{\mathsf{H}\sqrt{\omega_{k}}},
  \label{clusterf}
\end{aligned}
\end{equation}
while $r_\text{s}$ depends on both the expansion history and photon-baryon sound speed
\begin{equation}
  r_\text{s}=\int_0^{t_{\text{rec}}}\frac{c_\text{s}\mathrm{d}t}{a}, \quad c_s=\frac{1}{\sqrt{3\left( 1+3\omega_\text{b}a/4\omega_\text{r} \right)}}.
  \label{sound}
\end{equation}
If $z_{\text{rec}}$ is held constant, a general increase in $H$ for $z<z_\text{rec}$ consistent with local observations will reduce $D_\text{A}$ as expressed in \eqref{clusterf}. In order to preserve $l_\text{a}$ in \eqref{shifts}, we will therefore need a decrease in $r_\text{s}$. This can in turn be achieved by increasing $H$ for $z_\text{rec}<z$ and thus reducing $t_\text{rec}$ by \eqref{sound}. This mechanism is traditionally favoured because it impinges on relatively few of \textLambda CDM's moving parts. Of these parts, perhaps the strongest constraints come from Big Bang nucleosynthesis (BBN): if photons decouple at an earlier time then neutrinos decouple at a higher temperature. Fortunately, the implications for for the ratios of light nuclei are thought to be (just) consistent \cite{2018JCAP...09..025M} with a tension-resolving tweak to the early expansion rate. On the other hand, recent work \cite{2019JCAP...10..029S} combining BBN and BAO constraints (which probes only the background evolution so long as neutrino drag is neglected) indicates that dark radiation may only reduce the tension to $2.6\sigma$.

A selective increase in the early expansion rate independent of other density parameters is qualitatively implied by our model: the relaxed or arbitrary-$\varpi_\text{r}$ soluton to \cosmicclass{null}. Many alternative methods have been employed in recent years, most of which fall under the umbrellas of early dark energy \cite{2018JCAP...09..025M}, dark-sector interactions \cite{2020arXiv200110852Y,2020arXiv200206127L,PhysRevD.101.035031} or varying $\Lambda$ models \cite{2007A&A...471...65E}. These tend to lie on a spectrum between data-driven searches and theoretically motivated proposals for an extra component. Such motivations arise, for example, in particle physics \cite{Pandey} and string theory \cite{2016PhRvD..94j3523K}, though they mostly bear fruit in the form of toy models. Our proposal has the advantage that the effect emerges from an independently motivated theory of gravity, and can be compared to (e.g.) similar applications of the ghost-free bimetric theory \cite{2018JCAP...09..025M}. A more obvious approach is to simply introduce additional ultrarelativistic species such as sterile neutrinos and so to alter $\Delta N_\text{eff}$ -- we stress again that the quantity $\Delta N_\text{dr,eff}$ is introduced in \cref{dt} for convenience only, and does not confer any such ad hoc species. This is significant as some BBN-oriented studies \cite{2020JCAP...01..004S} specifically assume thermal particles in equilibrium with the standard-model plasma, while the Rayleigh-Jeans tail of the CMB can constrain some dark electromagnetism models \cite{2020arXiv200208942B}. The term `dark radiation' is also something of a misnomer, since our theory makes a clear prediction as to the evolution and present intensity of the pseudoscalar torsion mode, which ought to be nearly constant for $z\ll z_\text{rec}$, and on the order of the Planck mass
\begin{equation}
  Q_0\sim M_\text{P}.
  \label{falsifiability}
\end{equation}
As we observed earlier, this is precisely the torsion mode which is expected to interact with matter, introducing the potential for detection and falsifiability. On the other hand it must be noted that \eqref{falsifiability} relies on a somewhat naive interpretation of PGT\textsuperscript{q+} in which the $\{\alpha_i\}$ and $\{\beta_i\}$ along with the $\{\sigma_i\}$ and $\{\upsilon_i\}$ are assumed to be of order unity. There is reason to believe \cite{ewgt_conformal} that in eWGT\textsuperscript{q+} any experiment would only be able to determine the quantitiy $\sigma_1Q_0^2$, and that $\sigma_1$ need \textit{not} be of order unity. It should also be noted that attempts at measureing torsion are generally specific to the theory, with most attention naturally granted to ECT. The series \cite{2010RPPh...73e6901N,2016IJMPS..4060010N} provides a current review of spin-gravity interaction in theory and practice. Some quite concrete proposals have been made \cite{2014IJMPD..2342004P} based on microstructured matter, but these require nonminimal couplings of $\mathcal{  T}^a_{\ \ \ bc}$ and $\mathcal{  R}^a_{\ \ bcd}$ to the matter fields $\varphi$, which are not present in ten-parameter PGT\textsuperscript{q+}.

If the quantum considerations in \cite{2019PhRvD..99f4001L,Lin2} are preliminary, our classical results are doubly so, since we have restricted our attention to background cosmology. 
Compared to GR, our gravity theory is not so much modified as completely rewritten, and its effect on perturbations will eventually require a dedicated study, indeed the authors of \cite{2019JCAP...10..029S} emphasise that extra perturbation ingredients are of interest to the resolution of the $H_0$ tension.
In the near future, we envisage only a small modification to a publicly available Markov-Chain Monte-Carlo (MCMC) engine such as \texttt{COSMOMC} \cite{2011ascl.soft06025L} or \texttt{CLASS} \cite{2011arXiv1104.2932L}, restricted to the extra-component model set out in \cref{dt}. This may be done with nothing more sophisticated than a spline approximation of the equation of state parameter set out in \eqref{proposal} in \cref{intro}. Depending on the state of the perturbation theory, a more rigorous modification may then be justified.
The same basic questions surround, for example, solar system tests. 
On this point however there may be cause for optimism, as we believe both \cosmicclass{16} and \cosmicclass{14} theories generically admit Schwarzschild-de Sitter vacuum solutions, in common with RST \cite{lasenby-doran-heineke-2005}. The extra torsion freedom admitted by \cosmicclass{2} or \cosmicclass{15} may be extremely useful when constructing spherically symmetric solutions. Although a study of \cosmicclass{2}, \cosmicclass{15} and \cosmicclass{12} is beyond the scope of the present work (see \cite{attack_letter}), we reiterate that they remain attractive.

Paradoxically, we have had nothing to say about the `elephants in the cosmos' such as inflation, dark matter or dark energy. We cannot dismiss the idea that $k$-screening may be of some relevance to the flatness problem, or that the general unpredictability of \cosmicclass{16} cosmology at turnover epochs may help explain the cosmic coincidence. At the classical level, \cosmicclass{null} gravity only offers us a concrete route out of the subtler problem of the $H_0$ tension, and in this sense it is economical. In particular, the absence of a massive particle in \criticalcase{16} remains in line with the consensus that the origins of dark matter are not purely gravitational, and that the origins of dark energy are not classical\footnote{See for example a new semiclassical origin for $\Lambda$ within GR \cite{2019PhRvD.100f3511B}, we anticipate this `quantum bias' methodology can be adapted to gauge gravity.}. We have not yet attended to inflation, but rather invoked a natural freedom on the boundary of the radiation-dominated Big Bang, which is eliminated by dark energy at the Future Conformal Boundary. This raises questions of compatability with the conformal cyclic cosmology (CCC) of Penrose \cite{Penrose:2006zz}, or its recent reinterpretation \cite{Las}, and has the advantage of extending \textLambda CDM by only \textit{one} parameter. The obvious \textit{zero} parameter grail may be addressed in future work: one would like to replace the classical singularity with a torsion-driven inflationary epoch which naturally exits to the correct dark radiation fraction.
\begin{acknowledgments}
  We are grateful to Antony Lewis for a helpful discussion at the 30\textsuperscript{th} Texas Symposium, Yun-Cherng Lin for his assistance in incorporating the new critical cases of PGT\textsuperscript{q+}, Steven Gratton for his insights into the minisuperspace approximation and Marc Kamionkowski for his useful comments on dark radiation at the KICC 10\textsuperscript{th} Anniversary Symposium. WEVB is supported by STFC, and WJH by the Gonville and Caius Research Fellowship.
\end{acknowledgments}

\appendix
\section{Spin projection operators}\label{spos}
The building blocks of the SPOs are two $k^a$-dependent projections
\begin{equation}
  \Omega^{ab}=\frac{k^ak^b}{k^2}, \quad \Theta^{ab}=\eta^{ab}-\frac{k^ak^b}{k^2}.
\end{equation}
For the $\mathcal{  A}_{abc}$-field, the diagonal SPOs then have the following fundamental definitions
\begin{align}
\begin{split}
  \grave{\mathcal{  P}}_{11}(0^-)_{ijkabc}&=\tfrac{2}{3}\Theta_{ic}\Theta_{ja}\Theta_{kb}+\tfrac{1}{3}\Theta_{ja}\Theta_{jb}\Theta_{kc},\\
  \grave{\mathcal{  P}}_{11}(0^+)_{ijkabc}&=\tfrac{2}{3}\Theta_{cb}\Theta_{kj}\Omega_{ia},\\
  \grave{\mathcal{  P}}_{11}(1^-)_{ijkabc}&=\tfrac{2}{3}\Theta_{cb}\Theta_{ia}\Theta_{kj},\\
  \grave{\mathcal{  P}}_{22}(1^-)_{ijkabc}&=2\Theta_{ia}\Theta_{cb}\Theta_{kj},\\
  \grave{\mathcal{  P}}_{11}(1^+)_{ijkabc}&=\Theta_{ic}\Theta_{kb}\Omega_{ja}+\Theta_{ia}\Theta_{kc}\Omega_{jb},\\
  \grave{\mathcal{  P}}_{22}(1^+)_{ijkabc}&=\Theta_{ia}\Theta_{jb}\Omega_{jb},\\
  \grave{\mathcal{  P}}_{11}(2^-)_{ijkabc}&=\tfrac{2}{3}\Theta_{ic}\Theta_{jb}\Omega_{ka}+\tfrac{2}{3}\Theta_{ia}\Theta_{jb}\Omega_{kc}\\
  &\phantom{=}-\Theta_{cb}\Theta_{ia}\Omega_{kj},\\
  \grave{\mathcal{  P}}_{11}(2^+)_{ijkabc}&=-\tfrac{2}{3}\Theta_{cb}\Theta_{kj}\Omega_{ia}+\Theta_{ic}\Theta_{ka}\Omega_{jb}\\
  &\phantom{=}+\Theta_{ia}\Theta_{kc}\Omega_{jb}.
\end{split}
\end{align}
Since the $\mathcal{  A}_{abc}$-field has two $1^+$ and $1^-$ sectors, there is the opportunity for internal mixing. In particular the following off-diagonal SPOs are relevant for this work
\begin{equation}
\begin{aligned}
  \grave{\mathcal{  P}}_{12}(1^+)_{ijkabc}&=-\sqrt{2}\Theta_{ja}\Theta_{kb}\Omega_{ic},\\
  \grave{\mathcal{  P}}_{21}(1^+)_{ijkabc}&=-\sqrt{2}\Theta_{bi}\Theta_{kj}\Omega_{ic}.
\end{aligned}
\end{equation}
The diagonal SPOs are complete, idempotent and orthogonal across $J^P$ sectors. 
The correctly symmetrised forms of all SPOs are given by
\begin{equation}
  \mathcal{  P}_{ij}(J^P)_{ijkabc}=\grave{\mathcal{  P}}_{ij}(J^P)_{[ij]k[ab]c}.
\end{equation}
For the complete list of SPOs, inculding the off-diagonal SPOs of the $1^-$ sector and the SPOs of the $\mathfrak{s}_{ab}$ and $\mathfrak{a}_{ab}$ fields, see \cite{2019PhRvD..99f4001L} and references therein.
\section{Comparison with the literature}\label{chao}
Given the popularity of ten-parameter PGT\textsuperscript{q+} cosmology mentioned in \cref{intro}, it is appropriate to attempt some comparison with the literatre, although such an attempt will naturally be inexhaustive. Particularly, we will not consider extension to the odd-parity sector discussed by \cite{2011IJMPD..20.2125H,2011JPhCS.330a2005H,2015arXiv151201202H,2011PhRvD..83b4001B,2011CQGra..28u5017B}.

The original paper by Minkevich \cite{1980PhLA...80..232M} only admits $U$, and not $Q$ on the grounds of spacetime parity -- an examination of \cref{tor1,tor2,S,R} indicates that $\sigma_1$ and $\sigma_2$ do not arise in this case, and so $k$-screening cannot meaningfully occur.
Furthermore, \cite{1980PhLA...80..232M} retains $\check\alpha_0$ in order to force the correspondence principle. We note that this situation is slightly complicated in \cite{2003gr.qc....10060M,2006CQGra..23.4237M,2000CQGra..17.3045M} by the extension to MAGT. In \cite{2009PhLB..678..423M,2011arXiv1107.1566G,2013JCAP...03..040M} it appears that both $U$ and $Q$ are incorporated, but we find that the two constraints imposed on \eqref{pgtaction} translate to \eqref{u20}, while $\check{\alpha}_0$ remains free. 

In comparing the present work to \cite{2019arXiv190403545Z,2019arXiv190604340Z}, we make use of the following identity
\begin{equation}
  (\epsilon_{abcd}\mathcal{  R}^{abcd})^2=4\mathcal{  R}_{abcd}(4\mathcal{  R}^{acbd}-\mathcal{  R}^{abcd}-\mathcal{  R}^{cdab}).
  \label{swizzle}
\end{equation}
Throughout \cite{2019arXiv190403545Z,2019arXiv190604340Z} we again believe $\check\alpha_0$ to be retained, while \eqref{u20} to be imposed at certain points. Within \cite{2019arXiv190403545Z} two further constraints are applied which reduce to
\begin{subequations}
  \begin{gather}
    \sigma_1-\sigma_3=0,\label{tinker}\\
    \sigma_2-\sigma_3=0.\label{taylor}
  \end{gather}
\end{subequations}
Thus, while $\sigma_3$ remains free, \eqref{tinker} and \eqref{taylor} together imply the final constraint \eqref{b_constraint} which separates \cosmicclass{null} from \cosmicclass{16}.

Precisely \cref{tinker,taylor} are applied in \cite{2008PhRvD..78b3522S}, along with the torsion constraint
\begin{equation}
  4\upsilon_1+\upsilon_2=0,
  \label{snyc}
\end{equation}
to define the original SNY lagrangian. We note that \eqref{snyc} itself features in \cref{map_plot} to distinguish \cosmicclass{23} from \cosmicclass{20}. The SNY generalisation studied in \cite{2009JCAP...10..027C} replaces \cref{tinker,taylor} with
\begin{equation}
  \sigma_2+2\sigma_1-3\sigma_3,
  \label{snyc2}
\end{equation}
though we do not believe the quadratic torsion sector to be constrained. Once again, \eqref{snyc2} features in \cref{map_plot} to distinguish \cosmicclass{34} from \cosmicclass{36}.

Finally, we will consider \cite{lasenby-doran-heineke-2005}, in which a mathematically attractive solution to the cosmological equations of RST was presented. Here we will show that the solution satisfies a much broader class of cosmologically NSI theories. Beginning from the original root theory, we restrict to Yang-Mills gravity by applying \eqref{yangmill}, and then to cosmologically NSI gravity by eliminating the torsion with \eqref{tor1cons} and \eqref{u20}. The quadratic Riemann sector is then refined with two new constraints
\begin{subequations}
  \begin{equation}
  \sigma_1=0,\\
  \label{hein1}
  \end{equation}
  \begin{equation}
  \sigma_2-3\sigma_3=0.
  \label{hein2}
  \end{equation}
\end{subequations}
This cosmic class, to which RST belongs, is not populated by any of the critical cases considered here, and as such it does not appear in \cref{map_plot}. Note however, that it can be considered a grandchild of \cosmicclass{34}, which appears only to contain critical cases with massive $0^-$ gravitons. The torsion equations \eqref{tor1} and \eqref{tor2} then take the form
\begin{subequations}
  \begin{align}
    \left(\delta \tilde{\mathcal{ L}}_T/\delta X\right)_{\text{F}}&\propto\partial_\tau^2X+2X(3Y^2/4-X^2-k),\\
    \left(\delta \tilde{\mathcal{ L}}_T/\delta Y\right)_{\text{F}}&\propto-\partial_\tau^2Y+2Y(3X^2-Y^2/4+k),
  \end{align}
\end{subequations}
in which their mutual symmetry -- first noted in \cref{ksc_sec} -- are brought into sharp relief. The methodology of \cite{lasenby-doran-heineke-2005} exploits this directly, by encapsulating both equations though the concept of complex torsion
\begin{equation}
  Z=X+iY/2, \quad \partial_\tau^2Z-2Z^3+2kZ=0.
  \label{anthony_eq}
\end{equation}
The single resulting equation can then be solved compactly for $Z$ in terms of the Weierstrass elliptic function, such that the material source $\varrho_\text{r}$ appears as a constant of integration. This compact solution describes an interesting universe, if one of limited utility, in which the Hubble number and torsion may evolve chaotically. Our preferred formalism of \cref{permitted} affords a more respectable picture however, if we set $U=Q=0$. The density equation analogous to \eqref{density} or \eqref{HeqnJ} then becomes
\begin{equation}
  \begin{aligned}
  \Omega_\text{r}+\tfrac{8}{3}\sigma_2\kappa\big(& \left( \partial_tH/H \right)^2+2\partial_tH\\
  &-H^2\Omega_k(\Omega_k-2) \big)=0,
\end{aligned}
\end{equation}
and $\partial_tH$ can then be eliminated by the observable form of \eqref{tor1}
\begin{equation}
  \partial_t^2H+4H\partial_tH+2H^3\Omega_k=0.
\end{equation}
By writing the implied integration constant as a modified cosmological constant, $\breve{\Lambda}$ of dimension $\si{\electronvolt}$, this becomes
\begin{equation}
  \partial_tH=H^2(\Omega_k-2)+\tfrac{2}{3}\breve{\Lambda}.
\end{equation}
The final density equation then looks quite familiar
\begin{equation}
  \tfrac{9}{8}\kappa^{-1}\breve\Lambda^{-1} \Omega_\text{r}+\Omega_{\breve{\Lambda}}+\Omega_k=1.
\end{equation}
as an effective cosmological constant emerges up to a renormalisation of the radiation density.
\section{Cosmological equations of \cosmicclass{16}}\label{cc16}
The modified gravitaional densities in \eqref{density} and the coefficients to the auxiliary torsion equation \eqref{aux} have the following forms
\begin{widetext}
  \begin{subequations}
\begin{align}
    \Omega_\Psi+\Omega_\Phi=&{\frac { \left(  \left( 16\,{\sigma_1}^{2}-4\,{\sigma_2}^{2} \right) {\kappa}^{2}{Q}^{2}+\kappa\sigma_2\,\upsilon_2 \right) {\partial_{t}Q}^{2}}{ \left( 4\,{Q}^{2}\sigma_2\,\kappa-\upsilon_2 \right) {H}
^{2}}}
+32\,{\frac {Q \left( \kappa \left( {\sigma_1}^{2}-1/4\,{\sigma_2}^{2} \right) {Q}^{2}-1/4\,\upsilon_2\, \left( \sigma_1-\sigma_2/4 \right)  \right) \kappa\partial_{t}Q}{ \left( 4\,{Q}^{2}\sigma_2\,
\kappa-\upsilon_2 \right) H}}
\nonumber\\
    &+16\,{\frac { \left( \kappa \left( {\sigma_1}^{2}-1/4\,{\sigma_2}^{2} \right) {Q}^{2}-1/2\, \left( \sigma_1-5/8\,\sigma_2 \right) \upsilon_2 \right) {Q}^{2}\kappa}{4\,{Q}^{2}\sigma_2\,\kappa-\upsilon_2}}
,\\
  f_1=&2\,Q \left( 4\,\sigma_2\,\kappa{Q}^{2}-\upsilon_2 \right)  \left( 16\,\kappa{Q}^{2}{\sigma_1}^{2}-4\,\kappa{Q}^{2}{\sigma_2}^{2}+\sigma_2\,\upsilon_2 \right) 
,\\
  f_2=&-32\,{\sigma_1}^{2}\upsilon_2\,\kappa{Q}^{3}
,\\
  f_3=&6\,Q \left( 4\,\sigma_2\,\kappa{Q}^{2}-\upsilon_2 \right)  \left( 16\,\kappa{Q}^{2}{\sigma_1}^{2}-4\,\kappa{Q}^{2}{\sigma_2}^{2}+\sigma_2\,\upsilon_2 \right) 
,\\
  f_4=&2\,Q \left( 4\,\sigma_2\,\kappa{Q}^{2}-\upsilon_2 \right)  \left( 16\,\kappa{Q}^{2}{\sigma_1}^{2}-4\,\kappa{Q}^{2}{\sigma_2}^{2}-4\,\upsilon_2\,\sigma_1+\sigma_2\,\upsilon_2 \right) 
,\\
  f_5=&256\,Q \left(  \left( \sigma_2\,{\kappa}^{2}{\sigma_1}^{2}-1/4\,{\sigma_2}^{3}{\kappa}^{2} \right) {Q}^{4}-1/8\, \left( {\sigma_1}^{2}+3\,\sigma_1\,\sigma_2-{\sigma_2}^{2} \right) \upsilon_2\,\kappa{Q}^
{2}+1/32\, \left( \sigma_1+\sigma_2/2 \right) {\upsilon_2}^{2} \right) 
.
\end{align}
\end{subequations}
\end{widetext}
\bibliographystyle{apsrev4-1}
\bibliography{prd_paper}

\begin{thebibliography}{99}%
\makeatletter
\providecommand \@ifxundefined [1]{%
 \@ifx{#1\undefined}
}%
\providecommand \@ifnum [1]{%
 \ifnum #1\expandafter \@firstoftwo
 \else \expandafter \@secondoftwo
 \fi
}%
\providecommand \@ifx [1]{%
 \ifx #1\expandafter \@firstoftwo
 \else \expandafter \@secondoftwo
 \fi
}%
\providecommand \natexlab [1]{#1}%
\providecommand \enquote  [1]{``#1''}%
\providecommand \bibnamefont  [1]{#1}%
\providecommand \bibfnamefont [1]{#1}%
\providecommand \citenamefont [1]{#1}%
\providecommand \href@noop [0]{\@secondoftwo}%
\providecommand \href [0]{\begingroup \@sanitize@url \@href}%
\providecommand \@href[1]{\@@startlink{#1}\@@href}%
\providecommand \@@href[1]{\endgroup#1\@@endlink}%
\providecommand \@sanitize@url [0]{\catcode `\\12\catcode `\$12\catcode
  `\&12\catcode `\#12\catcode `\^12\catcode `\_12\catcode `\%12\relax}%
\providecommand \@@startlink[1]{}%
\providecommand \@@endlink[0]{}%
\providecommand \url  [0]{\begingroup\@sanitize@url \@url }%
\providecommand \@url [1]{\endgroup\@href {#1}{\urlprefix }}%
\providecommand \urlprefix  [0]{URL }%
\providecommand \Eprint [0]{\href }%
\providecommand \doibase [0]{http://dx.doi.org/}%
\providecommand \selectlanguage [0]{\@gobble}%
\providecommand \bibinfo  [0]{\@secondoftwo}%
\providecommand \bibfield  [0]{\@secondoftwo}%
\providecommand \translation [1]{[#1]}%
\providecommand \BibitemOpen [0]{}%
\providecommand \bibitemStop [0]{}%
\providecommand \bibitemNoStop [0]{.\EOS\space}%
\providecommand \EOS [0]{\spacefactor3000\relax}%
\providecommand \BibitemShut  [1]{\csname bibitem#1\endcsname}%
\let\auto@bib@innerbib\@empty
\bibitem [{\citenamefont {{Scott}}(2018)}]{2018arXiv180401318S}%
  \BibitemOpen
  \bibfield  {author} {\bibinfo {author} {\bibfnamefont {D.}~\bibnamefont
  {{Scott}}},\ }\href@noop {} {\bibfield  {journal} {\bibinfo  {journal} {arXiv
  e-prints}\ ,\ \bibinfo {eid} {arXiv:1804.01318}} (\bibinfo {year} {2018})},\
  \Eprint {http://arxiv.org/abs/1804.01318} {arXiv:1804.01318 [astro-ph.CO]}
  \BibitemShut {NoStop}%
\bibitem [{\citenamefont {{Barker}}\ \emph {et~al.}(2019)\citenamefont
  {{Barker}}, \citenamefont {{Lasenby}}, \citenamefont {{Hobson}},\ and\
  \citenamefont {{Handley}}}]{2019JMP....60e2504B}%
  \BibitemOpen
  \bibfield  {author} {\bibinfo {author} {\bibfnamefont {W.~E.~V.}\
  \bibnamefont {{Barker}}}, \bibinfo {author} {\bibfnamefont {A.~N.}\
  \bibnamefont {{Lasenby}}}, \bibinfo {author} {\bibfnamefont {M.~P.}\
  \bibnamefont {{Hobson}}}, \ and\ \bibinfo {author} {\bibfnamefont {W.~J.}\
  \bibnamefont {{Handley}}},\ }\href {\doibase 10.1063/1.5082730} {\bibfield
  {journal} {\bibinfo  {journal} {Journal of Mathematical Physics}\ }\textbf
  {\bibinfo {volume} {60}},\ \bibinfo {eid} {052504} (\bibinfo {year}
  {2019})},\ \Eprint {http://arxiv.org/abs/1811.09844} {arXiv:1811.09844
  [gr-qc]} \BibitemShut {NoStop}%
\bibitem [{\citenamefont {{Bull}}\ \emph {et~al.}(2016)\citenamefont {{Bull}},
  \citenamefont {{Akrami}}, \citenamefont {{Adamek}}, \citenamefont {{Baker}},
  \citenamefont {{Bellini}}, \citenamefont {{Beltr{\'a}n Jim{\'e}nez}},
  \citenamefont {{Bentivegna}}, \citenamefont {{Camera}}, \citenamefont
  {{Clesse}}, \citenamefont {{Davis}}, \citenamefont {{Di Dio}}, \citenamefont
  {{Enander}}, \citenamefont {{Heavens}}, \citenamefont {{Heisenberg}},
  \citenamefont {{Hu}}, \citenamefont {{Llinares}}, \citenamefont {{Maartens}},
  \citenamefont {{M{\"o}rtsell}}, \citenamefont {{Nadathur}}, \citenamefont
  {{Noller}}, \citenamefont {{Pasechnik}}, \citenamefont {{Pawlowski}},
  \citenamefont {{Pereira}}, \citenamefont {{Quartin}}, \citenamefont
  {{Ricciardone}}, \citenamefont {{Riemer-S{\o}rensen}}, \citenamefont
  {{Rinaldi}}, \citenamefont {{Sakstein}}, \citenamefont {{Saltas}},
  \citenamefont {{Salzano}}, \citenamefont {{Sawicki}}, \citenamefont
  {{Solomon}}, \citenamefont {{Spolyar}}, \citenamefont {{Starkman}},
  \citenamefont {{Steer}}, \citenamefont {{Tereno}}, \citenamefont {{Verde}},
  \citenamefont {{Villaescusa-Navarro}}, \citenamefont {{von Strauss}},\ and\
  \citenamefont {{Winther}}}]{2016PDU....12...56B}%
  \BibitemOpen
  \bibfield  {author} {\bibinfo {author} {\bibfnamefont {P.}~\bibnamefont
  {{Bull}}}, \bibinfo {author} {\bibfnamefont {Y.}~\bibnamefont {{Akrami}}},
  \bibinfo {author} {\bibfnamefont {J.}~\bibnamefont {{Adamek}}}, \bibinfo
  {author} {\bibfnamefont {T.}~\bibnamefont {{Baker}}}, \bibinfo {author}
  {\bibfnamefont {E.}~\bibnamefont {{Bellini}}}, \bibinfo {author}
  {\bibfnamefont {J.}~\bibnamefont {{Beltr{\'a}n Jim{\'e}nez}}}, \bibinfo
  {author} {\bibfnamefont {E.}~\bibnamefont {{Bentivegna}}}, \bibinfo {author}
  {\bibfnamefont {S.}~\bibnamefont {{Camera}}}, \bibinfo {author}
  {\bibfnamefont {S.}~\bibnamefont {{Clesse}}}, \bibinfo {author}
  {\bibfnamefont {J.~H.}\ \bibnamefont {{Davis}}}, \bibinfo {author}
  {\bibfnamefont {E.}~\bibnamefont {{Di Dio}}}, \bibinfo {author}
  {\bibfnamefont {J.}~\bibnamefont {{Enander}}}, \bibinfo {author}
  {\bibfnamefont {A.}~\bibnamefont {{Heavens}}}, \bibinfo {author}
  {\bibfnamefont {L.}~\bibnamefont {{Heisenberg}}}, \bibinfo {author}
  {\bibfnamefont {B.}~\bibnamefont {{Hu}}}, \bibinfo {author} {\bibfnamefont
  {C.}~\bibnamefont {{Llinares}}}, \bibinfo {author} {\bibfnamefont
  {R.}~\bibnamefont {{Maartens}}}, \bibinfo {author} {\bibfnamefont
  {E.}~\bibnamefont {{M{\"o}rtsell}}}, \bibinfo {author} {\bibfnamefont
  {S.}~\bibnamefont {{Nadathur}}}, \bibinfo {author} {\bibfnamefont
  {J.}~\bibnamefont {{Noller}}}, \bibinfo {author} {\bibfnamefont
  {R.}~\bibnamefont {{Pasechnik}}}, \bibinfo {author} {\bibfnamefont {M.~S.}\
  \bibnamefont {{Pawlowski}}}, \bibinfo {author} {\bibfnamefont {T.~S.}\
  \bibnamefont {{Pereira}}}, \bibinfo {author} {\bibfnamefont {M.}~\bibnamefont
  {{Quartin}}}, \bibinfo {author} {\bibfnamefont {A.}~\bibnamefont
  {{Ricciardone}}}, \bibinfo {author} {\bibfnamefont {S.}~\bibnamefont
  {{Riemer-S{\o}rensen}}}, \bibinfo {author} {\bibfnamefont {M.}~\bibnamefont
  {{Rinaldi}}}, \bibinfo {author} {\bibfnamefont {J.}~\bibnamefont
  {{Sakstein}}}, \bibinfo {author} {\bibfnamefont {I.~D.}\ \bibnamefont
  {{Saltas}}}, \bibinfo {author} {\bibfnamefont {V.}~\bibnamefont {{Salzano}}},
  \bibinfo {author} {\bibfnamefont {I.}~\bibnamefont {{Sawicki}}}, \bibinfo
  {author} {\bibfnamefont {A.~R.}\ \bibnamefont {{Solomon}}}, \bibinfo {author}
  {\bibfnamefont {D.}~\bibnamefont {{Spolyar}}}, \bibinfo {author}
  {\bibfnamefont {G.~D.}\ \bibnamefont {{Starkman}}}, \bibinfo {author}
  {\bibfnamefont {D.}~\bibnamefont {{Steer}}}, \bibinfo {author} {\bibfnamefont
  {I.}~\bibnamefont {{Tereno}}}, \bibinfo {author} {\bibfnamefont
  {L.}~\bibnamefont {{Verde}}}, \bibinfo {author} {\bibfnamefont
  {F.}~\bibnamefont {{Villaescusa-Navarro}}}, \bibinfo {author} {\bibfnamefont
  {M.}~\bibnamefont {{von Strauss}}}, \ and\ \bibinfo {author} {\bibfnamefont
  {H.~A.}\ \bibnamefont {{Winther}}},\ }\href {\doibase
  10.1016/j.dark.2016.02.001} {\bibfield  {journal} {\bibinfo  {journal}
  {Physics of the Dark Universe}\ }\textbf {\bibinfo {volume} {12}},\ \bibinfo
  {pages} {56} (\bibinfo {year} {2016})},\ \Eprint
  {http://arxiv.org/abs/1512.05356} {arXiv:1512.05356 [astro-ph.CO]}
  \BibitemShut {NoStop}%
\bibitem [{\citenamefont {{Velten}}\ \emph {et~al.}(2014)\citenamefont
  {{Velten}}, \citenamefont {{vom Marttens}},\ and\ \citenamefont
  {{Zimdahl}}}]{2014EPJC...74.3160V}%
  \BibitemOpen
  \bibfield  {author} {\bibinfo {author} {\bibfnamefont {H.~E.~S.}\
  \bibnamefont {{Velten}}}, \bibinfo {author} {\bibfnamefont {R.~F.}\
  \bibnamefont {{vom Marttens}}}, \ and\ \bibinfo {author} {\bibfnamefont
  {W.}~\bibnamefont {{Zimdahl}}},\ }\href {\doibase
  10.1140/epjc/s10052-014-3160-4} {\bibfield  {journal} {\bibinfo  {journal}
  {European Physical Journal C}\ }\textbf {\bibinfo {volume} {74}},\ \bibinfo
  {eid} {3160} (\bibinfo {year} {2014})},\ \Eprint
  {http://arxiv.org/abs/1410.2509} {arXiv:1410.2509 [astro-ph.CO]} \BibitemShut
  {NoStop}%
\bibitem [{\citenamefont {{Handley}}\ \emph {et~al.}(2014)\citenamefont
  {{Handley}}, \citenamefont {{Brechet}}, \citenamefont {{Lasenby}},\ and\
  \citenamefont {{Hobson}}}]{2014PhRvD..89f3505H}%
  \BibitemOpen
  \bibfield  {author} {\bibinfo {author} {\bibfnamefont {W.~J.}\ \bibnamefont
  {{Handley}}}, \bibinfo {author} {\bibfnamefont {S.~D.}\ \bibnamefont
  {{Brechet}}}, \bibinfo {author} {\bibfnamefont {A.~N.}\ \bibnamefont
  {{Lasenby}}}, \ and\ \bibinfo {author} {\bibfnamefont {M.~P.}\ \bibnamefont
  {{Hobson}}},\ }\href {\doibase 10.1103/PhysRevD.89.063505} {\bibfield
  {journal} {\bibinfo  {journal} {Phys. Rev. D}\ }\textbf {\bibinfo {volume}
  {89}},\ \bibinfo {eid} {063505} (\bibinfo {year} {2014})},\ \Eprint
  {http://arxiv.org/abs/1401.2253} {arXiv:1401.2253 [astro-ph.CO]} \BibitemShut
  {NoStop}%
\bibitem [{\citenamefont {{Verde}}\ \emph {et~al.}(2019)\citenamefont
  {{Verde}}, \citenamefont {{Treu}},\ and\ \citenamefont
  {{Riess}}}]{2019NatAs...3..891V}%
  \BibitemOpen
  \bibfield  {author} {\bibinfo {author} {\bibfnamefont {L.}~\bibnamefont
  {{Verde}}}, \bibinfo {author} {\bibfnamefont {T.}~\bibnamefont {{Treu}}}, \
  and\ \bibinfo {author} {\bibfnamefont {A.~G.}\ \bibnamefont {{Riess}}},\
  }\href {\doibase 10.1038/s41550-019-0902-0} {\bibfield  {journal} {\bibinfo
  {journal} {Nature Astronomy}\ }\textbf {\bibinfo {volume} {3}},\ \bibinfo
  {pages} {891} (\bibinfo {year} {2019})},\ \Eprint
  {http://arxiv.org/abs/1907.10625} {arXiv:1907.10625 [astro-ph.CO]}
  \BibitemShut {NoStop}%
\bibitem [{\citenamefont {{Handley}}(2019)}]{2019arXiv190809139H}%
  \BibitemOpen
  \bibfield  {author} {\bibinfo {author} {\bibfnamefont {W.}~\bibnamefont
  {{Handley}}},\ }\href@noop {} {\bibfield  {journal} {\bibinfo  {journal}
  {arXiv e-prints}\ ,\ \bibinfo {eid} {arXiv:1908.09139}} (\bibinfo {year}
  {2019})},\ \Eprint {http://arxiv.org/abs/1908.09139} {arXiv:1908.09139
  [astro-ph.CO]} \BibitemShut {NoStop}%
\bibitem [{\citenamefont {{Di Valentino}}\ \emph {et~al.}(2019)\citenamefont
  {{Di Valentino}}, \citenamefont {{Melchiorri}},\ and\ \citenamefont
  {{Silk}}}]{2019arXiv191102087D}%
  \BibitemOpen
  \bibfield  {author} {\bibinfo {author} {\bibfnamefont {E.}~\bibnamefont {{Di
  Valentino}}}, \bibinfo {author} {\bibfnamefont {A.}~\bibnamefont
  {{Melchiorri}}}, \ and\ \bibinfo {author} {\bibfnamefont {J.}~\bibnamefont
  {{Silk}}},\ }\href@noop {} {\bibfield  {journal} {\bibinfo  {journal} {arXiv
  e-prints}\ ,\ \bibinfo {eid} {arXiv:1911.02087}} (\bibinfo {year} {2019})},\
  \Eprint {http://arxiv.org/abs/1911.02087} {arXiv:1911.02087 [astro-ph.CO]}
  \BibitemShut {NoStop}%
\bibitem [{\citenamefont {{Bullock}}\ and\ \citenamefont
  {{Boylan-Kolchin}}(2017)}]{2017ARA&A..55..343B}%
  \BibitemOpen
  \bibfield  {author} {\bibinfo {author} {\bibfnamefont {J.~S.}\ \bibnamefont
  {{Bullock}}}\ and\ \bibinfo {author} {\bibfnamefont {M.}~\bibnamefont
  {{Boylan-Kolchin}}},\ }\href {\doibase 10.1146/annurev-astro-091916-055313}
  {\bibfield  {journal} {\bibinfo  {journal} {Annual Review of Astronomy and
  Astrophysics}\ }\textbf {\bibinfo {volume} {55}},\ \bibinfo {pages} {343}
  (\bibinfo {year} {2017})},\ \Eprint {http://arxiv.org/abs/1707.04256}
  {arXiv:1707.04256 [astro-ph.CO]} \BibitemShut {NoStop}%
\bibitem [{\citenamefont {{Tegmark}}\ \emph {et~al.}(2004)\citenamefont
  {{Tegmark}}, \citenamefont {{Strauss}}, \citenamefont {{Blanton}},
  \citenamefont {{Abazajian}}, \citenamefont {{Dodelson}}, \citenamefont
  {{Sandvik}}, \citenamefont {{Wang}}, \citenamefont {{Weinberg}},
  \citenamefont {{Zehavi}}, \citenamefont {{Bahcall}}, \citenamefont {{Hoyle}},
  \citenamefont {{Schlegel}}, \citenamefont {{Scoccimarro}}, \citenamefont
  {{Vogeley}}, \citenamefont {{Berlind}}, \citenamefont {{Budavari}},
  \citenamefont {{Connolly}}, \citenamefont {{Eisenstein}}, \citenamefont
  {{Finkbeiner}}, \citenamefont {{Frieman}}, \citenamefont {{Gunn}},
  \citenamefont {{Hui}}, \citenamefont {{Jain}}, \citenamefont {{Johnston}},
  \citenamefont {{Kent}}, \citenamefont {{Lin}}, \citenamefont {{Nakajima}},
  \citenamefont {{Nichol}}, \citenamefont {{Ostriker}}, \citenamefont {{Pope}},
  \citenamefont {{Scranton}}, \citenamefont {{Seljak}}, \citenamefont
  {{Sheth}}, \citenamefont {{Stebbins}}, \citenamefont {{Szalay}},
  \citenamefont {{Szapudi}}, \citenamefont {{Xu}}, \citenamefont {{Annis}},
  \citenamefont {{Brinkmann}}, \citenamefont {{Burles}}, \citenamefont
  {{Castand er}}, \citenamefont {{Csabai}}, \citenamefont {{Loveday}},
  \citenamefont {{Doi}}, \citenamefont {{Fukugita}}, \citenamefont
  {{Gillespie}}, \citenamefont {{Hennessy}}, \citenamefont {{Hogg}},
  \citenamefont {{Ivezi{\'c}}}, \citenamefont {{Knapp}}, \citenamefont
  {{Lamb}}, \citenamefont {{Lee}}, \citenamefont {{Lupton}}, \citenamefont
  {{McKay}}, \citenamefont {{Kunszt}}, \citenamefont {{Munn}}, \citenamefont
  {{O'Connell}}, \citenamefont {{Peoples}}, \citenamefont {{Pier}},
  \citenamefont {{Richmond}}, \citenamefont {{Rockosi}}, \citenamefont
  {{Schneider}}, \citenamefont {{Stoughton}}, \citenamefont {{Tucker}},
  \citenamefont {{vanden Berk}}, \citenamefont {{Yanny}},\ and\ \citenamefont
  {{York}}}]{2004PhRvD..69j3501T}%
  \BibitemOpen
  \bibfield  {author} {\bibinfo {author} {\bibfnamefont {M.}~\bibnamefont
  {{Tegmark}}}, \bibinfo {author} {\bibfnamefont {M.~A.}\ \bibnamefont
  {{Strauss}}}, \bibinfo {author} {\bibfnamefont {M.~R.}\ \bibnamefont
  {{Blanton}}}, \bibinfo {author} {\bibfnamefont {K.}~\bibnamefont
  {{Abazajian}}}, \bibinfo {author} {\bibfnamefont {S.}~\bibnamefont
  {{Dodelson}}}, \bibinfo {author} {\bibfnamefont {H.}~\bibnamefont
  {{Sandvik}}}, \bibinfo {author} {\bibfnamefont {X.}~\bibnamefont {{Wang}}},
  \bibinfo {author} {\bibfnamefont {D.~H.}\ \bibnamefont {{Weinberg}}},
  \bibinfo {author} {\bibfnamefont {I.}~\bibnamefont {{Zehavi}}}, \bibinfo
  {author} {\bibfnamefont {N.~A.}\ \bibnamefont {{Bahcall}}}, \bibinfo {author}
  {\bibfnamefont {F.}~\bibnamefont {{Hoyle}}}, \bibinfo {author} {\bibfnamefont
  {D.}~\bibnamefont {{Schlegel}}}, \bibinfo {author} {\bibfnamefont
  {R.}~\bibnamefont {{Scoccimarro}}}, \bibinfo {author} {\bibfnamefont {M.~S.}\
  \bibnamefont {{Vogeley}}}, \bibinfo {author} {\bibfnamefont {A.}~\bibnamefont
  {{Berlind}}}, \bibinfo {author} {\bibfnamefont {T.}~\bibnamefont
  {{Budavari}}}, \bibinfo {author} {\bibfnamefont {A.}~\bibnamefont
  {{Connolly}}}, \bibinfo {author} {\bibfnamefont {D.~J.}\ \bibnamefont
  {{Eisenstein}}}, \bibinfo {author} {\bibfnamefont {D.}~\bibnamefont
  {{Finkbeiner}}}, \bibinfo {author} {\bibfnamefont {J.~A.}\ \bibnamefont
  {{Frieman}}}, \bibinfo {author} {\bibfnamefont {J.~E.}\ \bibnamefont
  {{Gunn}}}, \bibinfo {author} {\bibfnamefont {L.}~\bibnamefont {{Hui}}},
  \bibinfo {author} {\bibfnamefont {B.}~\bibnamefont {{Jain}}}, \bibinfo
  {author} {\bibfnamefont {D.}~\bibnamefont {{Johnston}}}, \bibinfo {author}
  {\bibfnamefont {S.}~\bibnamefont {{Kent}}}, \bibinfo {author} {\bibfnamefont
  {H.}~\bibnamefont {{Lin}}}, \bibinfo {author} {\bibfnamefont
  {R.}~\bibnamefont {{Nakajima}}}, \bibinfo {author} {\bibfnamefont {R.~C.}\
  \bibnamefont {{Nichol}}}, \bibinfo {author} {\bibfnamefont {J.~P.}\
  \bibnamefont {{Ostriker}}}, \bibinfo {author} {\bibfnamefont
  {A.}~\bibnamefont {{Pope}}}, \bibinfo {author} {\bibfnamefont
  {R.}~\bibnamefont {{Scranton}}}, \bibinfo {author} {\bibfnamefont
  {U.}~\bibnamefont {{Seljak}}}, \bibinfo {author} {\bibfnamefont {R.~K.}\
  \bibnamefont {{Sheth}}}, \bibinfo {author} {\bibfnamefont {A.}~\bibnamefont
  {{Stebbins}}}, \bibinfo {author} {\bibfnamefont {A.~S.}\ \bibnamefont
  {{Szalay}}}, \bibinfo {author} {\bibfnamefont {I.}~\bibnamefont {{Szapudi}}},
  \bibinfo {author} {\bibfnamefont {Y.}~\bibnamefont {{Xu}}}, \bibinfo {author}
  {\bibfnamefont {J.}~\bibnamefont {{Annis}}}, \bibinfo {author} {\bibfnamefont
  {J.}~\bibnamefont {{Brinkmann}}}, \bibinfo {author} {\bibfnamefont
  {S.}~\bibnamefont {{Burles}}}, \bibinfo {author} {\bibfnamefont {F.~J.}\
  \bibnamefont {{Castand er}}}, \bibinfo {author} {\bibfnamefont
  {I.}~\bibnamefont {{Csabai}}}, \bibinfo {author} {\bibfnamefont
  {J.}~\bibnamefont {{Loveday}}}, \bibinfo {author} {\bibfnamefont
  {M.}~\bibnamefont {{Doi}}}, \bibinfo {author} {\bibfnamefont
  {M.}~\bibnamefont {{Fukugita}}}, \bibinfo {author} {\bibfnamefont
  {B.}~\bibnamefont {{Gillespie}}}, \bibinfo {author} {\bibfnamefont
  {G.}~\bibnamefont {{Hennessy}}}, \bibinfo {author} {\bibfnamefont {D.~W.}\
  \bibnamefont {{Hogg}}}, \bibinfo {author} {\bibfnamefont
  {{\v{Z}}.}~\bibnamefont {{Ivezi{\'c}}}}, \bibinfo {author} {\bibfnamefont
  {G.~R.}\ \bibnamefont {{Knapp}}}, \bibinfo {author} {\bibfnamefont {D.~Q.}\
  \bibnamefont {{Lamb}}}, \bibinfo {author} {\bibfnamefont {B.~C.}\
  \bibnamefont {{Lee}}}, \bibinfo {author} {\bibfnamefont {R.~H.}\ \bibnamefont
  {{Lupton}}}, \bibinfo {author} {\bibfnamefont {T.~A.}\ \bibnamefont
  {{McKay}}}, \bibinfo {author} {\bibfnamefont {P.}~\bibnamefont {{Kunszt}}},
  \bibinfo {author} {\bibfnamefont {J.~A.}\ \bibnamefont {{Munn}}}, \bibinfo
  {author} {\bibfnamefont {L.}~\bibnamefont {{O'Connell}}}, \bibinfo {author}
  {\bibfnamefont {J.}~\bibnamefont {{Peoples}}}, \bibinfo {author}
  {\bibfnamefont {J.~R.}\ \bibnamefont {{Pier}}}, \bibinfo {author}
  {\bibfnamefont {M.}~\bibnamefont {{Richmond}}}, \bibinfo {author}
  {\bibfnamefont {C.}~\bibnamefont {{Rockosi}}}, \bibinfo {author}
  {\bibfnamefont {D.~P.}\ \bibnamefont {{Schneider}}}, \bibinfo {author}
  {\bibfnamefont {C.}~\bibnamefont {{Stoughton}}}, \bibinfo {author}
  {\bibfnamefont {D.~L.}\ \bibnamefont {{Tucker}}}, \bibinfo {author}
  {\bibfnamefont {D.~E.}\ \bibnamefont {{vanden Berk}}}, \bibinfo {author}
  {\bibfnamefont {B.}~\bibnamefont {{Yanny}}}, \ and\ \bibinfo {author}
  {\bibfnamefont {D.~G.}\ \bibnamefont {{York}}},\ }\href {\doibase
  10.1103/PhysRevD.69.103501} {\bibfield  {journal} {\bibinfo  {journal} {Phys.
  Rev. D}\ }\textbf {\bibinfo {volume} {69}},\ \bibinfo {eid} {103501}
  (\bibinfo {year} {2004})},\ \Eprint {http://arxiv.org/abs/astro-ph/0310723}
  {arXiv:astro-ph/0310723 [astro-ph]} \BibitemShut {NoStop}%
\bibitem [{\citenamefont {{Planck Collaboration}}\ \emph
  {et~al.}(2018)\citenamefont {{Planck Collaboration}}, \citenamefont
  {{Aghanim}}, \citenamefont {{Akrami}}, \citenamefont {{Ashdown}},
  \citenamefont {{Aumont}}, \citenamefont {{Baccigalupi}}, \citenamefont
  {{Ballardini}}, \citenamefont {{Banday}}, \citenamefont {{Barreiro}},
  \citenamefont {{Bartolo}}, \citenamefont {{Basak}}, \citenamefont {{Battye}},
  \citenamefont {{Benabed}}, \citenamefont {{Bernard}}, \citenamefont
  {{Bersanelli}}, \citenamefont {{Bielewicz}}, \citenamefont {{Bock}},
  \citenamefont {{Bond}}, \citenamefont {{Borrill}}, \citenamefont {{Bouchet}},
  \citenamefont {{Boulanger}}, \citenamefont {{Bucher}}, \citenamefont
  {{Burigana}}, \citenamefont {{Butler}}, \citenamefont {{Calabrese}},
  \citenamefont {{Cardoso}}, \citenamefont {{Carron}}, \citenamefont
  {{Challinor}}, \citenamefont {{Chiang}}, \citenamefont {{Chluba}},
  \citenamefont {{Colombo}}, \citenamefont {{Combet}}, \citenamefont
  {{Contreras}}, \citenamefont {{Crill}}, \citenamefont {{Cuttaia}},
  \citenamefont {{de Bernardis}}, \citenamefont {{de Zotti}}, \citenamefont
  {{Delabrouille}}, \citenamefont {{Delouis}}, \citenamefont {{Di Valentino}},
  \citenamefont {{Diego}}, \citenamefont {{Dor{\'e}}}, \citenamefont
  {{Douspis}}, \citenamefont {{Ducout}}, \citenamefont {{Dupac}}, \citenamefont
  {{Dusini}}, \citenamefont {{Efstathiou}}, \citenamefont {{Elsner}},
  \citenamefont {{En{\ss}lin}}, \citenamefont {{Eriksen}}, \citenamefont
  {{Fantaye}}, \citenamefont {{Farhang}}, \citenamefont {{Fergusson}},
  \citenamefont {{Fernandez-Cobos}}, \citenamefont {{Finelli}}, \citenamefont
  {{Forastieri}}, \citenamefont {{Frailis}}, \citenamefont {{Fraisse}},
  \citenamefont {{Franceschi}}, \citenamefont {{Frolov}}, \citenamefont
  {{Galeotta}}, \citenamefont {{Galli}}, \citenamefont {{Ganga}}, \citenamefont
  {{G{\'e}nova-Santos}}, \citenamefont {{Gerbino}}, \citenamefont {{Ghosh}},
  \citenamefont {{Gonz{\'a}lez-Nuevo}}, \citenamefont {{G{\'o}rski}},
  \citenamefont {{Gratton}}, \citenamefont {{Gruppuso}}, \citenamefont
  {{Gudmundsson}}, \citenamefont {{Hamann}}, \citenamefont {{Handley}},
  \citenamefont {{Hansen}}, \citenamefont {{Herranz}}, \citenamefont
  {{Hildebrandt}}, \citenamefont {{Hivon}}, \citenamefont {{Huang}},
  \citenamefont {{Jaffe}}, \citenamefont {{Jones}}, \citenamefont {{Karakci}},
  \citenamefont {{Keih{\"a}nen}}, \citenamefont {{Keskitalo}}, \citenamefont
  {{Kiiveri}}, \citenamefont {{Kim}}, \citenamefont {{Kisner}}, \citenamefont
  {{Knox}}, \citenamefont {{Krachmalnicoff}}, \citenamefont {{Kunz}},
  \citenamefont {{Kurki-Suonio}}, \citenamefont {{Lagache}}, \citenamefont
  {{Lamarre}}, \citenamefont {{Lasenby}}, \citenamefont {{Lattanzi}},
  \citenamefont {{Lawrence}}, \citenamefont {{Le Jeune}}, \citenamefont
  {{Lemos}}, \citenamefont {{Lesgourgues}}, \citenamefont {{Levrier}},
  \citenamefont {{Lewis}}, \citenamefont {{Liguori}}, \citenamefont {{Lilje}},
  \citenamefont {{Lilley}}, \citenamefont {{Lindholm}}, \citenamefont
  {{L{\'o}pez-Caniego}}, \citenamefont {{Lubin}}, \citenamefont {{Ma}},
  \citenamefont {{Mac{\'\i}as-P{\'e}rez}}, \citenamefont {{Maggio}},
  \citenamefont {{Maino}}, \citenamefont {{Mandolesi}}, \citenamefont
  {{Mangilli}}, \citenamefont {{Marcos-Caballero}}, \citenamefont {{Maris}},
  \citenamefont {{Martin}}, \citenamefont {{Martinelli}}, \citenamefont
  {{Mart{\'\i}nez-Gonz{\'a}lez}}, \citenamefont {{Matarrese}}, \citenamefont
  {{Mauri}}, \citenamefont {{McEwen}}, \citenamefont {{Meinhold}},
  \citenamefont {{Melchiorri}}, \citenamefont {{Mennella}}, \citenamefont
  {{Migliaccio}}, \citenamefont {{Millea}}, \citenamefont {{Mitra}},
  \citenamefont {{Miville-Desch{\^e}nes}}, \citenamefont {{Molinari}},
  \citenamefont {{Montier}}, \citenamefont {{Morgante}}, \citenamefont
  {{Moss}}, \citenamefont {{Natoli}}, \citenamefont {{N{\o}rgaard-Nielsen}},
  \citenamefont {{Pagano}}, \citenamefont {{Paoletti}}, \citenamefont
  {{Partridge}}, \citenamefont {{Patanchon}}, \citenamefont {{Peiris}},
  \citenamefont {{Perrotta}}, \citenamefont {{Pettorino}}, \citenamefont
  {{Piacentini}}, \citenamefont {{Polastri}}, \citenamefont {{Polenta}},
  \citenamefont {{Puget}}, \citenamefont {{Rachen}}, \citenamefont
  {{Reinecke}}, \citenamefont {{Remazeilles}}, \citenamefont {{Renzi}},
  \citenamefont {{Rocha}}, \citenamefont {{Rosset}}, \citenamefont {{Roudier}},
  \citenamefont {{Rubi{\~n}o-Mart{\'\i}n}}, \citenamefont {{Ruiz-Granados}},
  \citenamefont {{Salvati}}, \citenamefont {{Sandri}}, \citenamefont
  {{Savelainen}}, \citenamefont {{Scott}}, \citenamefont {{Shellard}},
  \citenamefont {{Sirignano}}, \citenamefont {{Sirri}}, \citenamefont
  {{Spencer}}, \citenamefont {{Sunyaev}}, \citenamefont {{Suur-Uski}},
  \citenamefont {{Tauber}}, \citenamefont {{Tavagnacco}}, \citenamefont
  {{Tenti}}, \citenamefont {{Toffolatti}}, \citenamefont {{Tomasi}},
  \citenamefont {{Trombetti}}, \citenamefont {{Valenziano}}, \citenamefont
  {{Valiviita}}, \citenamefont {{Van Tent}}, \citenamefont {{Vibert}},
  \citenamefont {{Vielva}}, \citenamefont {{Villa}}, \citenamefont
  {{Vittorio}}, \citenamefont {{Wand elt}}, \citenamefont {{Wehus}},
  \citenamefont {{White}}, \citenamefont {{White}}, \citenamefont {{Zacchei}},\
  and\ \citenamefont {{Zonca}}}]{2018arXiv180706209P}%
  \BibitemOpen
  \bibfield  {author} {\bibinfo {author} {\bibnamefont {{Planck
  Collaboration}}}, \bibinfo {author} {\bibfnamefont {N.}~\bibnamefont
  {{Aghanim}}}, \bibinfo {author} {\bibfnamefont {Y.}~\bibnamefont {{Akrami}}},
  \bibinfo {author} {\bibfnamefont {M.}~\bibnamefont {{Ashdown}}}, \bibinfo
  {author} {\bibfnamefont {J.}~\bibnamefont {{Aumont}}}, \bibinfo {author}
  {\bibfnamefont {C.}~\bibnamefont {{Baccigalupi}}}, \bibinfo {author}
  {\bibfnamefont {M.}~\bibnamefont {{Ballardini}}}, \bibinfo {author}
  {\bibfnamefont {A.~J.}\ \bibnamefont {{Banday}}}, \bibinfo {author}
  {\bibfnamefont {R.~B.}\ \bibnamefont {{Barreiro}}}, \bibinfo {author}
  {\bibfnamefont {N.}~\bibnamefont {{Bartolo}}}, \bibinfo {author}
  {\bibfnamefont {S.}~\bibnamefont {{Basak}}}, \bibinfo {author} {\bibfnamefont
  {R.}~\bibnamefont {{Battye}}}, \bibinfo {author} {\bibfnamefont
  {K.}~\bibnamefont {{Benabed}}}, \bibinfo {author} {\bibfnamefont {J.~P.}\
  \bibnamefont {{Bernard}}}, \bibinfo {author} {\bibfnamefont {M.}~\bibnamefont
  {{Bersanelli}}}, \bibinfo {author} {\bibfnamefont {P.}~\bibnamefont
  {{Bielewicz}}}, \bibinfo {author} {\bibfnamefont {J.~J.}\ \bibnamefont
  {{Bock}}}, \bibinfo {author} {\bibfnamefont {J.~R.}\ \bibnamefont {{Bond}}},
  \bibinfo {author} {\bibfnamefont {J.}~\bibnamefont {{Borrill}}}, \bibinfo
  {author} {\bibfnamefont {F.~R.}\ \bibnamefont {{Bouchet}}}, \bibinfo {author}
  {\bibfnamefont {F.}~\bibnamefont {{Boulanger}}}, \bibinfo {author}
  {\bibfnamefont {M.}~\bibnamefont {{Bucher}}}, \bibinfo {author}
  {\bibfnamefont {C.}~\bibnamefont {{Burigana}}}, \bibinfo {author}
  {\bibfnamefont {R.~C.}\ \bibnamefont {{Butler}}}, \bibinfo {author}
  {\bibfnamefont {E.}~\bibnamefont {{Calabrese}}}, \bibinfo {author}
  {\bibfnamefont {J.~F.}\ \bibnamefont {{Cardoso}}}, \bibinfo {author}
  {\bibfnamefont {J.}~\bibnamefont {{Carron}}}, \bibinfo {author}
  {\bibfnamefont {A.}~\bibnamefont {{Challinor}}}, \bibinfo {author}
  {\bibfnamefont {H.~C.}\ \bibnamefont {{Chiang}}}, \bibinfo {author}
  {\bibfnamefont {J.}~\bibnamefont {{Chluba}}}, \bibinfo {author}
  {\bibfnamefont {L.~P.~L.}\ \bibnamefont {{Colombo}}}, \bibinfo {author}
  {\bibfnamefont {C.}~\bibnamefont {{Combet}}}, \bibinfo {author}
  {\bibfnamefont {D.}~\bibnamefont {{Contreras}}}, \bibinfo {author}
  {\bibfnamefont {B.~P.}\ \bibnamefont {{Crill}}}, \bibinfo {author}
  {\bibfnamefont {F.}~\bibnamefont {{Cuttaia}}}, \bibinfo {author}
  {\bibfnamefont {P.}~\bibnamefont {{de Bernardis}}}, \bibinfo {author}
  {\bibfnamefont {G.}~\bibnamefont {{de Zotti}}}, \bibinfo {author}
  {\bibfnamefont {J.}~\bibnamefont {{Delabrouille}}}, \bibinfo {author}
  {\bibfnamefont {J.~M.}\ \bibnamefont {{Delouis}}}, \bibinfo {author}
  {\bibfnamefont {E.}~\bibnamefont {{Di Valentino}}}, \bibinfo {author}
  {\bibfnamefont {J.~M.}\ \bibnamefont {{Diego}}}, \bibinfo {author}
  {\bibfnamefont {O.}~\bibnamefont {{Dor{\'e}}}}, \bibinfo {author}
  {\bibfnamefont {M.}~\bibnamefont {{Douspis}}}, \bibinfo {author}
  {\bibfnamefont {A.}~\bibnamefont {{Ducout}}}, \bibinfo {author}
  {\bibfnamefont {X.}~\bibnamefont {{Dupac}}}, \bibinfo {author} {\bibfnamefont
  {S.}~\bibnamefont {{Dusini}}}, \bibinfo {author} {\bibfnamefont
  {G.}~\bibnamefont {{Efstathiou}}}, \bibinfo {author} {\bibfnamefont
  {F.}~\bibnamefont {{Elsner}}}, \bibinfo {author} {\bibfnamefont {T.~A.}\
  \bibnamefont {{En{\ss}lin}}}, \bibinfo {author} {\bibfnamefont {H.~K.}\
  \bibnamefont {{Eriksen}}}, \bibinfo {author} {\bibfnamefont {Y.}~\bibnamefont
  {{Fantaye}}}, \bibinfo {author} {\bibfnamefont {M.}~\bibnamefont
  {{Farhang}}}, \bibinfo {author} {\bibfnamefont {J.}~\bibnamefont
  {{Fergusson}}}, \bibinfo {author} {\bibfnamefont {R.}~\bibnamefont
  {{Fernandez-Cobos}}}, \bibinfo {author} {\bibfnamefont {F.}~\bibnamefont
  {{Finelli}}}, \bibinfo {author} {\bibfnamefont {F.}~\bibnamefont
  {{Forastieri}}}, \bibinfo {author} {\bibfnamefont {M.}~\bibnamefont
  {{Frailis}}}, \bibinfo {author} {\bibfnamefont {A.~A.}\ \bibnamefont
  {{Fraisse}}}, \bibinfo {author} {\bibfnamefont {E.}~\bibnamefont
  {{Franceschi}}}, \bibinfo {author} {\bibfnamefont {A.}~\bibnamefont
  {{Frolov}}}, \bibinfo {author} {\bibfnamefont {S.}~\bibnamefont
  {{Galeotta}}}, \bibinfo {author} {\bibfnamefont {S.}~\bibnamefont {{Galli}}},
  \bibinfo {author} {\bibfnamefont {K.}~\bibnamefont {{Ganga}}}, \bibinfo
  {author} {\bibfnamefont {R.~T.}\ \bibnamefont {{G{\'e}nova-Santos}}},
  \bibinfo {author} {\bibfnamefont {M.}~\bibnamefont {{Gerbino}}}, \bibinfo
  {author} {\bibfnamefont {T.}~\bibnamefont {{Ghosh}}}, \bibinfo {author}
  {\bibfnamefont {J.}~\bibnamefont {{Gonz{\'a}lez-Nuevo}}}, \bibinfo {author}
  {\bibfnamefont {K.~M.}\ \bibnamefont {{G{\'o}rski}}}, \bibinfo {author}
  {\bibfnamefont {S.}~\bibnamefont {{Gratton}}}, \bibinfo {author}
  {\bibfnamefont {A.}~\bibnamefont {{Gruppuso}}}, \bibinfo {author}
  {\bibfnamefont {J.~E.}\ \bibnamefont {{Gudmundsson}}}, \bibinfo {author}
  {\bibfnamefont {J.}~\bibnamefont {{Hamann}}}, \bibinfo {author}
  {\bibfnamefont {W.}~\bibnamefont {{Handley}}}, \bibinfo {author}
  {\bibfnamefont {F.~K.}\ \bibnamefont {{Hansen}}}, \bibinfo {author}
  {\bibfnamefont {D.}~\bibnamefont {{Herranz}}}, \bibinfo {author}
  {\bibfnamefont {S.~R.}\ \bibnamefont {{Hildebrandt}}}, \bibinfo {author}
  {\bibfnamefont {E.}~\bibnamefont {{Hivon}}}, \bibinfo {author} {\bibfnamefont
  {Z.}~\bibnamefont {{Huang}}}, \bibinfo {author} {\bibfnamefont {A.~H.}\
  \bibnamefont {{Jaffe}}}, \bibinfo {author} {\bibfnamefont {W.~C.}\
  \bibnamefont {{Jones}}}, \bibinfo {author} {\bibfnamefont {A.}~\bibnamefont
  {{Karakci}}}, \bibinfo {author} {\bibfnamefont {E.}~\bibnamefont
  {{Keih{\"a}nen}}}, \bibinfo {author} {\bibfnamefont {R.}~\bibnamefont
  {{Keskitalo}}}, \bibinfo {author} {\bibfnamefont {K.}~\bibnamefont
  {{Kiiveri}}}, \bibinfo {author} {\bibfnamefont {J.}~\bibnamefont {{Kim}}},
  \bibinfo {author} {\bibfnamefont {T.~S.}\ \bibnamefont {{Kisner}}}, \bibinfo
  {author} {\bibfnamefont {L.}~\bibnamefont {{Knox}}}, \bibinfo {author}
  {\bibfnamefont {N.}~\bibnamefont {{Krachmalnicoff}}}, \bibinfo {author}
  {\bibfnamefont {M.}~\bibnamefont {{Kunz}}}, \bibinfo {author} {\bibfnamefont
  {H.}~\bibnamefont {{Kurki-Suonio}}}, \bibinfo {author} {\bibfnamefont
  {G.}~\bibnamefont {{Lagache}}}, \bibinfo {author} {\bibfnamefont {J.~M.}\
  \bibnamefont {{Lamarre}}}, \bibinfo {author} {\bibfnamefont {A.}~\bibnamefont
  {{Lasenby}}}, \bibinfo {author} {\bibfnamefont {M.}~\bibnamefont
  {{Lattanzi}}}, \bibinfo {author} {\bibfnamefont {C.~R.}\ \bibnamefont
  {{Lawrence}}}, \bibinfo {author} {\bibfnamefont {M.}~\bibnamefont {{Le
  Jeune}}}, \bibinfo {author} {\bibfnamefont {P.}~\bibnamefont {{Lemos}}},
  \bibinfo {author} {\bibfnamefont {J.}~\bibnamefont {{Lesgourgues}}}, \bibinfo
  {author} {\bibfnamefont {F.}~\bibnamefont {{Levrier}}}, \bibinfo {author}
  {\bibfnamefont {A.}~\bibnamefont {{Lewis}}}, \bibinfo {author} {\bibfnamefont
  {M.}~\bibnamefont {{Liguori}}}, \bibinfo {author} {\bibfnamefont {P.~B.}\
  \bibnamefont {{Lilje}}}, \bibinfo {author} {\bibfnamefont {M.}~\bibnamefont
  {{Lilley}}}, \bibinfo {author} {\bibfnamefont {V.}~\bibnamefont
  {{Lindholm}}}, \bibinfo {author} {\bibfnamefont {M.}~\bibnamefont
  {{L{\'o}pez-Caniego}}}, \bibinfo {author} {\bibfnamefont {P.~M.}\
  \bibnamefont {{Lubin}}}, \bibinfo {author} {\bibfnamefont {Y.~Z.}\
  \bibnamefont {{Ma}}}, \bibinfo {author} {\bibfnamefont {J.~F.}\ \bibnamefont
  {{Mac{\'\i}as-P{\'e}rez}}}, \bibinfo {author} {\bibfnamefont
  {G.}~\bibnamefont {{Maggio}}}, \bibinfo {author} {\bibfnamefont
  {D.}~\bibnamefont {{Maino}}}, \bibinfo {author} {\bibfnamefont
  {N.}~\bibnamefont {{Mandolesi}}}, \bibinfo {author} {\bibfnamefont
  {A.}~\bibnamefont {{Mangilli}}}, \bibinfo {author} {\bibfnamefont
  {A.}~\bibnamefont {{Marcos-Caballero}}}, \bibinfo {author} {\bibfnamefont
  {M.}~\bibnamefont {{Maris}}}, \bibinfo {author} {\bibfnamefont {P.~G.}\
  \bibnamefont {{Martin}}}, \bibinfo {author} {\bibfnamefont {M.}~\bibnamefont
  {{Martinelli}}}, \bibinfo {author} {\bibfnamefont {E.}~\bibnamefont
  {{Mart{\'\i}nez-Gonz{\'a}lez}}}, \bibinfo {author} {\bibfnamefont
  {S.}~\bibnamefont {{Matarrese}}}, \bibinfo {author} {\bibfnamefont
  {N.}~\bibnamefont {{Mauri}}}, \bibinfo {author} {\bibfnamefont {J.~D.}\
  \bibnamefont {{McEwen}}}, \bibinfo {author} {\bibfnamefont {P.~R.}\
  \bibnamefont {{Meinhold}}}, \bibinfo {author} {\bibfnamefont
  {A.}~\bibnamefont {{Melchiorri}}}, \bibinfo {author} {\bibfnamefont
  {A.}~\bibnamefont {{Mennella}}}, \bibinfo {author} {\bibfnamefont
  {M.}~\bibnamefont {{Migliaccio}}}, \bibinfo {author} {\bibfnamefont
  {M.}~\bibnamefont {{Millea}}}, \bibinfo {author} {\bibfnamefont
  {S.}~\bibnamefont {{Mitra}}}, \bibinfo {author} {\bibfnamefont {M.~A.}\
  \bibnamefont {{Miville-Desch{\^e}nes}}}, \bibinfo {author} {\bibfnamefont
  {D.}~\bibnamefont {{Molinari}}}, \bibinfo {author} {\bibfnamefont
  {L.}~\bibnamefont {{Montier}}}, \bibinfo {author} {\bibfnamefont
  {G.}~\bibnamefont {{Morgante}}}, \bibinfo {author} {\bibfnamefont
  {A.}~\bibnamefont {{Moss}}}, \bibinfo {author} {\bibfnamefont
  {P.}~\bibnamefont {{Natoli}}}, \bibinfo {author} {\bibfnamefont {H.~U.}\
  \bibnamefont {{N{\o}rgaard-Nielsen}}}, \bibinfo {author} {\bibfnamefont
  {L.}~\bibnamefont {{Pagano}}}, \bibinfo {author} {\bibfnamefont
  {D.}~\bibnamefont {{Paoletti}}}, \bibinfo {author} {\bibfnamefont
  {B.}~\bibnamefont {{Partridge}}}, \bibinfo {author} {\bibfnamefont
  {G.}~\bibnamefont {{Patanchon}}}, \bibinfo {author} {\bibfnamefont {H.~V.}\
  \bibnamefont {{Peiris}}}, \bibinfo {author} {\bibfnamefont {F.}~\bibnamefont
  {{Perrotta}}}, \bibinfo {author} {\bibfnamefont {V.}~\bibnamefont
  {{Pettorino}}}, \bibinfo {author} {\bibfnamefont {F.}~\bibnamefont
  {{Piacentini}}}, \bibinfo {author} {\bibfnamefont {L.}~\bibnamefont
  {{Polastri}}}, \bibinfo {author} {\bibfnamefont {G.}~\bibnamefont
  {{Polenta}}}, \bibinfo {author} {\bibfnamefont {J.~L.}\ \bibnamefont
  {{Puget}}}, \bibinfo {author} {\bibfnamefont {J.~P.}\ \bibnamefont
  {{Rachen}}}, \bibinfo {author} {\bibfnamefont {M.}~\bibnamefont
  {{Reinecke}}}, \bibinfo {author} {\bibfnamefont {M.}~\bibnamefont
  {{Remazeilles}}}, \bibinfo {author} {\bibfnamefont {A.}~\bibnamefont
  {{Renzi}}}, \bibinfo {author} {\bibfnamefont {G.}~\bibnamefont {{Rocha}}},
  \bibinfo {author} {\bibfnamefont {C.}~\bibnamefont {{Rosset}}}, \bibinfo
  {author} {\bibfnamefont {G.}~\bibnamefont {{Roudier}}}, \bibinfo {author}
  {\bibfnamefont {J.~A.}\ \bibnamefont {{Rubi{\~n}o-Mart{\'\i}n}}}, \bibinfo
  {author} {\bibfnamefont {B.}~\bibnamefont {{Ruiz-Granados}}}, \bibinfo
  {author} {\bibfnamefont {L.}~\bibnamefont {{Salvati}}}, \bibinfo {author}
  {\bibfnamefont {M.}~\bibnamefont {{Sandri}}}, \bibinfo {author}
  {\bibfnamefont {M.}~\bibnamefont {{Savelainen}}}, \bibinfo {author}
  {\bibfnamefont {D.}~\bibnamefont {{Scott}}}, \bibinfo {author} {\bibfnamefont
  {E.~P.~S.}\ \bibnamefont {{Shellard}}}, \bibinfo {author} {\bibfnamefont
  {C.}~\bibnamefont {{Sirignano}}}, \bibinfo {author} {\bibfnamefont
  {G.}~\bibnamefont {{Sirri}}}, \bibinfo {author} {\bibfnamefont {L.~D.}\
  \bibnamefont {{Spencer}}}, \bibinfo {author} {\bibfnamefont {R.}~\bibnamefont
  {{Sunyaev}}}, \bibinfo {author} {\bibfnamefont {A.~S.}\ \bibnamefont
  {{Suur-Uski}}}, \bibinfo {author} {\bibfnamefont {J.~A.}\ \bibnamefont
  {{Tauber}}}, \bibinfo {author} {\bibfnamefont {D.}~\bibnamefont
  {{Tavagnacco}}}, \bibinfo {author} {\bibfnamefont {M.}~\bibnamefont
  {{Tenti}}}, \bibinfo {author} {\bibfnamefont {L.}~\bibnamefont
  {{Toffolatti}}}, \bibinfo {author} {\bibfnamefont {M.}~\bibnamefont
  {{Tomasi}}}, \bibinfo {author} {\bibfnamefont {T.}~\bibnamefont
  {{Trombetti}}}, \bibinfo {author} {\bibfnamefont {L.}~\bibnamefont
  {{Valenziano}}}, \bibinfo {author} {\bibfnamefont {J.}~\bibnamefont
  {{Valiviita}}}, \bibinfo {author} {\bibfnamefont {B.}~\bibnamefont {{Van
  Tent}}}, \bibinfo {author} {\bibfnamefont {L.}~\bibnamefont {{Vibert}}},
  \bibinfo {author} {\bibfnamefont {P.}~\bibnamefont {{Vielva}}}, \bibinfo
  {author} {\bibfnamefont {F.}~\bibnamefont {{Villa}}}, \bibinfo {author}
  {\bibfnamefont {N.}~\bibnamefont {{Vittorio}}}, \bibinfo {author}
  {\bibfnamefont {B.~D.}\ \bibnamefont {{Wand elt}}}, \bibinfo {author}
  {\bibfnamefont {I.~K.}\ \bibnamefont {{Wehus}}}, \bibinfo {author}
  {\bibfnamefont {M.}~\bibnamefont {{White}}}, \bibinfo {author} {\bibfnamefont
  {S.~D.~M.}\ \bibnamefont {{White}}}, \bibinfo {author} {\bibfnamefont
  {A.}~\bibnamefont {{Zacchei}}}, \ and\ \bibinfo {author} {\bibfnamefont
  {A.}~\bibnamefont {{Zonca}}},\ }\href@noop {} {\bibfield  {journal} {\bibinfo
   {journal} {arXiv e-prints}\ ,\ \bibinfo {eid} {arXiv:1807.06209}} (\bibinfo
  {year} {2018})},\ \Eprint {http://arxiv.org/abs/1807.06209} {arXiv:1807.06209
  [astro-ph.CO]} \BibitemShut {NoStop}%
\bibitem [{\citenamefont {{Riess}}\ \emph {et~al.}(2016)\citenamefont
  {{Riess}}, \citenamefont {{Macri}}, \citenamefont {{Hoffmann}}, \citenamefont
  {{Scolnic}}, \citenamefont {{Casertano}}, \citenamefont {{Filippenko}},
  \citenamefont {{Tucker}}, \citenamefont {{Reid}}, \citenamefont {{Jones}},
  \citenamefont {{Silverman}}, \citenamefont {{Chornock}}, \citenamefont
  {{Challis}}, \citenamefont {{Yuan}}, \citenamefont {{Brown}},\ and\
  \citenamefont {{Foley}}}]{2016ApJ...826...56R}%
  \BibitemOpen
  \bibfield  {author} {\bibinfo {author} {\bibfnamefont {A.~G.}\ \bibnamefont
  {{Riess}}}, \bibinfo {author} {\bibfnamefont {L.~M.}\ \bibnamefont
  {{Macri}}}, \bibinfo {author} {\bibfnamefont {S.~L.}\ \bibnamefont
  {{Hoffmann}}}, \bibinfo {author} {\bibfnamefont {D.}~\bibnamefont
  {{Scolnic}}}, \bibinfo {author} {\bibfnamefont {S.}~\bibnamefont
  {{Casertano}}}, \bibinfo {author} {\bibfnamefont {A.~V.}\ \bibnamefont
  {{Filippenko}}}, \bibinfo {author} {\bibfnamefont {B.~E.}\ \bibnamefont
  {{Tucker}}}, \bibinfo {author} {\bibfnamefont {M.~J.}\ \bibnamefont
  {{Reid}}}, \bibinfo {author} {\bibfnamefont {D.~O.}\ \bibnamefont {{Jones}}},
  \bibinfo {author} {\bibfnamefont {J.~M.}\ \bibnamefont {{Silverman}}},
  \bibinfo {author} {\bibfnamefont {R.}~\bibnamefont {{Chornock}}}, \bibinfo
  {author} {\bibfnamefont {P.}~\bibnamefont {{Challis}}}, \bibinfo {author}
  {\bibfnamefont {W.}~\bibnamefont {{Yuan}}}, \bibinfo {author} {\bibfnamefont
  {P.~J.}\ \bibnamefont {{Brown}}}, \ and\ \bibinfo {author} {\bibfnamefont
  {R.~J.}\ \bibnamefont {{Foley}}},\ }\href {\doibase
  10.3847/0004-637X/826/1/56} {\bibfield  {journal} {\bibinfo  {journal}
  {\apj}\ }\textbf {\bibinfo {volume} {826}},\ \bibinfo {eid} {56} (\bibinfo
  {year} {2016})},\ \Eprint {http://arxiv.org/abs/1604.01424} {arXiv:1604.01424
  [astro-ph.CO]} \BibitemShut {NoStop}%
\bibitem [{\citenamefont {{Yuan}}\ \emph {et~al.}(2019)\citenamefont {{Yuan}},
  \citenamefont {{Riess}}, \citenamefont {{Macri}}, \citenamefont
  {{Casertano}},\ and\ \citenamefont {{Scolnic}}}]{2019arXiv190800993Y}%
  \BibitemOpen
  \bibfield  {author} {\bibinfo {author} {\bibfnamefont {W.}~\bibnamefont
  {{Yuan}}}, \bibinfo {author} {\bibfnamefont {A.~G.}\ \bibnamefont {{Riess}}},
  \bibinfo {author} {\bibfnamefont {L.~M.}\ \bibnamefont {{Macri}}}, \bibinfo
  {author} {\bibfnamefont {S.}~\bibnamefont {{Casertano}}}, \ and\ \bibinfo
  {author} {\bibfnamefont {D.}~\bibnamefont {{Scolnic}}},\ }\href@noop {}
  {\bibfield  {journal} {\bibinfo  {journal} {arXiv e-prints}\ ,\ \bibinfo
  {eid} {arXiv:1908.00993}} (\bibinfo {year} {2019})},\ \Eprint
  {http://arxiv.org/abs/1908.00993} {arXiv:1908.00993 [astro-ph.GA]}
  \BibitemShut {NoStop}%
\bibitem [{\citenamefont {{Freedman}}\ \emph {et~al.}(2019)\citenamefont
  {{Freedman}}, \citenamefont {{Madore}}, \citenamefont {{Hatt}}, \citenamefont
  {{Hoyt}}, \citenamefont {{Jang}}, \citenamefont {{Beaton}}, \citenamefont
  {{Burns}}, \citenamefont {{Lee}}, \citenamefont {{Monson}}, \citenamefont
  {{Neeley}}, \citenamefont {{Phillips}}, \citenamefont {{Rich}},\ and\
  \citenamefont {{Seibert}}}]{2019ApJ...882...34F}%
  \BibitemOpen
  \bibfield  {author} {\bibinfo {author} {\bibfnamefont {W.~L.}\ \bibnamefont
  {{Freedman}}}, \bibinfo {author} {\bibfnamefont {B.~F.}\ \bibnamefont
  {{Madore}}}, \bibinfo {author} {\bibfnamefont {D.}~\bibnamefont {{Hatt}}},
  \bibinfo {author} {\bibfnamefont {T.~J.}\ \bibnamefont {{Hoyt}}}, \bibinfo
  {author} {\bibfnamefont {I.~S.}\ \bibnamefont {{Jang}}}, \bibinfo {author}
  {\bibfnamefont {R.~L.}\ \bibnamefont {{Beaton}}}, \bibinfo {author}
  {\bibfnamefont {C.~R.}\ \bibnamefont {{Burns}}}, \bibinfo {author}
  {\bibfnamefont {M.~G.}\ \bibnamefont {{Lee}}}, \bibinfo {author}
  {\bibfnamefont {A.~J.}\ \bibnamefont {{Monson}}}, \bibinfo {author}
  {\bibfnamefont {J.~R.}\ \bibnamefont {{Neeley}}}, \bibinfo {author}
  {\bibfnamefont {M.~M.}\ \bibnamefont {{Phillips}}}, \bibinfo {author}
  {\bibfnamefont {J.~A.}\ \bibnamefont {{Rich}}}, \ and\ \bibinfo {author}
  {\bibfnamefont {M.}~\bibnamefont {{Seibert}}},\ }\href {\doibase
  10.3847/1538-4357/ab2f73} {\bibfield  {journal} {\bibinfo  {journal} {The
  Astrophysical Journal}\ }\textbf {\bibinfo {volume} {882}},\ \bibinfo {eid}
  {34} (\bibinfo {year} {2019})},\ \Eprint {http://arxiv.org/abs/1907.05922}
  {arXiv:1907.05922 [astro-ph.CO]} \BibitemShut {NoStop}%
\bibitem [{\citenamefont {{Dietrich}}\ \emph {et~al.}(2020)\citenamefont
  {{Dietrich}}, \citenamefont {{Coughlin}}, \citenamefont {{Pang}},
  \citenamefont {{Bulla}}, \citenamefont {{Heinzel}}, \citenamefont {{Issa}},
  \citenamefont {{Tews}},\ and\ \citenamefont
  {{Antier}}}]{2020arXiv200211355D}%
  \BibitemOpen
  \bibfield  {author} {\bibinfo {author} {\bibfnamefont {T.}~\bibnamefont
  {{Dietrich}}}, \bibinfo {author} {\bibfnamefont {M.~W.}\ \bibnamefont
  {{Coughlin}}}, \bibinfo {author} {\bibfnamefont {P.~T.~H.}\ \bibnamefont
  {{Pang}}}, \bibinfo {author} {\bibfnamefont {M.}~\bibnamefont {{Bulla}}},
  \bibinfo {author} {\bibfnamefont {J.}~\bibnamefont {{Heinzel}}}, \bibinfo
  {author} {\bibfnamefont {L.}~\bibnamefont {{Issa}}}, \bibinfo {author}
  {\bibfnamefont {I.}~\bibnamefont {{Tews}}}, \ and\ \bibinfo {author}
  {\bibfnamefont {S.}~\bibnamefont {{Antier}}},\ }\href@noop {} {\bibfield
  {journal} {\bibinfo  {journal} {arXiv e-prints}\ ,\ \bibinfo {eid}
  {arXiv:2002.11355}} (\bibinfo {year} {2020})},\ \Eprint
  {http://arxiv.org/abs/2002.11355} {arXiv:2002.11355 [astro-ph.HE]}
  \BibitemShut {NoStop}%
\bibitem [{\citenamefont {{Wong}}\ \emph {et~al.}(2019)\citenamefont {{Wong}},
  \citenamefont {{Suyu}}, \citenamefont {{Chen}}, \citenamefont {{Rusu}},
  \citenamefont {{Millon}}, \citenamefont {{Sluse}}, \citenamefont {{Bonvin}},
  \citenamefont {{Fassnacht}}, \citenamefont {{Taubenberger}}, \citenamefont
  {{Auger}}, \citenamefont {{Birrer}}, \citenamefont {{Chan}}, \citenamefont
  {{Courbin}}, \citenamefont {{Hilbert}}, \citenamefont {{Tihhonova}},
  \citenamefont {{Treu}}, \citenamefont {{Agnello}}, \citenamefont {{Ding}},
  \citenamefont {{Jee}}, \citenamefont {{Komatsu}}, \citenamefont {{Shajib}},
  \citenamefont {{Sonnenfeld}}, \citenamefont {{Bland ford}}, \citenamefont
  {{Koopmans}}, \citenamefont {{Marshall}},\ and\ \citenamefont
  {{Meylan}}}]{2019arXiv190704869W}%
  \BibitemOpen
  \bibfield  {author} {\bibinfo {author} {\bibfnamefont {K.~C.}\ \bibnamefont
  {{Wong}}}, \bibinfo {author} {\bibfnamefont {S.~H.}\ \bibnamefont {{Suyu}}},
  \bibinfo {author} {\bibfnamefont {G.~C.~F.}\ \bibnamefont {{Chen}}}, \bibinfo
  {author} {\bibfnamefont {C.~E.}\ \bibnamefont {{Rusu}}}, \bibinfo {author}
  {\bibfnamefont {M.}~\bibnamefont {{Millon}}}, \bibinfo {author}
  {\bibfnamefont {D.}~\bibnamefont {{Sluse}}}, \bibinfo {author} {\bibfnamefont
  {V.}~\bibnamefont {{Bonvin}}}, \bibinfo {author} {\bibfnamefont {C.~D.}\
  \bibnamefont {{Fassnacht}}}, \bibinfo {author} {\bibfnamefont
  {S.}~\bibnamefont {{Taubenberger}}}, \bibinfo {author} {\bibfnamefont
  {M.~W.}\ \bibnamefont {{Auger}}}, \bibinfo {author} {\bibfnamefont
  {S.}~\bibnamefont {{Birrer}}}, \bibinfo {author} {\bibfnamefont {J.~H.~H.}\
  \bibnamefont {{Chan}}}, \bibinfo {author} {\bibfnamefont {F.}~\bibnamefont
  {{Courbin}}}, \bibinfo {author} {\bibfnamefont {S.}~\bibnamefont
  {{Hilbert}}}, \bibinfo {author} {\bibfnamefont {O.}~\bibnamefont
  {{Tihhonova}}}, \bibinfo {author} {\bibfnamefont {T.}~\bibnamefont {{Treu}}},
  \bibinfo {author} {\bibfnamefont {A.}~\bibnamefont {{Agnello}}}, \bibinfo
  {author} {\bibfnamefont {X.}~\bibnamefont {{Ding}}}, \bibinfo {author}
  {\bibfnamefont {I.}~\bibnamefont {{Jee}}}, \bibinfo {author} {\bibfnamefont
  {E.}~\bibnamefont {{Komatsu}}}, \bibinfo {author} {\bibfnamefont {A.~J.}\
  \bibnamefont {{Shajib}}}, \bibinfo {author} {\bibfnamefont {A.}~\bibnamefont
  {{Sonnenfeld}}}, \bibinfo {author} {\bibfnamefont {R.~D.}\ \bibnamefont
  {{Bland ford}}}, \bibinfo {author} {\bibfnamefont {L.~V.~E.}\ \bibnamefont
  {{Koopmans}}}, \bibinfo {author} {\bibfnamefont {P.~J.}\ \bibnamefont
  {{Marshall}}}, \ and\ \bibinfo {author} {\bibfnamefont {G.}~\bibnamefont
  {{Meylan}}},\ }\href@noop {} {\bibfield  {journal} {\bibinfo  {journal}
  {arXiv e-prints}\ ,\ \bibinfo {eid} {arXiv:1907.04869}} (\bibinfo {year}
  {2019})},\ \Eprint {http://arxiv.org/abs/1907.04869} {arXiv:1907.04869
  [astro-ph.CO]} \BibitemShut {NoStop}%
\bibitem [{\citenamefont {{Riess}}(2019)}]{2019NatRP...2...10R}%
  \BibitemOpen
  \bibfield  {author} {\bibinfo {author} {\bibfnamefont {A.~G.}\ \bibnamefont
  {{Riess}}},\ }\href {\doibase 10.1038/s42254-019-0137-0} {\bibfield
  {journal} {\bibinfo  {journal} {Nature Reviews Physics}\ }\textbf {\bibinfo
  {volume} {2}},\ \bibinfo {pages} {10} (\bibinfo {year} {2019})},\ \Eprint
  {http://arxiv.org/abs/2001.03624} {arXiv:2001.03624 [astro-ph.CO]}
  \BibitemShut {NoStop}%
\bibitem [{\citenamefont {{Riess}}\ \emph {et~al.}(2019)\citenamefont
  {{Riess}}, \citenamefont {{Casertano}}, \citenamefont {{Yuan}}, \citenamefont
  {{Macri}},\ and\ \citenamefont {{Scolnic}}}]{2019ApJ...876...85R}%
  \BibitemOpen
  \bibfield  {author} {\bibinfo {author} {\bibfnamefont {A.~G.}\ \bibnamefont
  {{Riess}}}, \bibinfo {author} {\bibfnamefont {S.}~\bibnamefont
  {{Casertano}}}, \bibinfo {author} {\bibfnamefont {W.}~\bibnamefont {{Yuan}}},
  \bibinfo {author} {\bibfnamefont {L.~M.}\ \bibnamefont {{Macri}}}, \ and\
  \bibinfo {author} {\bibfnamefont {D.}~\bibnamefont {{Scolnic}}},\ }\href
  {\doibase 10.3847/1538-4357/ab1422} {\bibfield  {journal} {\bibinfo
  {journal} {The Astrophysical Journal}\ }\textbf {\bibinfo {volume} {876}},\
  \bibinfo {eid} {85} (\bibinfo {year} {2019})},\ \Eprint
  {http://arxiv.org/abs/1903.07603} {arXiv:1903.07603 [astro-ph.CO]}
  \BibitemShut {NoStop}%
\bibitem [{\citenamefont {{Pandey}}\ \emph {et~al.}(2019)\citenamefont
  {{Pandey}}, \citenamefont {{Karwal}},\ and\ \citenamefont {{Das}}}]{Pandey}%
  \BibitemOpen
  \bibfield  {author} {\bibinfo {author} {\bibfnamefont {K.~L.}\ \bibnamefont
  {{Pandey}}}, \bibinfo {author} {\bibfnamefont {T.}~\bibnamefont {{Karwal}}},
  \ and\ \bibinfo {author} {\bibfnamefont {S.}~\bibnamefont {{Das}}},\
  }\href@noop {} {\bibfield  {journal} {\bibinfo  {journal} {arXiv e-prints}\
  ,\ \bibinfo {eid} {arXiv:1902.10636}} (\bibinfo {year} {2019})},\ \Eprint
  {http://arxiv.org/abs/1902.10636} {arXiv:1902.10636 [astro-ph.CO]}
  \BibitemShut {NoStop}%
\bibitem [{\citenamefont {{Poulin}}\ \emph {et~al.}(2018)\citenamefont
  {{Poulin}}, \citenamefont {{Boddy}}, \citenamefont {{Bird}},\ and\
  \citenamefont {{Kamionkowski}}}]{2018PhRvD..97l3504P}%
  \BibitemOpen
  \bibfield  {author} {\bibinfo {author} {\bibfnamefont {V.}~\bibnamefont
  {{Poulin}}}, \bibinfo {author} {\bibfnamefont {K.~K.}\ \bibnamefont
  {{Boddy}}}, \bibinfo {author} {\bibfnamefont {S.}~\bibnamefont {{Bird}}}, \
  and\ \bibinfo {author} {\bibfnamefont {M.}~\bibnamefont {{Kamionkowski}}},\
  }\href {\doibase 10.1103/PhysRevD.97.123504} {\bibfield  {journal} {\bibinfo
  {journal} {Phys. Rev. D}\ }\textbf {\bibinfo {volume} {97}},\ \bibinfo {eid}
  {123504} (\bibinfo {year} {2018})},\ \Eprint
  {http://arxiv.org/abs/1803.02474} {arXiv:1803.02474 [astro-ph.CO]}
  \BibitemShut {NoStop}%
\bibitem [{\citenamefont {{M{\"o}rtsell}}\ and\ \citenamefont
  {{Dhawan}}(2018)}]{2018JCAP...09..025M}%
  \BibitemOpen
  \bibfield  {author} {\bibinfo {author} {\bibfnamefont {E.}~\bibnamefont
  {{M{\"o}rtsell}}}\ and\ \bibinfo {author} {\bibfnamefont {S.}~\bibnamefont
  {{Dhawan}}},\ }\href {\doibase 10.1088/1475-7516/2018/09/025} {\bibfield
  {journal} {\bibinfo  {journal} {Journal of Cosmology and Astroparticle
  Physics}\ }\textbf {\bibinfo {volume} {2018}},\ \bibinfo {eid} {025}
  (\bibinfo {year} {2018})},\ \Eprint {http://arxiv.org/abs/1801.07260}
  {arXiv:1801.07260 [astro-ph.CO]} \BibitemShut {NoStop}%
\bibitem [{\citenamefont {{Elgar{\o}y}}\ and\ \citenamefont
  {{Multam{\"a}ki}}(2007)}]{2007A&A...471...65E}%
  \BibitemOpen
  \bibfield  {author} {\bibinfo {author} {\bibfnamefont {O.}~\bibnamefont
  {{Elgar{\o}y}}}\ and\ \bibinfo {author} {\bibfnamefont {T.}~\bibnamefont
  {{Multam{\"a}ki}}},\ }\href {\doibase 10.1051/0004-6361:20077292} {\bibfield
  {journal} {\bibinfo  {journal} {Astronomy and Astrophysics}\ }\textbf
  {\bibinfo {volume} {471}},\ \bibinfo {pages} {65} (\bibinfo {year} {2007})},\
  \Eprint {http://arxiv.org/abs/astro-ph/0702343} {arXiv:astro-ph/0702343
  [astro-ph]} \BibitemShut {NoStop}%
\bibitem [{\citenamefont {{Sch{\"o}neberg}}\ \emph {et~al.}(2019)\citenamefont
  {{Sch{\"o}neberg}}, \citenamefont {{Lesgourgues}},\ and\ \citenamefont
  {{Hooper}}}]{2019JCAP...10..029S}%
  \BibitemOpen
  \bibfield  {author} {\bibinfo {author} {\bibfnamefont {N.}~\bibnamefont
  {{Sch{\"o}neberg}}}, \bibinfo {author} {\bibfnamefont {J.}~\bibnamefont
  {{Lesgourgues}}}, \ and\ \bibinfo {author} {\bibfnamefont {D.~C.}\
  \bibnamefont {{Hooper}}},\ }\href {\doibase 10.1088/1475-7516/2019/10/029}
  {\bibfield  {journal} {\bibinfo  {journal} {Journal of Cosmology and
  Astroparticle Physics}\ }\textbf {\bibinfo {volume} {2019}},\ \bibinfo {eid}
  {029} (\bibinfo {year} {2019})},\ \Eprint {http://arxiv.org/abs/1907.11594}
  {arXiv:1907.11594 [astro-ph.CO]} \BibitemShut {NoStop}%
\bibitem [{\citenamefont {{Bondarenko}}\ \emph {et~al.}(2020)\citenamefont
  {{Bondarenko}}, \citenamefont {{Pradler}},\ and\ \citenamefont
  {{Sokolenko}}}]{2020arXiv200208942B}%
  \BibitemOpen
  \bibfield  {author} {\bibinfo {author} {\bibfnamefont {K.}~\bibnamefont
  {{Bondarenko}}}, \bibinfo {author} {\bibfnamefont {J.}~\bibnamefont
  {{Pradler}}}, \ and\ \bibinfo {author} {\bibfnamefont {A.}~\bibnamefont
  {{Sokolenko}}},\ }\href@noop {} {\bibfield  {journal} {\bibinfo  {journal}
  {arXiv e-prints}\ ,\ \bibinfo {eid} {arXiv:2002.08942}} (\bibinfo {year}
  {2020})},\ \Eprint {http://arxiv.org/abs/2002.08942} {arXiv:2002.08942
  [astro-ph.CO]} \BibitemShut {NoStop}%
\bibitem [{\citenamefont {Desai}\ \emph {et~al.}(2020)\citenamefont {Desai},
  \citenamefont {Dienes},\ and\ \citenamefont {Thomas}}]{PhysRevD.101.035031}%
  \BibitemOpen
  \bibfield  {author} {\bibinfo {author} {\bibfnamefont {A.}~\bibnamefont
  {Desai}}, \bibinfo {author} {\bibfnamefont {K.~R.}\ \bibnamefont {Dienes}}, \
  and\ \bibinfo {author} {\bibfnamefont {B.}~\bibnamefont {Thomas}},\ }\href
  {\doibase 10.1103/PhysRevD.101.035031} {\bibfield  {journal} {\bibinfo
  {journal} {Phys. Rev. D}\ }\textbf {\bibinfo {volume} {101}},\ \bibinfo
  {pages} {035031} (\bibinfo {year} {2020})}\BibitemShut {NoStop}%
\bibitem [{\citenamefont {{Sabti}}\ \emph {et~al.}(2020)\citenamefont
  {{Sabti}}, \citenamefont {{Alvey}}, \citenamefont {{Escudero}}, \citenamefont
  {{Fairbairn}},\ and\ \citenamefont {{Blas}}}]{2020JCAP...01..004S}%
  \BibitemOpen
  \bibfield  {author} {\bibinfo {author} {\bibfnamefont {N.}~\bibnamefont
  {{Sabti}}}, \bibinfo {author} {\bibfnamefont {J.}~\bibnamefont {{Alvey}}},
  \bibinfo {author} {\bibfnamefont {M.}~\bibnamefont {{Escudero}}}, \bibinfo
  {author} {\bibfnamefont {M.}~\bibnamefont {{Fairbairn}}}, \ and\ \bibinfo
  {author} {\bibfnamefont {D.}~\bibnamefont {{Blas}}},\ }\href {\doibase
  10.1088/1475-7516/2020/01/004} {\bibfield  {journal} {\bibinfo  {journal}
  {Journal of Cosmology and Astroparticle Physics}\ }\textbf {\bibinfo {volume}
  {2020}},\ \bibinfo {eid} {004} (\bibinfo {year} {2020})},\ \Eprint
  {http://arxiv.org/abs/1910.01649} {arXiv:1910.01649 [hep-ph]} \BibitemShut
  {NoStop}%
\bibitem [{\citenamefont {{Verde}}\ \emph {et~al.}(2013)\citenamefont
  {{Verde}}, \citenamefont {{Feeney}}, \citenamefont {{Mortlock}},\ and\
  \citenamefont {{Peiris}}}]{2013JCAP...09..013V}%
  \BibitemOpen
  \bibfield  {author} {\bibinfo {author} {\bibfnamefont {L.}~\bibnamefont
  {{Verde}}}, \bibinfo {author} {\bibfnamefont {S.~M.}\ \bibnamefont
  {{Feeney}}}, \bibinfo {author} {\bibfnamefont {D.~J.}\ \bibnamefont
  {{Mortlock}}}, \ and\ \bibinfo {author} {\bibfnamefont {H.~V.}\ \bibnamefont
  {{Peiris}}},\ }\href {\doibase 10.1088/1475-7516/2013/09/013} {\bibfield
  {journal} {\bibinfo  {journal} {Journal of Cosmology and Astroparticle
  Physics}\ }\textbf {\bibinfo {volume} {2013}},\ \bibinfo {eid} {013}
  (\bibinfo {year} {2013})},\ \Eprint {http://arxiv.org/abs/1307.2904}
  {arXiv:1307.2904 [astro-ph.CO]} \BibitemShut {NoStop}%
\bibitem [{\citenamefont {{Buoninfante}}(2016)}]{2016arXiv161008744B}%
  \BibitemOpen
  \bibfield  {author} {\bibinfo {author} {\bibfnamefont {L.}~\bibnamefont
  {{Buoninfante}}},\ }\href@noop {} {\bibfield  {journal} {\bibinfo  {journal}
  {arXiv e-prints}\ ,\ \bibinfo {eid} {arXiv:1610.08744}} (\bibinfo {year}
  {2016})},\ \Eprint {http://arxiv.org/abs/1610.08744} {arXiv:1610.08744
  [gr-qc]} \BibitemShut {NoStop}%
\bibitem [{\citenamefont {{Ni}}(2010)}]{2010RPPh...73e6901N}%
  \BibitemOpen
  \bibfield  {author} {\bibinfo {author} {\bibfnamefont {W.-T.}\ \bibnamefont
  {{Ni}}},\ }\href {\doibase 10.1088/0034-4885/73/5/056901} {\bibfield
  {journal} {\bibinfo  {journal} {Reports on Progress in Physics}\ }\textbf
  {\bibinfo {volume} {73}},\ \bibinfo {eid} {056901} (\bibinfo {year}
  {2010})},\ \Eprint {http://arxiv.org/abs/0912.5057} {arXiv:0912.5057 [gr-qc]}
  \BibitemShut {NoStop}%
\bibitem [{\citenamefont {{Lin}}\ \emph
  {et~al.}(2019{\natexlab{a}})\citenamefont {{Lin}}, \citenamefont {{Hobson}},\
  and\ \citenamefont {{Lasenby}}}]{2019PhRvD..99f4001L}%
  \BibitemOpen
  \bibfield  {author} {\bibinfo {author} {\bibfnamefont {Y.-C.}\ \bibnamefont
  {{Lin}}}, \bibinfo {author} {\bibfnamefont {M.~P.}\ \bibnamefont {{Hobson}}},
  \ and\ \bibinfo {author} {\bibfnamefont {A.~N.}\ \bibnamefont {{Lasenby}}},\
  }\href {\doibase 10.1103/PhysRevD.99.064001} {\bibfield  {journal} {\bibinfo
  {journal} {Phys. Rev. D}\ }\textbf {\bibinfo {volume} {99}},\ \bibinfo {eid}
  {064001} (\bibinfo {year} {2019}{\natexlab{a}})},\ \Eprint
  {http://arxiv.org/abs/1812.02675} {arXiv:1812.02675 [gr-qc]} \BibitemShut
  {NoStop}%
\bibitem [{\citenamefont {{Lin}}\ \emph
  {et~al.}(2019{\natexlab{b}})\citenamefont {{Lin}}, \citenamefont {{Hobson}},\
  and\ \citenamefont {{Lasenby}}}]{Lin2}%
  \BibitemOpen
  \bibfield  {author} {\bibinfo {author} {\bibfnamefont {Y.-C.}\ \bibnamefont
  {{Lin}}}, \bibinfo {author} {\bibfnamefont {M.~P.}\ \bibnamefont {{Hobson}}},
  \ and\ \bibinfo {author} {\bibfnamefont {A.~N.}\ \bibnamefont {{Lasenby}}},\
  }\href@noop {} {\bibfield  {journal} {\bibinfo  {journal} {arXiv e-prints}\
  ,\ \bibinfo {eid} {arXiv:1910.14197}} (\bibinfo {year}
  {2019}{\natexlab{b}})},\ \Eprint {http://arxiv.org/abs/1910.14197}
  {arXiv:1910.14197 [gr-qc]} \BibitemShut {NoStop}%
\bibitem [{\citenamefont {{Yo}}\ \emph {et~al.}(2002)\citenamefont {{Yo}},
  \citenamefont {{Nester}},\ and\ \citenamefont {{Ni}}}]{2002IJMPD..11..747Y}%
  \BibitemOpen
  \bibfield  {author} {\bibinfo {author} {\bibfnamefont {H.-J.}\ \bibnamefont
  {{Yo}}}, \bibinfo {author} {\bibfnamefont {J.~M.}\ \bibnamefont {{Nester}}},
  \ and\ \bibinfo {author} {\bibfnamefont {W.~T.}\ \bibnamefont {{Ni}}},\
  }\href {\doibase 10.1142/S0218271802001998} {\bibfield  {journal} {\bibinfo
  {journal} {International Journal of Modern Physics D}\ }\textbf {\bibinfo
  {volume} {11}},\ \bibinfo {pages} {747} (\bibinfo {year} {2002})},\ \Eprint
  {http://arxiv.org/abs/gr-qc/0112030} {arXiv:gr-qc/0112030 [gr-qc]}
  \BibitemShut {NoStop}%
\bibitem [{\citenamefont {{Yo}}\ and\ \citenamefont
  {{Nester}}(1999)}]{1999IJMPD...8..459Y}%
  \BibitemOpen
  \bibfield  {author} {\bibinfo {author} {\bibfnamefont {H.-J.}\ \bibnamefont
  {{Yo}}}\ and\ \bibinfo {author} {\bibfnamefont {J.~M.}\ \bibnamefont
  {{Nester}}},\ }\href {\doibase 10.1142/S021827189900033X} {\bibfield
  {journal} {\bibinfo  {journal} {International Journal of Modern Physics D}\
  }\textbf {\bibinfo {volume} {8}},\ \bibinfo {pages} {459} (\bibinfo {year}
  {1999})},\ \Eprint {http://arxiv.org/abs/gr-qc/9902032} {arXiv:gr-qc/9902032
  [gr-qc]} \BibitemShut {NoStop}%
\bibitem [{\citenamefont {Maplesoft}(2018)}]{maple}%
  \BibitemOpen
  \bibfield  {author} {\bibinfo {author} {\bibfnamefont {W.~O.}\ \bibnamefont
  {Maplesoft}, \bibfnamefont {a~division of Waterloo Maple~Inc.}},\ }\href@noop
  {} {\enquote {\bibinfo {title} {{Maple 2018}},}\ } (\bibinfo {year}
  {2018})\BibitemShut {NoStop}%
\bibitem [{\citenamefont {{Wald}}(1984)}]{1984ucp..book.....W}%
  \BibitemOpen
  \bibfield  {author} {\bibinfo {author} {\bibfnamefont {R.~M.}\ \bibnamefont
  {{Wald}}},\ }\href@noop {} {\emph {\bibinfo {title} {{General relativity}}}}\
  (\bibinfo {year} {1984})\BibitemShut {NoStop}%
\bibitem [{\citenamefont {{Kibble}}(1961)}]{1961JMP.....2..212K}%
  \BibitemOpen
  \bibfield  {author} {\bibinfo {author} {\bibfnamefont {T.~W.~B.}\
  \bibnamefont {{Kibble}}},\ }\href {\doibase 10.1063/1.1703702} {\bibfield
  {journal} {\bibinfo  {journal} {Journal of Mathematical Physics}\ }\textbf
  {\bibinfo {volume} {2}},\ \bibinfo {pages} {212} (\bibinfo {year}
  {1961})}\BibitemShut {NoStop}%
\bibitem [{\citenamefont {Utiyama}(1956)}]{PhysRev.101.1597}%
  \BibitemOpen
  \bibfield  {author} {\bibinfo {author} {\bibfnamefont {R.}~\bibnamefont
  {Utiyama}},\ }\href {\doibase 10.1103/PhysRev.101.1597} {\bibfield  {journal}
  {\bibinfo  {journal} {Phys. Rev.}\ }\textbf {\bibinfo {volume} {101}},\
  \bibinfo {pages} {1597} (\bibinfo {year} {1956})}\BibitemShut {NoStop}%
\bibitem [{\citenamefont {SCIAMA}(1964)}]{RevModPhys.36.463}%
  \BibitemOpen
  \bibfield  {author} {\bibinfo {author} {\bibfnamefont {D.~W.}\ \bibnamefont
  {SCIAMA}},\ }\href {\doibase 10.1103/RevModPhys.36.463} {\bibfield  {journal}
  {\bibinfo  {journal} {Rev. Mod. Phys.}\ }\textbf {\bibinfo {volume} {36}},\
  \bibinfo {pages} {463} (\bibinfo {year} {1964})}\BibitemShut {NoStop}%
\bibitem [{\citenamefont {Blagojevi{\'c}}(2002)}]{blagojevic2002gravitation}%
  \BibitemOpen
  \bibfield  {author} {\bibinfo {author} {\bibfnamefont {M.}~\bibnamefont
  {Blagojevi{\'c}}},\ }\href@noop {} {\emph {\bibinfo {title} {Gravitation and
  Gauge Symmetries}}}\ (\bibinfo  {publisher} {Institute of Physics Pub.},\
  \bibinfo {year} {2002})\BibitemShut {NoStop}%
\bibitem [{\citenamefont {{Kasuya}}(1975)}]{1975NCimB..28..127K}%
  \BibitemOpen
  \bibfield  {author} {\bibinfo {author} {\bibfnamefont {M.}~\bibnamefont
  {{Kasuya}}},\ }\href {\doibase 10.1007/BF02722810} {\bibfield  {journal}
  {\bibinfo  {journal} {Nuovo Cimento B Serie}\ }\textbf {\bibinfo {volume}
  {28}},\ \bibinfo {pages} {127} (\bibinfo {year} {1975})}\BibitemShut
  {NoStop}%
\bibitem [{\citenamefont {{Lasenby}}\ and\ \citenamefont
  {{Hobson}}(2016)}]{lasenby-hobson-2016}%
  \BibitemOpen
  \bibfield  {author} {\bibinfo {author} {\bibfnamefont {A.~N.}\ \bibnamefont
  {{Lasenby}}}\ and\ \bibinfo {author} {\bibfnamefont {M.~P.}\ \bibnamefont
  {{Hobson}}},\ }\href {\doibase 10.1063/1.4963143} {\bibfield  {journal}
  {\bibinfo  {journal} {Journal of Mathematical Physics}\ }\textbf {\bibinfo
  {volume} {57}},\ \bibinfo {eid} {092505} (\bibinfo {year} {2016})},\ \Eprint
  {http://arxiv.org/abs/1510.06699} {arXiv:1510.06699 [gr-qc]} \BibitemShut
  {NoStop}%
\bibitem [{\citenamefont {{Aoki}}\ and\ \citenamefont
  {{Mukohyama}}(2020)}]{2020arXiv200300664A}%
  \BibitemOpen
  \bibfield  {author} {\bibinfo {author} {\bibfnamefont {K.}~\bibnamefont
  {{Aoki}}}\ and\ \bibinfo {author} {\bibfnamefont {S.}~\bibnamefont
  {{Mukohyama}}},\ }\href@noop {} {\bibfield  {journal} {\bibinfo  {journal}
  {arXiv e-prints}\ ,\ \bibinfo {eid} {arXiv:2003.00664}} (\bibinfo {year}
  {2020})},\ \Eprint {http://arxiv.org/abs/2003.00664} {arXiv:2003.00664
  [hep-th]} \BibitemShut {NoStop}%
\bibitem [{\citenamefont {{Baekler}}\ \emph {et~al.}(2011)\citenamefont
  {{Baekler}}, \citenamefont {{Hehl}},\ and\ \citenamefont
  {{Nester}}}]{2011PhRvD..83b4001B}%
  \BibitemOpen
  \bibfield  {author} {\bibinfo {author} {\bibfnamefont {P.}~\bibnamefont
  {{Baekler}}}, \bibinfo {author} {\bibfnamefont {F.~W.}\ \bibnamefont
  {{Hehl}}}, \ and\ \bibinfo {author} {\bibfnamefont {J.~M.}\ \bibnamefont
  {{Nester}}},\ }\href {\doibase 10.1103/PhysRevD.83.024001} {\bibfield
  {journal} {\bibinfo  {journal} {Phys. Rev. D}\ }\textbf {\bibinfo {volume}
  {83}},\ \bibinfo {eid} {024001} (\bibinfo {year} {2011})},\ \Eprint
  {http://arxiv.org/abs/1009.5112} {arXiv:1009.5112 [gr-qc]} \BibitemShut
  {NoStop}%
\bibitem [{\citenamefont {{Kopczy{\'n}ski}}(1972)}]{1972PhLA...39..219K}%
  \BibitemOpen
  \bibfield  {author} {\bibinfo {author} {\bibfnamefont {W.}~\bibnamefont
  {{Kopczy{\'n}ski}}},\ }\href {\doibase 10.1016/0375-9601(72)90714-1}
  {\bibfield  {journal} {\bibinfo  {journal} {Physics Letters A}\ }\textbf
  {\bibinfo {volume} {39}},\ \bibinfo {pages} {219} (\bibinfo {year}
  {1972})}\BibitemShut {NoStop}%
\bibitem [{\citenamefont {{Tsamparlis}}(1979)}]{1979PhLA...75...27T}%
  \BibitemOpen
  \bibfield  {author} {\bibinfo {author} {\bibfnamefont {M.}~\bibnamefont
  {{Tsamparlis}}},\ }\href {\doibase 10.1016/0375-9601(79)90265-2} {\bibfield
  {journal} {\bibinfo  {journal} {Physics Letters A}\ }\textbf {\bibinfo
  {volume} {75}},\ \bibinfo {pages} {27} (\bibinfo {year} {1979})}\BibitemShut
  {NoStop}%
\bibitem [{\citenamefont {{Minkevich}}(1980)}]{1980PhLA...80..232M}%
  \BibitemOpen
  \bibfield  {author} {\bibinfo {author} {\bibfnamefont {A.~V.}\ \bibnamefont
  {{Minkevich}}},\ }\href {\doibase 10.1016/0375-9601(80)90008-0} {\bibfield
  {journal} {\bibinfo  {journal} {Physics Letters A}\ }\textbf {\bibinfo
  {volume} {80}},\ \bibinfo {pages} {232} (\bibinfo {year} {1980})}\BibitemShut
  {NoStop}%
\bibitem [{\citenamefont {{Minkevich}}\ \emph {et~al.}(2003)\citenamefont
  {{Minkevich}}, \citenamefont {{Garkun}},\ and\ \citenamefont
  {{Vasilevski}}}]{2003gr.qc....10060M}%
  \BibitemOpen
  \bibfield  {author} {\bibinfo {author} {\bibfnamefont {A.~V.}\ \bibnamefont
  {{Minkevich}}}, \bibinfo {author} {\bibfnamefont {A.~S.}\ \bibnamefont
  {{Garkun}}}, \ and\ \bibinfo {author} {\bibfnamefont {Y.~G.}\ \bibnamefont
  {{Vasilevski}}},\ }\href@noop {} {\bibfield  {journal} {\bibinfo  {journal}
  {arXiv e-prints}\ ,\ \bibinfo {eid} {gr-qc/0310060}} (\bibinfo {year}
  {2003})},\ \Eprint {http://arxiv.org/abs/gr-qc/0310060} {arXiv:gr-qc/0310060
  [gr-qc]} \BibitemShut {NoStop}%
\bibitem [{\citenamefont {{Minkevich}}(2007)}]{2007AcPPB..38...61M}%
  \BibitemOpen
  \bibfield  {author} {\bibinfo {author} {\bibfnamefont {A.~V.}\ \bibnamefont
  {{Minkevich}}},\ }\href@noop {} {\bibfield  {journal} {\bibinfo  {journal}
  {Acta Physica Polonica B}\ }\textbf {\bibinfo {volume} {38}},\ \bibinfo
  {pages} {61} (\bibinfo {year} {2007})},\ \Eprint
  {http://arxiv.org/abs/gr-qc/0512123} {arXiv:gr-qc/0512123 [gr-qc]}
  \BibitemShut {NoStop}%
\bibitem [{\citenamefont {{Minkevich}}\ and\ \citenamefont
  {{Garkun}}(2006)}]{2006CQGra..23.4237M}%
  \BibitemOpen
  \bibfield  {author} {\bibinfo {author} {\bibfnamefont {A.~V.}\ \bibnamefont
  {{Minkevich}}}\ and\ \bibinfo {author} {\bibfnamefont {A.~S.}\ \bibnamefont
  {{Garkun}}},\ }\href {\doibase 10.1088/0264-9381/23/12/018} {\bibfield
  {journal} {\bibinfo  {journal} {Classical and Quantum Gravity}\ }\textbf
  {\bibinfo {volume} {23}},\ \bibinfo {pages} {4237} (\bibinfo {year}
  {2006})},\ \Eprint {http://arxiv.org/abs/gr-qc/0512130} {arXiv:gr-qc/0512130
  [gr-qc]} \BibitemShut {NoStop}%
\bibitem [{\citenamefont {{Minkevich}}(2009)}]{2009PhLB..678..423M}%
  \BibitemOpen
  \bibfield  {author} {\bibinfo {author} {\bibfnamefont {A.~V.}\ \bibnamefont
  {{Minkevich}}},\ }\href {\doibase 10.1016/j.physletb.2009.06.050} {\bibfield
  {journal} {\bibinfo  {journal} {Physics Letters B}\ }\textbf {\bibinfo
  {volume} {678}},\ \bibinfo {pages} {423} (\bibinfo {year} {2009})},\ \Eprint
  {http://arxiv.org/abs/0902.2860} {arXiv:0902.2860 [gr-qc]} \BibitemShut
  {NoStop}%
\bibitem [{\citenamefont {{Garkun}}\ \emph {et~al.}(2011)\citenamefont
  {{Garkun}}, \citenamefont {{Kudin}}, \citenamefont {{Minkevich}},\ and\
  \citenamefont {{Vasilevsky}}}]{2011arXiv1107.1566G}%
  \BibitemOpen
  \bibfield  {author} {\bibinfo {author} {\bibfnamefont {A.~S.}\ \bibnamefont
  {{Garkun}}}, \bibinfo {author} {\bibfnamefont {V.~I.}\ \bibnamefont
  {{Kudin}}}, \bibinfo {author} {\bibfnamefont {A.~V.}\ \bibnamefont
  {{Minkevich}}}, \ and\ \bibinfo {author} {\bibfnamefont {Y.~G.}\ \bibnamefont
  {{Vasilevsky}}},\ }\href@noop {} {\bibfield  {journal} {\bibinfo  {journal}
  {arXiv e-prints}\ ,\ \bibinfo {eid} {arXiv:1107.1566}} (\bibinfo {year}
  {2011})},\ \Eprint {http://arxiv.org/abs/1107.1566} {arXiv:1107.1566 [gr-qc]}
  \BibitemShut {NoStop}%
\bibitem [{\citenamefont {{Minkevich}}\ \emph {et~al.}(2013)\citenamefont
  {{Minkevich}}, \citenamefont {{Garkun}},\ and\ \citenamefont
  {{Kudin}}}]{2013JCAP...03..040M}%
  \BibitemOpen
  \bibfield  {author} {\bibinfo {author} {\bibfnamefont {A.~V.}\ \bibnamefont
  {{Minkevich}}}, \bibinfo {author} {\bibfnamefont {A.~S.}\ \bibnamefont
  {{Garkun}}}, \ and\ \bibinfo {author} {\bibfnamefont {V.~I.}\ \bibnamefont
  {{Kudin}}},\ }\href {\doibase 10.1088/1475-7516/2013/03/040} {\bibfield
  {journal} {\bibinfo  {journal} {Journal of Cosmology and Astro-Particle
  Physics}\ }\textbf {\bibinfo {volume} {2013}},\ \bibinfo {eid} {040}
  (\bibinfo {year} {2013})},\ \Eprint {http://arxiv.org/abs/1302.2578}
  {arXiv:1302.2578 [gr-qc]} \BibitemShut {NoStop}%
\bibitem [{\citenamefont {{Minkevich}}\ and\ \citenamefont
  {{Garkun}}(2000)}]{2000CQGra..17.3045M}%
  \BibitemOpen
  \bibfield  {author} {\bibinfo {author} {\bibfnamefont {A.~V.}\ \bibnamefont
  {{Minkevich}}}\ and\ \bibinfo {author} {\bibfnamefont {A.~S.}\ \bibnamefont
  {{Garkun}}},\ }\href {\doibase 10.1088/0264-9381/17/15/312} {\bibfield
  {journal} {\bibinfo  {journal} {Classical and Quantum Gravity}\ }\textbf
  {\bibinfo {volume} {17}},\ \bibinfo {pages} {3045} (\bibinfo {year}
  {2000})}\BibitemShut {NoStop}%
\bibitem [{\citenamefont {{Goenner}}\ and\ \citenamefont
  {{Mueller-Hoissen}}(1984)}]{1984CQGra...1..651G}%
  \BibitemOpen
  \bibfield  {author} {\bibinfo {author} {\bibfnamefont {H.}~\bibnamefont
  {{Goenner}}}\ and\ \bibinfo {author} {\bibfnamefont {F.}~\bibnamefont
  {{Mueller-Hoissen}}},\ }\href {\doibase 10.1088/0264-9381/1/6/010} {\bibfield
   {journal} {\bibinfo  {journal} {Classical and Quantum Gravity}\ }\textbf
  {\bibinfo {volume} {1}},\ \bibinfo {pages} {651} (\bibinfo {year}
  {1984})}\BibitemShut {NoStop}%
\bibitem [{\citenamefont {{Puetzfeld}}(2005)}]{2005NewAR..49...59P}%
  \BibitemOpen
  \bibfield  {author} {\bibinfo {author} {\bibfnamefont {D.}~\bibnamefont
  {{Puetzfeld}}},\ }\href {\doibase 10.1016/j.newar.2005.01.022} {\bibfield
  {journal} {\bibinfo  {journal} {nar}\ }\textbf {\bibinfo {volume} {49}},\
  \bibinfo {pages} {59} (\bibinfo {year} {2005})},\ \Eprint
  {http://arxiv.org/abs/gr-qc/0404119} {arXiv:gr-qc/0404119 [gr-qc]}
  \BibitemShut {NoStop}%
\bibitem [{\citenamefont {{Lasenby}}\ \emph {et~al.}(2005)\citenamefont
  {{Lasenby}}, \citenamefont {{Doran}},\ and\ \citenamefont
  {{Heineke}}}]{lasenby-doran-heineke-2005}%
  \BibitemOpen
  \bibfield  {author} {\bibinfo {author} {\bibfnamefont {A.~N.}\ \bibnamefont
  {{Lasenby}}}, \bibinfo {author} {\bibfnamefont {C.~J.~L.}\ \bibnamefont
  {{Doran}}}, \ and\ \bibinfo {author} {\bibfnamefont {R.}~\bibnamefont
  {{Heineke}}},\ }\href@noop {} {\bibfield  {journal} {\bibinfo  {journal}
  {arXiv e-prints}\ ,\ \bibinfo {eid} {gr-qc/0509014}} (\bibinfo {year}
  {2005})},\ \Eprint {http://arxiv.org/abs/gr-qc/0509014} {arXiv:gr-qc/0509014
  [gr-qc]} \BibitemShut {NoStop}%
\bibitem [{\citenamefont {{Shie}}\ \emph {et~al.}(2008)\citenamefont {{Shie}},
  \citenamefont {{Nester}},\ and\ \citenamefont {{Yo}}}]{2008PhRvD..78b3522S}%
  \BibitemOpen
  \bibfield  {author} {\bibinfo {author} {\bibfnamefont {K.-F.}\ \bibnamefont
  {{Shie}}}, \bibinfo {author} {\bibfnamefont {J.~M.}\ \bibnamefont
  {{Nester}}}, \ and\ \bibinfo {author} {\bibfnamefont {H.-J.}\ \bibnamefont
  {{Yo}}},\ }\href {\doibase 10.1103/PhysRevD.78.023522} {\bibfield  {journal}
  {\bibinfo  {journal} {Phys. Rev. D}\ }\textbf {\bibinfo {volume} {78}},\
  \bibinfo {eid} {023522} (\bibinfo {year} {2008})},\ \Eprint
  {http://arxiv.org/abs/0805.3834} {arXiv:0805.3834 [gr-qc]} \BibitemShut
  {NoStop}%
\bibitem [{\citenamefont {{Li}}\ \emph {et~al.}(2009)\citenamefont {{Li}},
  \citenamefont {{Sun}},\ and\ \citenamefont {{Xi}}}]{2009PhRvD..79b7301L}%
  \BibitemOpen
  \bibfield  {author} {\bibinfo {author} {\bibfnamefont {X.-Z.}\ \bibnamefont
  {{Li}}}, \bibinfo {author} {\bibfnamefont {C.-B.}\ \bibnamefont {{Sun}}}, \
  and\ \bibinfo {author} {\bibfnamefont {P.}~\bibnamefont {{Xi}}},\ }\href
  {\doibase 10.1103/PhysRevD.79.027301} {\bibfield  {journal} {\bibinfo
  {journal} {Phys. Rev. D}\ }\textbf {\bibinfo {volume} {79}},\ \bibinfo {eid}
  {027301} (\bibinfo {year} {2009})},\ \Eprint {http://arxiv.org/abs/0903.3088}
  {arXiv:0903.3088 [gr-qc]} \BibitemShut {NoStop}%
\bibitem [{\citenamefont {{Chen}}\ \emph {et~al.}(2009)\citenamefont {{Chen}},
  \citenamefont {{Ho}}, \citenamefont {{Nester}}, \citenamefont {{Wang}},\ and\
  \citenamefont {{Yo}}}]{2009JCAP...10..027C}%
  \BibitemOpen
  \bibfield  {author} {\bibinfo {author} {\bibfnamefont {H.}~\bibnamefont
  {{Chen}}}, \bibinfo {author} {\bibfnamefont {F.-H.}\ \bibnamefont {{Ho}}},
  \bibinfo {author} {\bibfnamefont {J.~M.}\ \bibnamefont {{Nester}}}, \bibinfo
  {author} {\bibfnamefont {C.-H.}\ \bibnamefont {{Wang}}}, \ and\ \bibinfo
  {author} {\bibfnamefont {H.-J.}\ \bibnamefont {{Yo}}},\ }\href {\doibase
  10.1088/1475-7516/2009/10/027} {\bibfield  {journal} {\bibinfo  {journal}
  {jcap}\ }\textbf {\bibinfo {volume} {2009}},\ \bibinfo {eid} {027} (\bibinfo
  {year} {2009})},\ \Eprint {http://arxiv.org/abs/0908.3323} {arXiv:0908.3323
  [gr-qc]} \BibitemShut {NoStop}%
\bibitem [{\citenamefont {{Ho}}\ and\ \citenamefont
  {{Nester}}(2011{\natexlab{a}})}]{2011JPhCS.330a2005H}%
  \BibitemOpen
  \bibfield  {author} {\bibinfo {author} {\bibfnamefont {F.-H.}\ \bibnamefont
  {{Ho}}}\ and\ \bibinfo {author} {\bibfnamefont {J.~M.}\ \bibnamefont
  {{Nester}}},\ }in\ \href {\doibase 10.1088/1742-6596/330/1/012005} {\emph
  {\bibinfo {booktitle} {Journal of Physics Conference Series}}},\ \bibinfo
  {series} {Journal of Physics Conference Series}, Vol.\ \bibinfo {volume}
  {330}\ (\bibinfo {year} {2011})\ p.\ \bibinfo {pages} {012005},\ \Eprint
  {http://arxiv.org/abs/1105.5001} {arXiv:1105.5001 [gr-qc]} \BibitemShut
  {NoStop}%
\bibitem [{\citenamefont {{Ho}}\ and\ \citenamefont
  {{Nester}}(2011{\natexlab{b}})}]{2011IJMPD..20.2125H}%
  \BibitemOpen
  \bibfield  {author} {\bibinfo {author} {\bibfnamefont {F.-H.}\ \bibnamefont
  {{Ho}}}\ and\ \bibinfo {author} {\bibfnamefont {J.~M.}\ \bibnamefont
  {{Nester}}},\ }\href {\doibase 10.1142/S0218271811020391} {\bibfield
  {journal} {\bibinfo  {journal} {International Journal of Modern Physics D}\
  }\textbf {\bibinfo {volume} {20}},\ \bibinfo {pages} {2125} (\bibinfo {year}
  {2011}{\natexlab{b}})}\BibitemShut {NoStop}%
\bibitem [{\citenamefont {{Baekler}}\ and\ \citenamefont
  {{Hehl}}(2011)}]{2011CQGra..28u5017B}%
  \BibitemOpen
  \bibfield  {author} {\bibinfo {author} {\bibfnamefont {P.}~\bibnamefont
  {{Baekler}}}\ and\ \bibinfo {author} {\bibfnamefont {F.~W.}\ \bibnamefont
  {{Hehl}}},\ }\href {\doibase 10.1088/0264-9381/28/21/215017} {\bibfield
  {journal} {\bibinfo  {journal} {Classical and Quantum Gravity}\ }\textbf
  {\bibinfo {volume} {28}},\ \bibinfo {eid} {215017} (\bibinfo {year}
  {2011})},\ \Eprint {http://arxiv.org/abs/1105.3504} {arXiv:1105.3504 [gr-qc]}
  \BibitemShut {NoStop}%
\bibitem [{\citenamefont {{Ho}}\ \emph {et~al.}(2015)\citenamefont {{Ho}},
  \citenamefont {{Chen}}, \citenamefont {{Nester}},\ and\ \citenamefont
  {{Yo}}}]{2015arXiv151201202H}%
  \BibitemOpen
  \bibfield  {author} {\bibinfo {author} {\bibfnamefont {F.-H.}\ \bibnamefont
  {{Ho}}}, \bibinfo {author} {\bibfnamefont {H.}~\bibnamefont {{Chen}}},
  \bibinfo {author} {\bibfnamefont {J.~M.}\ \bibnamefont {{Nester}}}, \ and\
  \bibinfo {author} {\bibfnamefont {H.-J.}\ \bibnamefont {{Yo}}},\ }\href@noop
  {} {\bibfield  {journal} {\bibinfo  {journal} {arXiv e-prints}\ ,\ \bibinfo
  {eid} {arXiv:1512.01202}} (\bibinfo {year} {2015})},\ \Eprint
  {http://arxiv.org/abs/1512.01202} {arXiv:1512.01202 [gr-qc]} \BibitemShut
  {NoStop}%
\bibitem [{\citenamefont {{Zhang}}\ and\ \citenamefont
  {{Xu}}(2019{\natexlab{a}})}]{2019arXiv190403545Z}%
  \BibitemOpen
  \bibfield  {author} {\bibinfo {author} {\bibfnamefont {H.}~\bibnamefont
  {{Zhang}}}\ and\ \bibinfo {author} {\bibfnamefont {L.}~\bibnamefont {{Xu}}},\
  }\href {\doibase 10.1088/1475-7516/2019/09/050} {\bibfield  {journal}
  {\bibinfo  {journal} {Journal of Cosmology and Astroparticle Physics}\
  }\textbf {\bibinfo {volume} {2019}},\ \bibinfo {eid} {050} (\bibinfo {year}
  {2019}{\natexlab{a}})},\ \Eprint {http://arxiv.org/abs/1904.03545}
  {arXiv:1904.03545 [gr-qc]} \BibitemShut {NoStop}%
\bibitem [{\citenamefont {{Zhang}}\ and\ \citenamefont
  {{Xu}}(2019{\natexlab{b}})}]{2019arXiv190604340Z}%
  \BibitemOpen
  \bibfield  {author} {\bibinfo {author} {\bibfnamefont {H.}~\bibnamefont
  {{Zhang}}}\ and\ \bibinfo {author} {\bibfnamefont {L.}~\bibnamefont {{Xu}}},\
  }\href@noop {} {\bibfield  {journal} {\bibinfo  {journal} {arXiv e-prints}\
  ,\ \bibinfo {eid} {arXiv:1906.04340}} (\bibinfo {year}
  {2019}{\natexlab{b}})},\ \Eprint {http://arxiv.org/abs/1906.04340}
  {arXiv:1906.04340 [gr-qc]} \BibitemShut {NoStop}%
\bibitem [{\citenamefont {{Pop{\l}awski}}(2018)}]{2018IJMPD..2747020P}%
  \BibitemOpen
  \bibfield  {author} {\bibinfo {author} {\bibfnamefont {N.}~\bibnamefont
  {{Pop{\l}awski}}},\ }\href {\doibase 10.1142/S021827181847020X} {\bibfield
  {journal} {\bibinfo  {journal} {International Journal of Modern Physics D}\
  }\textbf {\bibinfo {volume} {27}},\ \bibinfo {eid} {1847020} (\bibinfo {year}
  {2018})},\ \Eprint {http://arxiv.org/abs/1801.08076} {arXiv:1801.08076
  [physics.pop-ph]} \BibitemShut {NoStop}%
\bibitem [{\citenamefont {{Marques}}\ and\ \citenamefont
  {{Martins}}(2019)}]{2019arXiv191108232M}%
  \BibitemOpen
  \bibfield  {author} {\bibinfo {author} {\bibfnamefont {C.~M.~J.}\
  \bibnamefont {{Marques}}}\ and\ \bibinfo {author} {\bibfnamefont
  {C.~J.~A.~P.}\ \bibnamefont {{Martins}}},\ }\href@noop {} {\bibfield
  {journal} {\bibinfo  {journal} {arXiv e-prints}\ ,\ \bibinfo {eid}
  {arXiv:1911.08232}} (\bibinfo {year} {2019})},\ \Eprint
  {http://arxiv.org/abs/1911.08232} {arXiv:1911.08232 [astro-ph.CO]}
  \BibitemShut {NoStop}%
\bibitem [{\citenamefont {{Lasenby}}\ \emph {et~al.}(1998)\citenamefont
  {{Lasenby}}, \citenamefont {{Doran}},\ and\ \citenamefont
  {{Gull}}}]{1998RSPTA.356..487L}%
  \BibitemOpen
  \bibfield  {author} {\bibinfo {author} {\bibfnamefont {A.}~\bibnamefont
  {{Lasenby}}}, \bibinfo {author} {\bibfnamefont {C.}~\bibnamefont {{Doran}}},
  \ and\ \bibinfo {author} {\bibfnamefont {S.}~\bibnamefont {{Gull}}},\ }\href
  {\doibase 10.1098/rsta.1998.0178} {\bibfield  {journal} {\bibinfo  {journal}
  {Philosophical Transactions of the Royal Society of London Series A}\
  }\textbf {\bibinfo {volume} {356}},\ \bibinfo {pages} {487} (\bibinfo {year}
  {1998})},\ \Eprint {http://arxiv.org/abs/gr-qc/0405033} {arXiv:gr-qc/0405033
  [gr-qc]} \BibitemShut {NoStop}%
\bibitem [{\citenamefont {{Remmen}}\ and\ \citenamefont
  {{Carroll}}(2013)}]{2013PhRvD..88h3518R}%
  \BibitemOpen
  \bibfield  {author} {\bibinfo {author} {\bibfnamefont {G.~N.}\ \bibnamefont
  {{Remmen}}}\ and\ \bibinfo {author} {\bibfnamefont {S.~M.}\ \bibnamefont
  {{Carroll}}},\ }\href {\doibase 10.1103/PhysRevD.88.083518} {\bibfield
  {journal} {\bibinfo  {journal} {Phys. Rev. D}\ }\textbf {\bibinfo {volume}
  {88}},\ \bibinfo {eid} {083518} (\bibinfo {year} {2013})},\ \Eprint
  {http://arxiv.org/abs/1309.2611} {arXiv:1309.2611 [gr-qc]} \BibitemShut
  {NoStop}%
\bibitem [{\citenamefont {Ashtekar}\ and\ \citenamefont
  {Samuel}(1991)}]{Ashtekar_1991}%
  \BibitemOpen
  \bibfield  {author} {\bibinfo {author} {\bibfnamefont {A.}~\bibnamefont
  {Ashtekar}}\ and\ \bibinfo {author} {\bibfnamefont {J.}~\bibnamefont
  {Samuel}},\ }\href {\doibase 10.1088/0264-9381/8/12/005} {\bibfield
  {journal} {\bibinfo  {journal} {Classical and Quantum Gravity}\ }\textbf
  {\bibinfo {volume} {8}},\ \bibinfo {pages} {2191} (\bibinfo {year}
  {1991})}\BibitemShut {NoStop}%
\bibitem [{\citenamefont {{Faraoni}}(2009)}]{faroni-2009}%
  \BibitemOpen
  \bibfield  {author} {\bibinfo {author} {\bibfnamefont {V.}~\bibnamefont
  {{Faraoni}}},\ }\href {\doibase 10.1103/PhysRevD.80.124040} {\bibfield
  {journal} {\bibinfo  {journal} {phys-rev-d}\ }\textbf {\bibinfo {volume}
  {80}},\ \bibinfo {eid} {124040} (\bibinfo {year} {2009})},\ \Eprint
  {http://arxiv.org/abs/0912.1249} {arXiv:0912.1249 [astro-ph.GA]} \BibitemShut
  {NoStop}%
\bibitem [{\citenamefont {{Brown}}(1993)}]{brown-1993}%
  \BibitemOpen
  \bibfield  {author} {\bibinfo {author} {\bibfnamefont {J.~D.}\ \bibnamefont
  {{Brown}}},\ }\href {\doibase 10.1088/0264-9381/10/8/017} {\bibfield
  {journal} {\bibinfo  {journal} {Classical and Quantum Gravity}\ }\textbf
  {\bibinfo {volume} {10}},\ \bibinfo {pages} {1579} (\bibinfo {year}
  {1993})},\ \Eprint {http://arxiv.org/abs/gr-qc/9304026} {gr-qc/9304026}
  \BibitemShut {NoStop}%
\bibitem [{\citenamefont {{Boehmer}}\ and\ \citenamefont
  {{Bronowski}}(2006)}]{2006gr.qc.....1089B}%
  \BibitemOpen
  \bibfield  {author} {\bibinfo {author} {\bibfnamefont {C.~G.}\ \bibnamefont
  {{Boehmer}}}\ and\ \bibinfo {author} {\bibfnamefont {P.}~\bibnamefont
  {{Bronowski}}},\ }\href@noop {} {\bibfield  {journal} {\bibinfo  {journal}
  {arXiv e-prints}\ ,\ \bibinfo {eid} {gr-qc/0601089}} (\bibinfo {year}
  {2006})},\ \Eprint {http://arxiv.org/abs/gr-qc/0601089} {arXiv:gr-qc/0601089
  [gr-qc]} \BibitemShut {NoStop}%
\bibitem [{\citenamefont {{Brechet}}\ \emph {et~al.}(2008)\citenamefont
  {{Brechet}}, \citenamefont {{Hobson}},\ and\ \citenamefont
  {{Lasenby}}}]{2008CQGra..25x5016B}%
  \BibitemOpen
  \bibfield  {author} {\bibinfo {author} {\bibfnamefont {S.~D.}\ \bibnamefont
  {{Brechet}}}, \bibinfo {author} {\bibfnamefont {M.~P.}\ \bibnamefont
  {{Hobson}}}, \ and\ \bibinfo {author} {\bibfnamefont {A.~N.}\ \bibnamefont
  {{Lasenby}}},\ }\href {\doibase 10.1088/0264-9381/25/24/245016} {\bibfield
  {journal} {\bibinfo  {journal} {Classical and Quantum Gravity}\ }\textbf
  {\bibinfo {volume} {25}},\ \bibinfo {eid} {245016} (\bibinfo {year}
  {2008})},\ \Eprint {http://arxiv.org/abs/0807.2523} {arXiv:0807.2523 [gr-qc]}
  \BibitemShut {NoStop}%
\bibitem [{\citenamefont {{Dirac}}(1973)}]{1973RSPSA.333..403D}%
  \BibitemOpen
  \bibfield  {author} {\bibinfo {author} {\bibfnamefont {P.~A.~M.}\
  \bibnamefont {{Dirac}}},\ }\href {\doibase 10.1098/rspa.1973.0070} {\bibfield
   {journal} {\bibinfo  {journal} {Proceedings of the Royal Society of London
  Series A}\ }\textbf {\bibinfo {volume} {333}},\ \bibinfo {pages} {403}
  (\bibinfo {year} {1973})}\BibitemShut {NoStop}%
\bibitem [{\citenamefont {{Mannheim}}(1990)}]{1990GReGr..22..289M}%
  \BibitemOpen
  \bibfield  {author} {\bibinfo {author} {\bibfnamefont {P.~D.}\ \bibnamefont
  {{Mannheim}}},\ }\href {\doibase 10.1007/BF00756278} {\bibfield  {journal}
  {\bibinfo  {journal} {General Relativity and Gravitation}\ }\textbf {\bibinfo
  {volume} {22}},\ \bibinfo {pages} {289} (\bibinfo {year} {1990})}\BibitemShut
  {NoStop}%
\bibitem [{\citenamefont {{Mannheim}}(1992)}]{1992ApJ...391..429M}%
  \BibitemOpen
  \bibfield  {author} {\bibinfo {author} {\bibfnamefont {P.~D.}\ \bibnamefont
  {{Mannheim}}},\ }\href {\doibase 10.1086/171358} {\bibfield  {journal}
  {\bibinfo  {journal} {The Astrophysical Journal}\ }\textbf {\bibinfo {volume}
  {391}},\ \bibinfo {pages} {429} (\bibinfo {year} {1992})}\BibitemShut
  {NoStop}%
\bibitem [{\citenamefont {{Bertolami}}\ \emph {et~al.}(2008)\citenamefont
  {{Bertolami}}, \citenamefont {{Lobo}},\ and\ \citenamefont
  {{P{\'a}ramos}}}]{bertolami-lobo-paramos-2008}%
  \BibitemOpen
  \bibfield  {author} {\bibinfo {author} {\bibfnamefont {O.}~\bibnamefont
  {{Bertolami}}}, \bibinfo {author} {\bibfnamefont {F.~S.~N.}\ \bibnamefont
  {{Lobo}}}, \ and\ \bibinfo {author} {\bibfnamefont {J.}~\bibnamefont
  {{P{\'a}ramos}}},\ }\href {\doibase 10.1103/PhysRevD.78.064036} {\bibfield
  {journal} {\bibinfo  {journal} {phys-rev-d}\ }\textbf {\bibinfo {volume}
  {78}},\ \bibinfo {eid} {064036} (\bibinfo {year} {2008})},\ \Eprint
  {http://arxiv.org/abs/0806.4434} {arXiv:0806.4434 [gr-qc]} \BibitemShut
  {NoStop}%
\bibitem [{\citenamefont {{Pitrou}}\ \emph {et~al.}(2018)\citenamefont
  {{Pitrou}}, \citenamefont {{Coc}}, \citenamefont {{Uzan}},\ and\
  \citenamefont {{Vangioni}}}]{2018PhR...754....1P}%
  \BibitemOpen
  \bibfield  {author} {\bibinfo {author} {\bibfnamefont {C.}~\bibnamefont
  {{Pitrou}}}, \bibinfo {author} {\bibfnamefont {A.}~\bibnamefont {{Coc}}},
  \bibinfo {author} {\bibfnamefont {J.-P.}\ \bibnamefont {{Uzan}}}, \ and\
  \bibinfo {author} {\bibfnamefont {E.}~\bibnamefont {{Vangioni}}},\ }\href
  {\doibase 10.1016/j.physrep.2018.04.005} {\bibfield  {journal} {\bibinfo
  {journal} {Physics Reports}\ }\textbf {\bibinfo {volume} {754}},\ \bibinfo
  {pages} {1} (\bibinfo {year} {2018})},\ \Eprint
  {http://arxiv.org/abs/1801.08023} {arXiv:1801.08023 [astro-ph.CO]}
  \BibitemShut {NoStop}%
\bibitem [{\citenamefont {{Yokoyama}}\ and\ \citenamefont
  {{Maeda}}(1988)}]{1988PhLB..207...31Y}%
  \BibitemOpen
  \bibfield  {author} {\bibinfo {author} {\bibfnamefont {J.}~\bibnamefont
  {{Yokoyama}}}\ and\ \bibinfo {author} {\bibfnamefont {K.-I.}\ \bibnamefont
  {{Maeda}}},\ }\href {\doibase 10.1016/0370-2693(88)90880-5} {\bibfield
  {journal} {\bibinfo  {journal} {Physics Letters B}\ }\textbf {\bibinfo
  {volume} {207}},\ \bibinfo {pages} {31} (\bibinfo {year} {1988})}\BibitemShut
  {NoStop}%
\bibitem [{\citenamefont {{Hestenes}}\ \emph {et~al.}(1985)\citenamefont
  {{Hestenes}}, \citenamefont {{Sobczyk}},\ and\ \citenamefont
  {{Marsh}}}]{1985AmJPh..53..510H}%
  \BibitemOpen
  \bibfield  {author} {\bibinfo {author} {\bibfnamefont {D.}~\bibnamefont
  {{Hestenes}}}, \bibinfo {author} {\bibfnamefont {G.}~\bibnamefont
  {{Sobczyk}}}, \ and\ \bibinfo {author} {\bibfnamefont {J.~S.}\ \bibnamefont
  {{Marsh}}},\ }\href {\doibase 10.1119/1.14223} {\bibfield  {journal}
  {\bibinfo  {journal} {American Journal of Physics}\ }\textbf {\bibinfo
  {volume} {53}},\ \bibinfo {pages} {510} (\bibinfo {year} {1985})}\BibitemShut
  {NoStop}%
\bibitem [{\citenamefont {{Doran}}\ and\ \citenamefont
  {{Lasenby}}(2007)}]{doran-lasenby}%
  \BibitemOpen
  \bibfield  {author} {\bibinfo {author} {\bibfnamefont {C.}~\bibnamefont
  {{Doran}}}\ and\ \bibinfo {author} {\bibfnamefont {A.}~\bibnamefont
  {{Lasenby}}},\ }\href@noop {} {\emph {\bibinfo {title} {Geometric Algebra for
  Physicists, by Chris Doran , Anthony Lasenby, Cambridge, UK: Cambridge
  University Press, 2007}}}\ (\bibinfo {year} {2007})\BibitemShut {NoStop}%
\bibitem [{\citenamefont {{Lewis}}\ \emph {et~al.}(2000)\citenamefont
  {{Lewis}}, \citenamefont {{Doran}},\ and\ \citenamefont {{Lasenby}}}]{lewis}%
  \BibitemOpen
  \bibfield  {author} {\bibinfo {author} {\bibfnamefont {A.}~\bibnamefont
  {{Lewis}}}, \bibinfo {author} {\bibfnamefont {C.}~\bibnamefont {{Doran}}}, \
  and\ \bibinfo {author} {\bibfnamefont {A.}~\bibnamefont {{Lasenby}}},\ }\href
  {\doibase 10.1023/A:1001856702156} {\bibfield  {journal} {\bibinfo  {journal}
  {General Relativity and Gravitation}\ }\textbf {\bibinfo {volume} {32}},\
  \bibinfo {pages} {161} (\bibinfo {year} {2000})},\ \Eprint
  {http://arxiv.org/abs/gr-qc/9910039} {arXiv:gr-qc/9910039 [gr-qc]}
  \BibitemShut {NoStop}%
\bibitem [{\citenamefont {{Lasenby et al.}}(2018)}]{ewgt_conformal}%
  \BibitemOpen
  \bibfield  {author} {\bibinfo {author} {\bibfnamefont {A.~N.}\ \bibnamefont
  {{Lasenby et al.}}},\ }\href@noop {} {\  (\bibinfo {year} {2018})},\ \bibinfo
  {note} {manuscript in preparation: eWGT in Geometric Algebra, With
  Applications to The Conformal Group}\BibitemShut {NoStop}%
\bibitem [{\citenamefont {{Lin}}\ \emph {et~al.}(2020)\citenamefont {{Lin}},
  \citenamefont {{Hobson}},\ and\ \citenamefont {{Lasenby}}}]{Lin3}%
  \BibitemOpen
  \bibfield  {author} {\bibinfo {author} {\bibfnamefont {Y.-C.}\ \bibnamefont
  {{Lin}}}, \bibinfo {author} {\bibfnamefont {M.~P.}\ \bibnamefont {{Hobson}}},
  \ and\ \bibinfo {author} {\bibfnamefont {A.~N.}\ \bibnamefont {{Lasenby}}},\
  }\href@noop {} {\bibfield  {journal} {\bibinfo  {journal} {arXiv e-prints}\
  ,\ \bibinfo {eid} {arXiv:2005.02228}} (\bibinfo {year} {2020})},\ \Eprint
  {http://arxiv.org/abs/2005.02228} {arXiv:2005.02228 [gr-qc]} \BibitemShut
  {NoStop}%
\bibitem [{\citenamefont {{Sezgin}}\ and\ \citenamefont {{van
  Nieuwenhuizen}}(1980)}]{1980PhRvD..21.3269S}%
  \BibitemOpen
  \bibfield  {author} {\bibinfo {author} {\bibfnamefont {E.}~\bibnamefont
  {{Sezgin}}}\ and\ \bibinfo {author} {\bibfnamefont {P.}~\bibnamefont {{van
  Nieuwenhuizen}}},\ }\href {\doibase 10.1103/PhysRevD.21.3269} {\bibfield
  {journal} {\bibinfo  {journal} {Phys. Rev. D}\ }\textbf {\bibinfo {volume}
  {21}},\ \bibinfo {pages} {3269} (\bibinfo {year} {1980})}\BibitemShut
  {NoStop}%
\bibitem [{\citenamefont {{Bernal}}\ \emph {et~al.}(2016)\citenamefont
  {{Bernal}}, \citenamefont {{Verde}},\ and\ \citenamefont
  {{Riess}}}]{2016JCAP...10..019B}%
  \BibitemOpen
  \bibfield  {author} {\bibinfo {author} {\bibfnamefont {J.~L.}\ \bibnamefont
  {{Bernal}}}, \bibinfo {author} {\bibfnamefont {L.}~\bibnamefont {{Verde}}}, \
  and\ \bibinfo {author} {\bibfnamefont {A.~G.}\ \bibnamefont {{Riess}}},\
  }\href {\doibase 10.1088/1475-7516/2016/10/019} {\bibfield  {journal}
  {\bibinfo  {journal} {Journal of Cosmology and Astroparticle Physics}\
  }\textbf {\bibinfo {volume} {2016}},\ \bibinfo {eid} {019} (\bibinfo {year}
  {2016})},\ \Eprint {http://arxiv.org/abs/1607.05617} {arXiv:1607.05617
  [astro-ph.CO]} \BibitemShut {NoStop}%
\bibitem [{\citenamefont {{Efstathiou}}\ and\ \citenamefont
  {{Bond}}(1999)}]{1999MNRAS.304...75E}%
  \BibitemOpen
  \bibfield  {author} {\bibinfo {author} {\bibfnamefont {G.}~\bibnamefont
  {{Efstathiou}}}\ and\ \bibinfo {author} {\bibfnamefont {J.~R.}\ \bibnamefont
  {{Bond}}},\ }\href {\doibase 10.1046/j.1365-8711.1999.02274.x} {\bibfield
  {journal} {\bibinfo  {journal} {Monthly Notices of the RAS}\ }\textbf
  {\bibinfo {volume} {304}},\ \bibinfo {pages} {75} (\bibinfo {year} {1999})},\
  \Eprint {http://arxiv.org/abs/astro-ph/9807103} {arXiv:astro-ph/9807103
  [astro-ph]} \BibitemShut {NoStop}%
\bibitem [{\citenamefont {{Yang}}\ \emph {et~al.}(2020)\citenamefont {{Yang}},
  \citenamefont {{Di Valentino}}, \citenamefont {{Mena}}, \citenamefont
  {{Pan}},\ and\ \citenamefont {{Nunes}}}]{2020arXiv200110852Y}%
  \BibitemOpen
  \bibfield  {author} {\bibinfo {author} {\bibfnamefont {W.}~\bibnamefont
  {{Yang}}}, \bibinfo {author} {\bibfnamefont {E.}~\bibnamefont {{Di
  Valentino}}}, \bibinfo {author} {\bibfnamefont {O.}~\bibnamefont {{Mena}}},
  \bibinfo {author} {\bibfnamefont {S.}~\bibnamefont {{Pan}}}, \ and\ \bibinfo
  {author} {\bibfnamefont {R.~C.}\ \bibnamefont {{Nunes}}},\ }\href@noop {}
  {\bibfield  {journal} {\bibinfo  {journal} {arXiv e-prints}\ ,\ \bibinfo
  {eid} {arXiv:2001.10852}} (\bibinfo {year} {2020})},\ \Eprint
  {http://arxiv.org/abs/2001.10852} {arXiv:2001.10852 [astro-ph.CO]}
  \BibitemShut {NoStop}%
\bibitem [{\citenamefont {{Lucca}}\ and\ \citenamefont
  {{Hooper}}(2020)}]{2020arXiv200206127L}%
  \BibitemOpen
  \bibfield  {author} {\bibinfo {author} {\bibfnamefont {M.}~\bibnamefont
  {{Lucca}}}\ and\ \bibinfo {author} {\bibfnamefont {D.~C.}\ \bibnamefont
  {{Hooper}}},\ }\href@noop {} {\bibfield  {journal} {\bibinfo  {journal}
  {arXiv e-prints}\ ,\ \bibinfo {eid} {arXiv:2002.06127}} (\bibinfo {year}
  {2020})},\ \Eprint {http://arxiv.org/abs/2002.06127} {arXiv:2002.06127
  [astro-ph.CO]} \BibitemShut {NoStop}%
\bibitem [{\citenamefont {{Karwal}}\ and\ \citenamefont
  {{Kamionkowski}}(2016)}]{2016PhRvD..94j3523K}%
  \BibitemOpen
  \bibfield  {author} {\bibinfo {author} {\bibfnamefont {T.}~\bibnamefont
  {{Karwal}}}\ and\ \bibinfo {author} {\bibfnamefont {M.}~\bibnamefont
  {{Kamionkowski}}},\ }\href {\doibase 10.1103/PhysRevD.94.103523} {\bibfield
  {journal} {\bibinfo  {journal} {Phys. Rev. D}\ }\textbf {\bibinfo {volume}
  {94}},\ \bibinfo {eid} {103523} (\bibinfo {year} {2016})},\ \Eprint
  {http://arxiv.org/abs/1608.01309} {arXiv:1608.01309 [astro-ph.CO]}
  \BibitemShut {NoStop}%
\bibitem [{\citenamefont {{Ni}}(2016)}]{2016IJMPS..4060010N}%
  \BibitemOpen
  \bibfield  {author} {\bibinfo {author} {\bibfnamefont {W.-T.}\ \bibnamefont
  {{Ni}}},\ }in\ \href {\doibase 10.1142/S2010194516600107} {\emph {\bibinfo
  {booktitle} {International Journal of Modern Physics Conference Series}}},\
  \bibinfo {series} {International Journal of Modern Physics Conference
  Series}, Vol.~\bibinfo {volume} {40}\ (\bibinfo {year} {2016})\ pp.\ \bibinfo
  {pages} {1660010--146},\ \Eprint {http://arxiv.org/abs/1501.07696}
  {arXiv:1501.07696 [hep-ph]} \BibitemShut {NoStop}%
\bibitem [{\citenamefont {{Puetzfeld}}\ and\ \citenamefont
  {{Obukhov}}(2014)}]{2014IJMPD..2342004P}%
  \BibitemOpen
  \bibfield  {author} {\bibinfo {author} {\bibfnamefont {D.}~\bibnamefont
  {{Puetzfeld}}}\ and\ \bibinfo {author} {\bibfnamefont {Y.~N.}\ \bibnamefont
  {{Obukhov}}},\ }\href {\doibase 10.1142/S0218271814420048} {\bibfield
  {journal} {\bibinfo  {journal} {International Journal of Modern Physics D}\
  }\textbf {\bibinfo {volume} {23}},\ \bibinfo {eid} {1442004} (\bibinfo {year}
  {2014})},\ \Eprint {http://arxiv.org/abs/1405.4137} {arXiv:1405.4137 [gr-qc]}
  \BibitemShut {NoStop}%
\bibitem [{\citenamefont {{Lewis}}\ and\ \citenamefont
  {{Bridle}}(2011)}]{2011ascl.soft06025L}%
  \BibitemOpen
  \bibfield  {author} {\bibinfo {author} {\bibfnamefont {A.}~\bibnamefont
  {{Lewis}}}\ and\ \bibinfo {author} {\bibfnamefont {S.}~\bibnamefont
  {{Bridle}}},\ }\href@noop {} {\enquote {\bibinfo {title} {{CosmoMC:
  Cosmological MonteCarlo}},}\ } (\bibinfo {year} {2011}),\ \Eprint
  {http://arxiv.org/abs/1106.025} {ascl:1106.025} \BibitemShut {NoStop}%
\bibitem [{\citenamefont {{Lesgourgues}}(2011)}]{2011arXiv1104.2932L}%
  \BibitemOpen
  \bibfield  {author} {\bibinfo {author} {\bibfnamefont {J.}~\bibnamefont
  {{Lesgourgues}}},\ }\href@noop {} {\bibfield  {journal} {\bibinfo  {journal}
  {arXiv e-prints}\ ,\ \bibinfo {eid} {arXiv:1104.2932}} (\bibinfo {year}
  {2011})},\ \Eprint {http://arxiv.org/abs/1104.2932} {arXiv:1104.2932
  [astro-ph.IM]} \BibitemShut {NoStop}%
\bibitem [{\citenamefont {{Barker}}\ \emph {et~al.}(2020)\citenamefont
  {{Barker}}, \citenamefont {{Lasenby}}, \citenamefont {{Hobson}},\ and\
  \citenamefont {{Hand ley}}}]{attack_letter}%
  \BibitemOpen
  \bibfield  {author} {\bibinfo {author} {\bibfnamefont {W.~E.~V.}\
  \bibnamefont {{Barker}}}, \bibinfo {author} {\bibfnamefont {A.~N.}\
  \bibnamefont {{Lasenby}}}, \bibinfo {author} {\bibfnamefont {M.~P.}\
  \bibnamefont {{Hobson}}}, \ and\ \bibinfo {author} {\bibfnamefont {W.~J.}\
  \bibnamefont {{Hand ley}}},\ }\href@noop {} {\bibfield  {journal} {\bibinfo
  {journal} {arXiv e-prints}\ ,\ \bibinfo {eid} {arXiv:2006.03581}} (\bibinfo
  {year} {2020})},\ \Eprint {http://arxiv.org/abs/2006.03581} {arXiv:2006.03581
  [gr-qc]} \BibitemShut {NoStop}%
\bibitem [{\citenamefont {{Butcher}}(2019)}]{2019PhRvD.100f3511B}%
  \BibitemOpen
  \bibfield  {author} {\bibinfo {author} {\bibfnamefont {L.~M.}\ \bibnamefont
  {{Butcher}}},\ }\href {\doibase 10.1103/PhysRevD.100.063511} {\bibfield
  {journal} {\bibinfo  {journal} {Phys. Rev. D}\ }\textbf {\bibinfo {volume}
  {100}},\ \bibinfo {eid} {063511} (\bibinfo {year} {2019})},\ \Eprint
  {http://arxiv.org/abs/1810.08616} {arXiv:1810.08616 [gr-qc]} \BibitemShut
  {NoStop}%
\bibitem [{\citenamefont {Penrose}(2006)}]{Penrose:2006zz}%
  \BibitemOpen
  \bibfield  {author} {\bibinfo {author} {\bibfnamefont {R.}~\bibnamefont
  {Penrose}},\ }\bibfield  {booktitle} {\emph {\bibinfo {booktitle} {{Particle
  accelerator. Proceedings, 10th European Conference, EPAC 2006, Edinburgh, UK,
  June 26-30, 2006}}},\ }\href@noop {} {\bibfield  {journal} {\bibinfo
  {journal} {Conf. Proc.}\ }\textbf {\bibinfo {volume} {C060626}},\ \bibinfo
  {pages} {2759} (\bibinfo {year} {2006})},\ \bibinfo {note}
  {[,2759(2006)]}\BibitemShut {NoStop}%
\bibitem [{\citenamefont {{Lasenby et al.}}(2019)}]{Las}%
  \BibitemOpen
  \bibfield  {author} {\bibinfo {author} {\bibfnamefont {A.~N.}\ \bibnamefont
  {{Lasenby et al.}}},\ }\href@noop {} {\  (\bibinfo {year} {2019})},\ \bibinfo
  {note} {manuscript in preparation: Radiation perturbations in a flat-Lambda
  universe}\BibitemShut {NoStop}%
\end{thebibliography}%

\end{document}